\documentclass{econ-paper-ecma}
\providecommand{\sym}[1]{\ifmmode^{#1}\else\(^{#1}\)\fi}
\providecommand{\contact}[1]{}
\providecommand{\thanknotes}[1]{}
\providecommand{\jelcodes}[1]{}
\usepackage{xcolor}
\usepackage{hyperref}
\hypersetup{colorlinks=true, urlcolor=[HTML]{006896},
            linkcolor=black, citecolor=black}

\title{Paying for Progress: Innovation Diffusion Under Regulation}
\author[]{By Zakaria El Amrani and Anastasiia Evdokimova}

\affil[]{Northwestern University, Yale University}

\date{\today\\[2pt]
\href{https://www.dropbox.com/scl/fi/pw0oiv30dxw82kk8k1rcn/TAVR.pdf?rlkey=8h5pwdosetzuamhfxpt6ung6c}{\small Latest version}}

\shortauthor{Z. El Amrani AND A. Evdokimova}
\contact{zakaria.elamrani@kellogg.northwestern.edu, anastasiia.evdokimova@yale.edu}
\thanknotes{We thank David Cutler, David Dranove, Gast\'on Illanes, Molly Schnell, and the Yale IO group, as well as participants at ASHEcon 2026 and the EARIE Annual Conference, for helpful comments and suggestions. We gratefully acknowledge financial support from the Agency for Healthcare Research and Quality (AHRQ) under grant U19HS024072. All errors are our own.}
\abstract{A regulated price breaks the link between the value a technology creates and the return to adopting it. We ask what finances adoption when the price does not reward it, and how a budget-constrained payer should intervene, using transcatheter aortic valve replacement (TAVR) under Medicare. For three years Medicare paid TAVR the rate of the procedure it replaced although it cost more, yet over two fifths of the hospitals that chose to adopt entered then. The mechanism is competition for patients. A TAVR program raises a hospital's value to patients across the risk distribution, more to those likely to receive TAVR but also to those with almost no chance of it, which suggests that patients value more than the procedure, so a program draws patients many of whom receive the surgical procedure. In an adoption model the patients a program draws account for nine tenths of the value of adopting, a fifth of it from patients who receive surgery, and without them half of the hospitals that entered before the payment revision would not have. The policy lesson is to pay at the margin. The rate update is paid on every case at every adopter, most of which enter without it: moving it two years earlier adds 0.9 hospitals for \$95 million. Delaying the rate identifies its inframarginal recipients, the hospitals that enter without it, so at the outlay of the observed revision a delayed rate with a lump sum paid from the revision to predicted non-adopters and a travel voucher reaches 671 hospitals instead of 619, and the voucher with the rise delayed treats 32 percent more TAVR patients for 9 percent less.}
\jelcodes{I110, I180, D83, D7}

\graphicspath{{figures/}{exhibits/fig/}{inputs/}}

\usepackage{booktabs}
\usepackage{threeparttable}

\usepackage[section]{placeins}
\makeatletter
\let\orig@table\table \let\endorig@table\endtable
\renewenvironment{table}[1][]{\orig@table[!htbp]}{\endorig@table}
\let\orig@figure\figure \let\endorig@figure\endfigure
\renewenvironment{figure}[1][]{\orig@figure[!htbp]}{\endorig@figure}
\makeatother

\begin{document}

\vspace*{-52pt}
\maketitle
\section{Introduction}

A regulated price breaks the link between the value a technology creates and the return to adopting it. Where prices are set in the market, the adopter can charge for part of the value it creates, so its return moves with that value and adoption follows profitability \citep{Griliches1957}; competition among adopters also shapes the timing, as when cannibalization delays incumbents' innovation \citep{Igami2017}. Under regulation the payment is set per case, is the same for every provider, and is revised with a lag. A new technology can therefore be paid for years at the rate of the procedure it replaces although it costs more, and while it earns less per case than the procedure it replaces, its adoption must be financed by something other than the per-case premium. We show that competition for patients provides that return: a regulated price fixes what a hospital earns per patient but not how many patients it treats, and offering the new technology raises that number, including patients who receive the old procedure.

The same return determines how a planner should intervene. The level and structure of regulated payment shape which technologies providers adopt \citep{Finkelstein2007, AcemogluFinkelstein2008, Chandra2012Technology, league2023catiii}, and raising the regulated price is the natural instrument for faster diffusion. A planner, however, faces a limited budget and limited information, and whether a higher rate induces adoption or is inframarginal, paid to providers that would have adopted anyway, depends on what each provider earns beyond the payment. The question is therefore not only how much to pay and when, but also whom to pay, hospitals or the patients who travel to them, and in what form; the adoption-subsidy literature finds large differences in cost-effectiveness across these choices \citep{XingLeardLi2021, DeGrooteVerboven2019, Springel2021, LangerLemoine2022}.

This paper asks two questions. First, what finances early adoption when the regulated price does not? Second, can a planner who spends what the rate revision cost reach more hospitals offering the technology, treat more patients, or raise patients' welfare more than the observed policy did? We answer both for transcatheter aortic valve replacement (TAVR) under Medicare. TAVR replaces the aortic valve through a catheter and substitutes for open-heart surgical replacement (SAVR); it costs a hospital more than SAVR, primarily because of the device \citep{reddy2023economic, McCarthy2017Margin}. Medicare covered TAVR from fiscal 2012 but paid roughly the surgical rate until fiscal 2015, when it created separate payment groups for transcatheter valve replacement \citep{Dranove2022TheInnovation}. Three features of the setting fit the two questions. Hospitals could bill TAVR throughout, so for three years it was paid the rate of the procedure it replaced while costing more, and entry in those years reveals a return the payment did not provide. The revision came on a date common to all hospitals, so the response to the rate can be read at hospitals that already operated a program. And 100 percent Medicare inpatient and outpatient claims for 2010--2020 record every hospital's TAVR cases, every diagnosed patient and every patient's choice of hospital, including the choice not to be treated. TAVR is performed almost entirely on patients 65 and older: the median patient in the national registry was 84 in 2011--13 and 80 in 2019 \citep{mack2013outcomes, carroll2020sts}. The claims omit Medicare Advantage patients, but fee-for-service patients are the majority of Medicare's, so a program's first fee-for-service case dates its entry closely. Because Medicare sets both what hospitals receive and what patients pay, neither price varies across hospitals, and travel time is the price a patient pays to reach one.

Consider the first question: what financed entry before the revision? More than two fifths of the hospitals that chose to adopt entered while TAVR was paid the SAVR rate, when a TAVR case earned less than a SAVR case.\footnote{Counting the 97 pivotal-trial sites, which ran TAVR programs before Medicare coverage, half of eventual adopters entered before the revision.} If the payment had held hospitals back, entry would have jumped when the rate rose; it did not. Hospitals respond to the rate, but through the procedure rather than through entry: at the revision, the TAVR volume of the programs already operating in 2012 rose by about 10 percent as they moved patients from SAVR to TAVR. Learning by doing is unlikely to carry early entry on its own. Experience shortens stays but cannot lower the cost of the device, and a hospital with no rival in its HRR, which keeps all of the local volume over which experience pays, enters no earlier than others. Competition for patients can. Patients travel farther to reach a hospital with a program, including patients who then receive SAVR, and a hospital's adoption hazard falls as rivals enter its own patient catchment. Hospitals that already treat most of their catchment enter earlier, which fits both learning and the defensive side of competition for patients.

To measure what competition for patients is worth, we estimate patient demand for hospitals. We model demand for the hospital rather than the procedure, because the heart team, not the patient, assigns the procedure after admission, and with prices fixed by Medicare a patient expresses her preference for an adopter through how far she travels \citep{GaynorPropperSeiler2016}. The value of an adopter varies with the patient's risk, which also governs the chance that the heart team assigns her to TAVR. We estimate a mixed logit with hospital fixed effects and random travel coefficients \citep{Berry1995}. Patients pay for proximity much as in other hospital-choice settings: bypassing the nearest hospital for one 20 minutes farther cuts the odds of choosing it to about a sixth, close to estimates for childbirth \citep{HoPakes2014, RavalRosenbaum2021}, and beyond the nearest hospital extra minutes cost little, as expected for a planned, referral-driven procedure. Measured against that price, the value of an adopter rises with the chance of TAVR and extends to patients who will almost surely have surgery: patients with a predicted chance of TAVR near 7 percent value an adopter at 6 to 20 minutes of travel, depending on the year. That patients with almost no chance of the procedure value an adopter suggests that they value more than the procedure, the hospital or its program, although the estimates do not separate the two. This is the central fact of the paper: a TAVR program raises demand for the hospital, not only for the procedure. Before the rate revision, 68 percent of patients who passed over a nearest hospital without a program to be treated at an adopter received SAVR, and adoption raises a hospital's SAVR volume as well as its TAVR volume. The advantage persists, because a program's value to patients grows with its age, so an early entrant keeps its draw ahead of later rivals.

The demand estimates give the return to adoption; an adoption model gives its cost. Each hospital chooses once when to enter, or never, under perfect foresight of the payment path and given its rivals' entry dates, and a program then assigns each admitted patient to TAVR or SAVR. A hospital can adopt only in years in which it meets a volume requirement modeled on Medicare's coverage rule \citep{CMS2012TAVRNCD}. The model recovers the fixed cost of adoption in dollars, \$6.5M for the median adopter and \$11.0M for the median hospital that never adopts, of the order of industry estimates of the gross cost of a program, about \$1M, and of the hybrid operating room it requires, \$3--5M \citep{tangorra2015roi, ecri2013hybrid, fornell2011hybrid}. The model answers the first question. A program raises its hospital's surgeries as well as its TAVR cases, so the value of adopting is the value of patients the hospital would not otherwise have treated. Without them the value of adopting falls by nine tenths, and predicted entry before the revision falls by half; the cost profile that favors 2012 accounts for less of the first wave than competition for patients does. Patients drawn to surgery account for a fifth of the value and a fifth of the 2012 wave; most of the return comes from patients a program brings in for TAVR.

These results move the question from mechanism to policy. Because the return to adoption runs through patients, a planner can act on it directly, and the rate is not the only lever. We trace the policy frontier over three instruments, at the outlay of the observed revision and at half to twice it, counting what the instruments pay above the unrevised schedule, with every policy evaluated at the Nash equilibrium in adoption dates it induces. The first is the timing of the rate increase, which acts on the procedure: an earlier update shifts patients from SAVR to TAVR at programs already operating, but a hospital whose fixed cost exceeds its return at the revised rate stays out whenever that rate arrives. The second is a lump sum paid at the update to hospitals predicted not to enter. The delayed rate has already separated hospitals with a return to early entry from the rest, so the transfer can avoid inframarginal payments to hospitals that would enter without it. The third acts on the mechanism itself: a travel voucher for patients who receive TAVR enlarges the pool an adopter can draw on, so it raises the return to adoption through demand, and it treats patients at existing programs rather than waiting for a local hospital to adopt. In equilibrium each hospital's entry responds to its rivals' entry under the policy.

The policy lesson is to pay at the margin. At the observed outlay of \$506M the observed schedule leaves 619 hospitals offering TAVR by 2019 and 19.5 thousand TAVR patients a year. The rate rise is almost entirely inframarginal for entry, because it is paid on every case at every adopter: advancing it by two years adds 0.9 hospitals for \$95M, about \$106M per hospital, and delaying it by two years saves \$169M and loses 2.7 hospitals. The money the delay saves buys more elsewhere. At the observed outlay, the rise delayed to 2018 with a \$9 million lump sum to predicted non-adopters and the voucher reaches 671 hospitals, 52 more, for \$489M. Each instrument also works alone, with the rise left in 2015: a lump sum to predicted non-adopters adds 24 to 54 hospitals at \$7--10M each, and the voucher adds 40 hospitals at about \$5M each. Patients come from the voucher, most of them patients who would otherwise not have been treated: with the rise delayed to 2017 and no lump sum it treats 25.7 thousand TAVR patients a year, 6.2 thousand or 32 percent more, for \$460M, 9 percent less than the observed outlay. At half the outlay, the rise in 2018 with a \$2 million lump sum keeps coverage within one hospital of the observed policy. Patients' welfare, which the demand model measures in travel time and so only as access, changes by at most \$0.75 million under the best policies.

\textit{Related literature.} The paper contributes to three literatures. The first studies how regulated reimbursement shapes hospitals' technology choices \citep{Finkelstein2007, AcemogluFinkelstein2008, Chandra2012Technology, Dranove2022TheInnovation} and how sharply providers respond to it \citep{Dafny2005PriceChanges, ClemensGottlieb2014, Eliason2018LTCH, EinavFinkelsteinMahoney2018, AlexanderSchnell2024, CabralCareyMiller2025}. Two papers in this literature study TAVR. \citet{yang2023experience} estimates a dynamic equilibrium model of learning by doing, hospital choice and adoption, and finds that Medicare's volume requirements for TAVR programs attain the consumer welfare of unrestricted entry at lower aggregate fixed cost, because the hospitals they exclude are those patients value least. \citet{Hoagland} shows that as cardiologists adopt TAVR they perform fewer coronary interventions, a loss concentrated among disadvantaged patients. In both, the return to a program and its spillovers run through the procedure, through experience with it in the first and through competition for operator capacity in the second. A clinical literature documents volume and outcomes, physician incentives, geographic access and hospital margins in TAVR \citep{Vemulapalli2019Volume, groeneveld2020effects, Strobel2019Competition, Nathan2021Programs, Damluji2020Rural, Benharash2025Competition, McCarthy2017Margin, reddy2025contribution}; closest to our steering result, \citet{Brennan2014Availability} find that surgical valve replacements rose by 22 percent at hospitals with TAVR programs over 2008--2013, mostly among low- and intermediate-risk patients. We study the margin this work leaves open, adoption while the regulated price paid TAVR no more than the procedure it replaces, and show that the return runs through the hospital: adoption raises demand even among patients with little chance of receiving TAVR, while learning, which shortens stays but does not lower the cost of the device, does not account for early entry on its own.

The second is non-price competition and structural models of adoption. Hospitals under regulated prices compete on quality and technology \citep{RobinsonLuft1985, Dranove1992Wasteful, KesslerMcClellan2000, Cooper2011Competition, GaynorMorenoSerraPropper2013, Bloom2015Competition, GaynorHoTown2015}, and new technologies reallocate patients toward adopters \citep{Aggarwal2017Choice, Chandra2016Exceptionalism, Horn2022RoboticSurgery}. The closest paper in method is \citet{schmidt2006timing}, who estimates a dynamic game of MRI adoption among US hospitals and decomposes the effect of competition on adoption timing into business stealing and preemption, of which business stealing is the larger. There business stealing is recovered from adoption timing and market structure. We measure it from patient choice, linking the hospital-choice literature \citep{Capps2003Option, HoPakes2014, GaynorPropperSeiler2016} to models of adoption and entry timing \citep{Reinganum1981, FudenbergTirole1985, AckerbergGowrisankaran2006, Ryan2012, Dunne2013Entry, Igami2017}, and show that under a regulated price business stealing is not only a cost competition imposes but the return that finances early adoption. The demand response extends to patients with little chance of receiving the new procedure, so the hospital-level return to innovation exceeds the value generated by its use.

The third is adoption-subsidy design, where the central problem is inframarginal spending and the form, targeting and timing of a subsidy determine its cost-effectiveness \citep{XingLeardLi2021, DeGrooteVerboven2019, SheldonDua2019, HughesPodolefsky2015, Springel2021, LangerLemoine2022}. We show that a delayed rate screens hospitals by private return, in the spirit of screening by ordeal and regulation under asymmetric information \citep{NicholsZeckhauser1982, Besley1992Workfare, BaronMyerson1982, LaffontTirole1993}, and quantify the adoption per dollar that targeted transfers buy relative to an earlier increase in the regulated price.

The remainder of the paper proceeds as follows. Section~\ref{data} describes the institutional setting and Section~\ref{sec:data} the data. Sections~\ref{sec:two-puzzles}--\ref{sec:puzzle2-tests} document the descriptive evidence and develop the two-stage model of entry under a delayed rate. Section~\ref{sec:demand} estimates patient demand for hospitals and Section~\ref{sec:demand-results} reports the estimates. Section~\ref{sec:dynamic_model} presents the adoption-timing model, Section~\ref{sec:counterfactuals} the planner's problem and the policy comparison, and Section~\ref{sec:conclusion} concludes.

\section{Institutional setting}\label{data}

Aortic stenosis is a narrowing of the aortic valve that, once severe and symptomatic, is fatal within a few years unless the valve is replaced. For decades the only treatment was surgical aortic valve replacement (SAVR), an open-heart operation under cardiopulmonary bypass, which many elderly or frail patients cannot withstand. Transcatheter aortic valve replacement (TAVR) delivers a replacement valve through a catheter, usually via the femoral artery, without opening the chest. Randomized trials established its benefit in the order of surgical risk: TAVR reduced mortality relative to medical management for patients who could not undergo surgery \citep{Leon2010TAVI}, matched surgery for high-risk patients \citep{Smith2011TAVR}, and later matched it for intermediate-risk patients \citep{Reardon2017TAVR}. It also shortens recovery: in our data the mean length of stay is about five days for TAVR against ten for SAVR (Table~\ref{tab:patient_chars}). The FDA approved the first device in late 2011 and Medicare covered the procedure from fiscal 2012; fiscal 2011 activity is confined to clinical trial sites.

\begin{figure}[htb]
    \centering
    \includegraphics[width=0.85\linewidth]{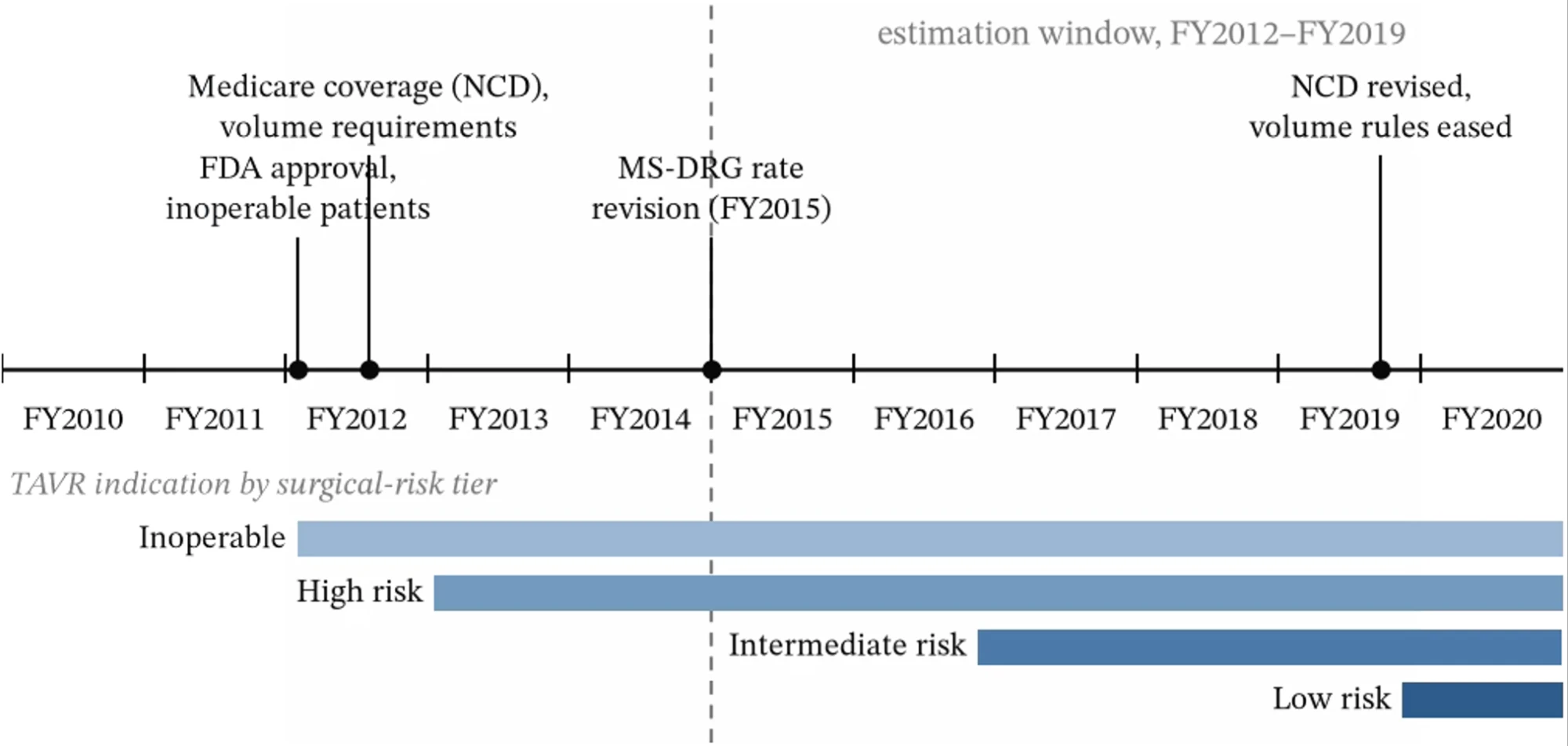}
    \caption{Coverage, coding and payment for TAVR under Medicare, fiscal years 2011--2019. The procedure was billable from the coverage decision under the payment groups for cardiac valve procedures; separate groups for endovascular valve replacement (MS-DRG 266/267) with a higher payment took effect in fiscal 2015.}
    \label{fig:timeline}
\end{figure}

TAVR costs a hospital more than SAVR, largely because of the valve, and Medicare's payment did not reflect this for three years. Hospitals could code and bill the procedure from the moment Medicare covered it, because TAVR was assigned to the existing payment groups for cardiac valve procedures, MS-DRGs 216--221. The payment attached to those groups reflected the surgical procedures already in them and not the cost of a transcatheter valve, so hospitals were reimbursed at approximately the surgical rate until fiscal 2015, when Medicare created separate payment groups for endovascular valve replacement, MS-DRGs 266 and 267, with a higher payment \citep{CMS2014IPPSFinal, Dranove2022TheInnovation}. Fiscal years 2012 through 2014 are therefore three years in which TAVR was billable and paid at about the surgical rate although it cost more, so a TAVR case earned a lower margin than a SAVR case; physicians, who billed under a dedicated CPT code, were paid for the procedure \citep{reddy2023economic, McCarthy2017Margin}. That the constraint was price and not coding matters for what follows: hospitals that did not adopt in this window and met the coverage rule's start-up requirements were able to bill for the procedure and chose not to perform it, so the window reveals which hospitals were willing to adopt before per-case margins cleared their fixed cost. Figure~\ref{fig:timeline} summarizes the timing.

Who could receive TAVR was set by Medicare coverage, which follows the FDA indication and expanded by surgical-risk tier: inoperable or extreme-risk patients from fiscal 2012 (Edwards SAPIEN), high-risk patients from fiscal 2013, intermediate-risk patients from the end of fiscal 2016 (SAPIEN~3) and in fiscal 2017 (Evolut), and low-risk patients from the end of fiscal 2019. Risk is measured by the Society of Thoracic Surgeons (STS) predicted operative mortality. We assign each fiscal year the threshold in force for most of it, 8 percent through fiscal 2016 and 4 percent from fiscal 2017; fiscal 2016 and 2019 end with an expansion, so in their last months the assigned threshold overstates the risk required for coverage. The indication bounds who can receive TAVR but does not determine who does: within it the heart team judges operability on criteria the score omits, and the score we reconstruct from claims measures risk with error, so some patients below the threshold receive TAVR (Table~\ref{tab:tavr_joint}). Two consequences follow. The pool of patients who could receive TAVR at a given hospital changed over the sample, but on dates common to all hospitals, so the change is not confounded with any hospital's adoption date. And the chance of receiving TAVR differed widely across patients at every date: in 2012--2014 most AVR patients had little chance of it, and as coverage widened the chance rose, mostly for higher-risk patients; Section~\ref{sec:demand} predicts this chance from each patient's age, risk score and conditions and relates it to the value patients place on a hospital with a program.

Finally, Medicare's national coverage determination conditions payment on program requirements \citep{CMS2012TAVRNCD, CMS2019TAVRNCD}. Until 2019 a hospital opening a program needed at least 50 aortic valve replacements in the prior year, including 10 high-risk patients, at least 1,000 catheterizations including 400 PCIs, and two surgeons with cardiac privileges, and an operating program had to sustain 20 AVRs a year; the 2019 revision lowered the requirements for new programs. The start-up requirement can bind for small hospitals. The adoption model of Section~\ref{sec:dynamic_model} lets a hospital adopt in a year only if it performed at least ten fee-for-service AVRs the year before, a lower bar than the rule's 50 because fee-for-service claims omit Medicare Advantage and other payers; the screen captures the volume part of the rule, while its staffing and procedure requirements are not observed and are absorbed in the adoption cost; and the counterfactuals of Section~\ref{sec:counterfactuals} offer a lump sum only to hospitals that would treat at least ten TAVR patients a year.

\section{Data}\label{sec:data}

The paper needs four elements from the data: every hospital's adoption date and volume, to study entry; every patient's diagnosis and choice of hospital, including the choice not to be treated, to estimate demand; hospital costs and Medicare payments, to value adoption in dollars; and the characteristics of hospitals and patients that enter the three models. Most of these come from 100 percent Medicare fee-for-service claims for fiscal years 2010--2020: the Inpatient file, which records for every stay the beneficiary's residential ZIP code, demographics, diagnoses and procedures, length of stay, total charges, DRG and Medicare payment, and a hospital identifier; the Outpatient file, which we use only to identify diagnosed patients who are not treated; and the Master Beneficiary Summary File. All aortic valve replacements for Medicare beneficiaries in this period occur in inpatient settings, and TAVR is performed almost entirely on patients aged 65 and older \citep{mack2013outcomes, carroll2020sts}, so the claims cover nearly every procedure outside Medicare Advantage. We supplement them with hospital characteristics from the American Hospital Association Annual Survey and the CMS Provider of Services file (beds, teaching status, ownership, system membership, location), with hospital-year cost-to-charge ratios from the Hospital Cost Reports, and with driving times from ZIP-code centroids to hospitals, computed for every patient--hospital pair used in the demand model.

An AVR episode is an inpatient stay with a TAVR or SAVR procedure code.\footnote{TAVR: ICD-9-CM 35.05 and 35.06; ICD-10-PCS 02RF38H, 02RF38Z, 02RF3KH, 02RF3KZ, 02RF37H, 02RF37Z, 02RF3JH and 02RF3JZ. SAVR: ICD-9-CM 35.21 and 35.22; ICD-10-PCS 02RF07Z, 02RF08Z, 02RF0JZ and 02RF0KZ.} A hospital's adoption year is the first fiscal year in which it performs at least one TAVR procedure, and TAVR and SAVR volumes are the number of procedures with each code in a hospital-year, including patients who die in hospital. Because the procedure codes distinguish the transcatheter from the surgical approach throughout the period, adoption and volume are measured consistently on both sides of the fiscal 2015 payment revision. The sample of hospitals is short-stay general acute-care hospitals that perform AVR, after excluding those with sporadic or incomplete cost-reporting records: $946$ hospitals, which we use throughout the descriptive evidence. The adoption-timing model of Section~\ref{sec:dynamic_model} uses $945$ of them: $97$ pivotal-trial sites that performed TAVR before fiscal 2012, and $848$ hospitals that choose when to enter, of which $521$ adopt by 2019. One of the $946$ has valve replacements only outside fiscal 2012--2019. The demand model is estimated on $932$ hospitals: $931$ of the $945$, and that one, which patients in the market still choose. The other $14$ sample hospitals have no driving time in the distance file, so they enter no patient's choice set, and their patients are among those dropped in Section~\ref{sec:demand} because the chosen hospital lies outside the set.

Table~\ref{tab:hospital_chars} describes the hospitals. They are large, a mean of $434$ beds; most are teaching ($70$ percent) and nonprofit ($70$ percent); and the typical hospital performed about $38$ SAVR cases a year before TAVR arrived. Markets are not thin: the average hospital shares its referral region with five other sample hospitals and has eight others within an hour's drive, so for most hospitals adoption takes place in the presence of rivals that patients can reach. Section~\ref{sec:descriptive} reports these characteristics by adoption cohort.


\begin{table}[htb]
\centering
\small
\caption{Hospital characteristics}
\label{tab:hospital_chars}
\begin{threeparttable}
\begin{tabular}{lc}
\toprule
 & All hospitals \\
\midrule
Hospitals (N) & 946 \\
Beds & 434 \\
Teaching (share) & 0.70 \\
Government-owned (share) & 0.10 \\
Nonprofit (share) & 0.70 \\
For-profit (share) & 0.16 \\
Ownership missing (share) & 0.05 \\
SAVR cases per year, FY2010--11 & 38 \\
Sample hospitals in the HRR & 6 \\
Sample hospitals within 60 minutes & 8 \\
\bottomrule
\end{tabular}
\begin{tablenotes}[flushleft]\footnotesize
\item \textit{Notes:} Means unless noted. Pre-period volumes are annual averages over FY2010--2011. Hospitals within 60 minutes counts other sample hospitals reachable within 60 minutes of driving from the hospital's ZIP.
\end{tablenotes}
\end{threeparttable}
\end{table}

Patients enter the analysis in two ways. For the descriptive evidence the unit is the AVR episode. For the demand model, and through it for the supply model, the unit is the patient at risk, because a diagnosed patient can choose not to be treated, and the outside option is what lets the model count patients that adoption brings into treatment rather than only patients it moves between hospitals. The population at risk in a fiscal year is therefore every fee-for-service beneficiary aged 65 or older with a diagnosis of aortic valve disease that year, defined as at least one inpatient claim or at least two outpatient claims carrying the diagnosis,\footnote{ICD-9 424.1 and ICD-10 I35, which cover all nonrheumatic aortic valve disease, not only severe symptomatic stenosis. The outside option absorbs the breadth of the pool in levels, but the pool includes patients with no indication for replacement, who would not respond to access, so the model may overstate how many untreated patients a new program brings into treatment.} together with every beneficiary treated that year. Over fiscal years 2012 to 2019 this population contains 2.70 million patient-years, of which 14 percent receive a valve replacement in the year, a share that rises from 11 percent in 2012 to 15 percent in 2019. Table~\ref{tab:patient_chars} describes the population.

\begin{table}[htb]
\centering
\small
\caption{The diagnosed population and treated patients}
\label{tab:patient_chars}
\begin{threeparttable}
\begin{tabular*}{\linewidth}{@{\extracolsep{\fill}} l r r r @{}}
\toprule
 & Untreated & SAVR & TAVR \\
\midrule
Patient-years & 2,325,816 & 208,505 & 161,787 \\
Age, mean & 81.8 & 74.0 & 81.8 \\
Female (percent) & 53.4 & 37.0 & 46.9 \\
\addlinespace
\multicolumn{4}{@{}l}{\textit{STS predicted mortality (percent)}} \\
\quad Mean & 3.24 & 1.72 & 3.31 \\
\quad Median & 2.45 & 1.44 & 2.73 \\
\quad Interquartile range & 1.46--4.26 & 0.97--2.08 & 1.82--4.35 \\
\quad Below 2 (percent) & 41.2 & 72.8 & 30.6 \\
\quad 2--4 (percent) & 30.8 & 22.4 & 39.5 \\
\quad 4--8 (percent) & 22.6 & 4.6 & 26.5 \\
\quad 8 or above (percent) & 5.4 & 0.1 & 3.3 \\
\addlinespace
\multicolumn{4}{@{}l}{\textit{Conditions in the prior twelve months (percent)}} \\
\quad Balloon aortic valvuloplasty & 0.5 & 0.4 & 5.2 \\
\quad Coronary bypass surgery & 15.8 & 6.8 & 20.4 \\
\quad Home oxygen or chronic respiratory failure & 16.7 & 5.1 & 13.2 \\
\quad Peripheral artery disease & 17.3 & 11.0 & 21.6 \\
\quad Aortic atherosclerosis & 4.9 & 7.9 & 17.0 \\
\quad Skilled-nursing or home-health use & 50.8 & 10.7 & 28.1 \\
\quad Dialysis or CKD stage 4--5 & 10.5 & 4.7 & 8.9 \\
\quad Prior valve surgery & 6.9 & 5.2 & 8.4 \\
\quad Any of the eight & 70.1 & 35.9 & 65.4 \\
\quad Two or more & 35.1 & 11.7 & 35.6 \\
\addlinespace
\multicolumn{4}{@{}l}{\textit{Travel time (minutes)}} \\
\quad Nearest hospital, median & 14.2 & 15.5 & 13.8 \\
\quad Nearest TAVR hospital, median & 20.9 & 23.5 & 17.6 \\
\quad Treating hospital, median & & 25.1 & 27.0 \\
\quad Treating hospital, mean & & 46.8 & 51.6 \\
\quad Treated at the nearest hospital (percent) & & 42.7 & 33.1 \\
\quad Treated at a TAVR hospital (percent) & & 75.4 & 100.0 \\
\bottomrule
\end{tabular*}
\begin{tablenotes}[flushleft]\footnotesize
\item \textit{Notes:} Demand estimation sample, fiscal years 2012--2019: fee-for-service beneficiaries aged 65 or older diagnosed with aortic valve disease, and every beneficiary treated with TAVR or SAVR, one observation per patient-year. STS predicted mortality from the age--sex--comorbidity grid. Conditions are read from inpatient, outpatient, DME, home-health and skilled-nursing claims in the 365 days before the AVR admission; for untreated patients the window ends on 30 September of the fiscal year, so it also covers care after the diagnosis. TAVR hospital: a hospital operating a TAVR program in that year; the nearest one is among the patient's 50 nearest hospitals. Travel time is driving time from the residential ZIP code.
\end{tablenotes}
\end{threeparttable}
\end{table}

Coverage, and hence the chance that a patient receives TAVR, was defined by surgical risk, so we approximate the STS predicted operative mortality from its key inputs. The claims do not carry the score, but they carry these inputs: predicted mortality is read from the STS risk grid, a table of predicted operative mortality by age, sex, dialysis, hypertension and heart failure with $1{,}600$ cells ranging from $0.3$ to $30$ percent, and each patient is matched to a cell on age, sex and the three comorbidity flags, read from the diagnosis codes on the beneficiary's inpatient and outpatient claims in the same fiscal year.

The score alone does not determine who receives TAVR. The heart team assesses operability on conditions the score omits, and the pivotal trials defined inoperability partly by them: a calcified (porcelain) aorta, prior sternotomy, frailty and chest radiation \citep{Leon2010TAVI}. We therefore record eight conditions from each patient's claims in the twelve months before the admission, or before the end of the fiscal year for untreated patients: prior balloon aortic valvuloplasty, prior coronary bypass surgery, home oxygen or chronic respiratory failure, peripheral artery disease, aortic atherosclerosis, the claims counterpart of a porcelain aorta, skilled-nursing or home-health use, a marker of frailty, dialysis or advanced chronic kidney disease, and prior valve surgery; the last two were exclusions in the first trials.\footnote{Codes, ICD-9 / ICD-10: valvuloplasty 35.96 / 027F, CPT 92986; coronary bypass V45.81, 414.02--414.05, 36.1 / Z95.1, I25.7, 021; home oxygen or respiratory failure V46.2, 518.83 / Z99.81, J96.1, and oxygen-equipment HCPCS codes; peripheral artery disease 440.2, 443.8--443.9, 447.1 / I70.2, I73.9, I77.1; aortic atherosclerosis 440.0 / I70.0; skilled-nursing or home-health use, any claim of either type; dialysis or chronic kidney disease stage 4--5 V45.11, V56, 585.4--585.6 / Z99.2, Z49, N18.4--N18.6; prior valve surgery V43.3, 35.1, 35.2 / Z95.2--Z95.4, 02RF--02RJ.}

Table~\ref{tab:tavr_predictors} shows that these conditions carry information the score does not. Among treated patients, year effects and STS tiers explain between 19 and 25 percent of the variation in modality within each policy period; adding the eight conditions raises this to between 29 and 35 percent and leaves the STS coefficients almost unchanged, so the conditions are close to orthogonal to the score. They enter with the signs the clinical criteria imply. Over 2012--2019, prior valvuloplasty and prior bypass surgery each raise the probability of TAVR by 26 percentage points, home oxygen by 13, aortic atherosclerosis by 10, skilled-nursing use by 9 and peripheral artery disease by 8, while dialysis lowers it by 6. Prior valve surgery has no effect before fiscal 2015 and raises the probability by 2 points afterwards, when TAVR inside a failed surgical valve became available. Within the score, the gradient flattens above 4 percent as the indication widens: relative to patients at 4 to 8 percent, those above 8 percent are 28 points more likely to receive TAVR in 2012--2014 and 5 points in 2017--2019. Table~\ref{tab:tavr_joint} shows the same pattern without functional form: among patients treated in 2012--2014 with a score below 2 percent, 3 percent of those with none of the conditions received TAVR, against 29 percent of those with two or more; by 2017--2019 the two shares are 28 and 72 percent.

\begin{table}[htb]
\centering
\small
\caption{Who receives TAVR: risk score and conditions}
\label{tab:tavr_predictors}
\begin{threeparttable}
\begin{tabular*}{\linewidth}{@{\extracolsep{\fill}} l cc cc cc cc @{}}
\toprule
 & \multicolumn{2}{c}{2012--2014} & \multicolumn{2}{c}{2015--2016} & \multicolumn{2}{c}{2017--2019} & \multicolumn{2}{c}{2012--2019} \\
\cmidrule(lr){2-3}\cmidrule(lr){4-5}\cmidrule(lr){6-7}\cmidrule(lr){8-9}
 & (1) & (2) & (3) & (4) & (5) & (6) & (7) & (8) \\
\midrule
STS below 2 & $-45.9$ & $-39.4$ & $-61.8$ & $-56.7$ & $-53.7$ & $-51.1$ & $-53.4$ & $-49.2$ \\
 & (0.4) & (0.4) & (0.4) & (0.4) & (0.2) & (0.2) & (0.2) & (0.2) \\
STS 2--4 & $-27.5$ & $-24.5$ & $-27.2$ & $-26.3$ & $-15.7$ & $-15.7$ & $-22.2$ & $-21.2$ \\
 & (0.5) & (0.5) & (0.4) & (0.4) & (0.2) & (0.2) & (0.2) & (0.2) \\
STS 8 or above & 28.3 & 24.5 & 12.8 & 11.3 & 4.7 & 4.7 & 11.1 & 10.0 \\
 & (1.2) & (1.2) & (0.6) & (0.7) & (0.3) & (0.3) & (0.3) & (0.4) \\
\addlinespace
Valvuloplasty & & 38.4 & & 26.3 & & 17.1 & & 26.5 \\
 & & (0.7) & & (0.7) & & (0.5) & & (0.4) \\
Coronary bypass & & 22.9 & & 30.6 & & 25.2 & & 26.0 \\
 & & (0.4) & & (0.4) & & (0.3) & & (0.2) \\
Home oxygen & & 13.1 & & 13.9 & & 11.7 & & 12.8 \\
 & & (0.5) & & (0.5) & & (0.4) & & (0.2) \\
Peripheral artery disease & & 7.6 & & 8.5 & & 7.6 & & 7.9 \\
 & & (0.3) & & (0.4) & & (0.3) & & (0.2) \\
Aortic atherosclerosis & & 11.6 & & 10.9 & & 9.0 & & 10.2 \\
 & & (0.4) & & (0.4) & & (0.3) & & (0.2) \\
Skilled-nursing use & & 9.8 & & 10.3 & & 7.6 & & 9.0 \\
 & & (0.3) & & (0.4) & & (0.3) & & (0.2) \\
Dialysis or CKD 4--5 & & $-4.0$ & & $-6.5$ & & $-7.6$ & & $-6.3$ \\
 & & (0.5) & & (0.6) & & (0.4) & & (0.3) \\
Prior valve surgery & & $-0.7$ & & 2.2 & & 2.3 & & 1.6 \\
 & & (0.5) & & (0.6) & & (0.4) & & (0.3) \\
\addlinespace
Year effects & Yes & Yes & Yes & Yes & Yes & Yes & Yes & Yes \\
$R^2$ & 0.194 & 0.319 & 0.249 & 0.347 & 0.224 & 0.285 & 0.319 & 0.392 \\
Patients & \multicolumn{2}{c}{111,778} & \multicolumn{2}{c}{95,543} & \multicolumn{2}{c}{162,971} & \multicolumn{2}{c}{370,292} \\
\bottomrule
\end{tabular*}
\begin{tablenotes}[flushleft]\footnotesize
\item \textit{Notes:} Linear probability models of receiving TAVR among patients treated with TAVR or SAVR, in percentage points. The omitted STS tier is 4--8 percent. Heteroskedasticity-robust standard errors in parentheses. Conditions are defined in Table~\ref{tab:patient_chars}.
\end{tablenotes}
\end{threeparttable}
\end{table}

\begin{table}[htb]
\centering
\small
\caption{Share of treated patients receiving TAVR, by risk score and number of conditions (percent)}
\label{tab:tavr_joint}
\begin{threeparttable}
\begin{tabular*}{\linewidth}{@{\extracolsep{\fill}} l ccc ccc ccc @{}}
\toprule
 & \multicolumn{3}{c}{2012--2014} & \multicolumn{3}{c}{2015--2016} & \multicolumn{3}{c}{2017--2019} \\
\cmidrule(lr){2-4}\cmidrule(lr){5-7}\cmidrule(lr){8-10}
STS (percent) & None & One & Two+ & None & One & Two+ & None & One & Two+ \\
\midrule
Below 2 & 2.9 & 8.8 & 29.4 & 10.5 & 26.3 & 53.4 & 27.8 & 48.4 & 71.5 \\
2--4 & 13.1 & 26.6 & 52.0 & 41.1 & 57.9 & 75.9 & 70.5 & 79.4 & 87.9 \\
4--8 & 40.0 & 54.0 & 71.0 & 78.0 & 82.5 & 89.1 & 93.3 & 93.8 & 94.4 \\
8 or above & 75.3 & 82.3 & 88.5 & 96.6 & 95.7 & 96.4 & $>98$ & $>98$ & 98.2 \\
\addlinespace
All & 8.5 & 20.8 & 46.7 & 25.8 & 46.9 & 71.1 & 47.6 & 68.0 & 83.9 \\
\bottomrule
\end{tabular*}
\begin{tablenotes}[flushleft]\footnotesize
\item \textit{Notes:} Patients treated with TAVR or SAVR. Columns count how many of the eight conditions of Table~\ref{tab:patient_chars} the patient carries. Cells with fewer than 11 SAVR patients are reported as bounds.
\end{tablenotes}
\end{threeparttable}
\end{table}

We therefore describe a patient by the STS score and the eight conditions wherever risk enters: in the demand model of Section~\ref{sec:demand}, where both shift the value of a TAVR hospital, and in the patient types of Section~\ref{sec:dynamic_model}. Only to describe the demand results, Section~\ref{sec:demand} summarizes the score and the conditions by each patient's predicted chance of receiving TAVR at an adopter, from a logit of TAVR on age, the risk score and the eight conditions estimated year by year among patients treated at adopters; the chance does not enter demand, because it reflects the heart team's selection as well as the patient. It is this chance, rather than the STS threshold in force, that identifies patients with a negligible chance of the procedure. For each patient we also compute driving time from the residential ZIP-code centroid to each hospital in the consideration set; travel time, not administrative market boundaries, defines choice sets, since about a third of AVR patients are treated outside their home hospital referral region.

Table~\ref{tab:patient_chars} describes the diagnosed population and the treated. TAVR patients are older than SAVR patients ($81.8$ against $74.0$ years), more often women, and riskier, with a median STS score of $2.7$ percent against $1.4$, as the risk-tiered indication implies; two thirds carry at least one of the eight conditions, against a third of SAVR patients. The untreated are as old as TAVR patients and have a similar distribution of the score. Despite the difference in case mix, the DRG weight and the Medicare payment pooled over 2012--2019 are almost identical across the two procedures, about \$50,000 per case (Appendix Table~\ref{tab:episode_outcomes}); the payments by year below show the delay described in Section~\ref{data} and the gap the revision opened; TAVR patients stay half as long ($4.9$ against $10.0$ days) and die in hospital half as often. Patients travel for aortic valve replacement: the mean trip to the treating hospital is about 50 minutes, against a median of 14 minutes to the nearest hospital, and two thirds of TAVR patients and more than half of SAVR patients are treated elsewhere than at their nearest hospital. Three quarters of SAVR patients are treated at a hospital that offers TAVR. The question for the demand model is how much of that travel adoption redirects.

\begin{figure}[htb]
    \centering
    \includegraphics[width=0.72\linewidth]{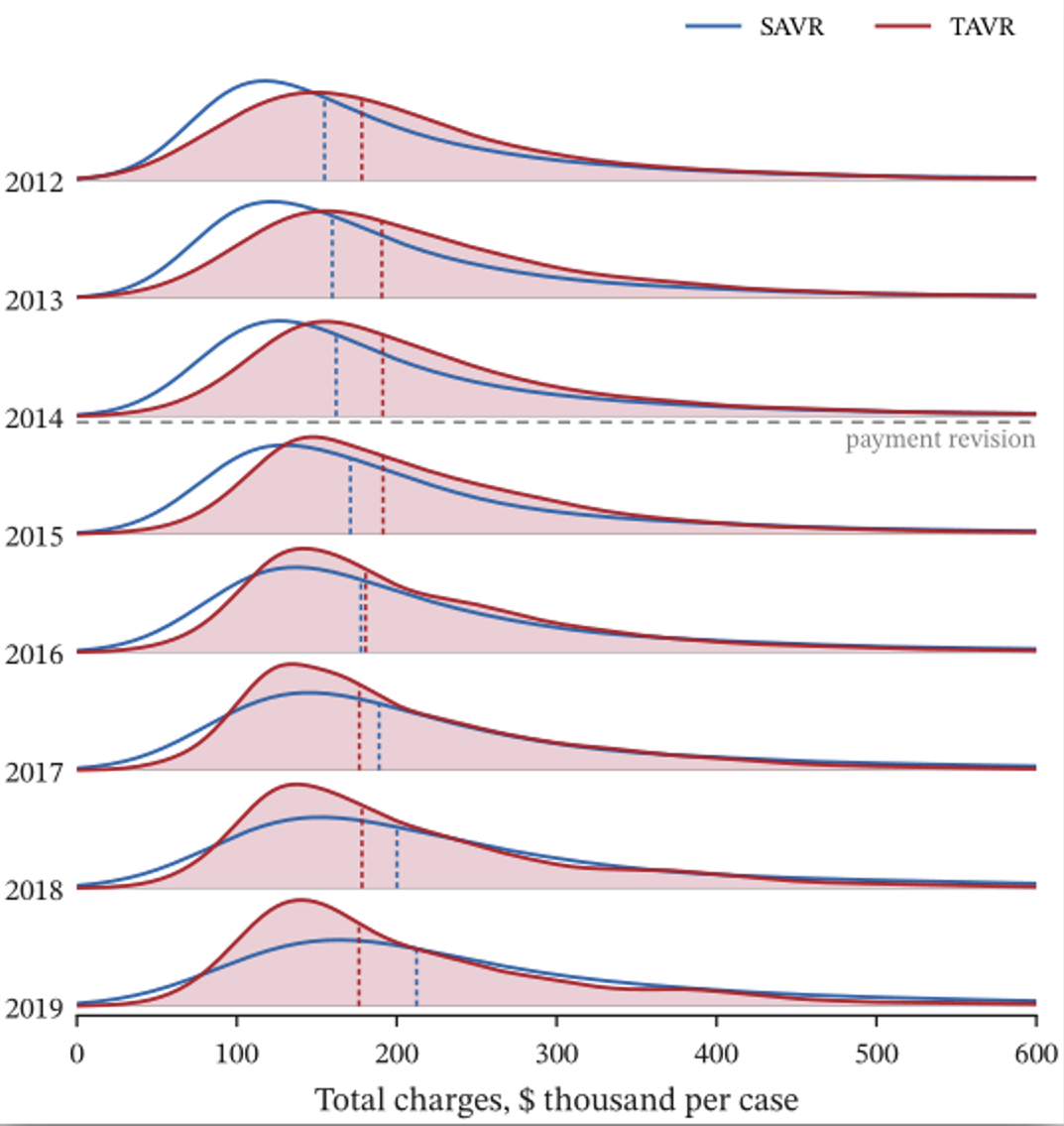}
    \caption{Total charges per case for TAVR and SAVR by fiscal year: kernel densities, one density scale for all years. TAVR shaded, SAVR as a line; dotted ticks mark the medians; the dashed rule marks the fiscal 2015 payment revision. Source: Medicare fee-for-service AVR episodes, FY2012--2019.}
    \label{fig:charges_density}
\end{figure}

\begin{figure}[htb]
    \centering
    \includegraphics[width=0.72\linewidth]{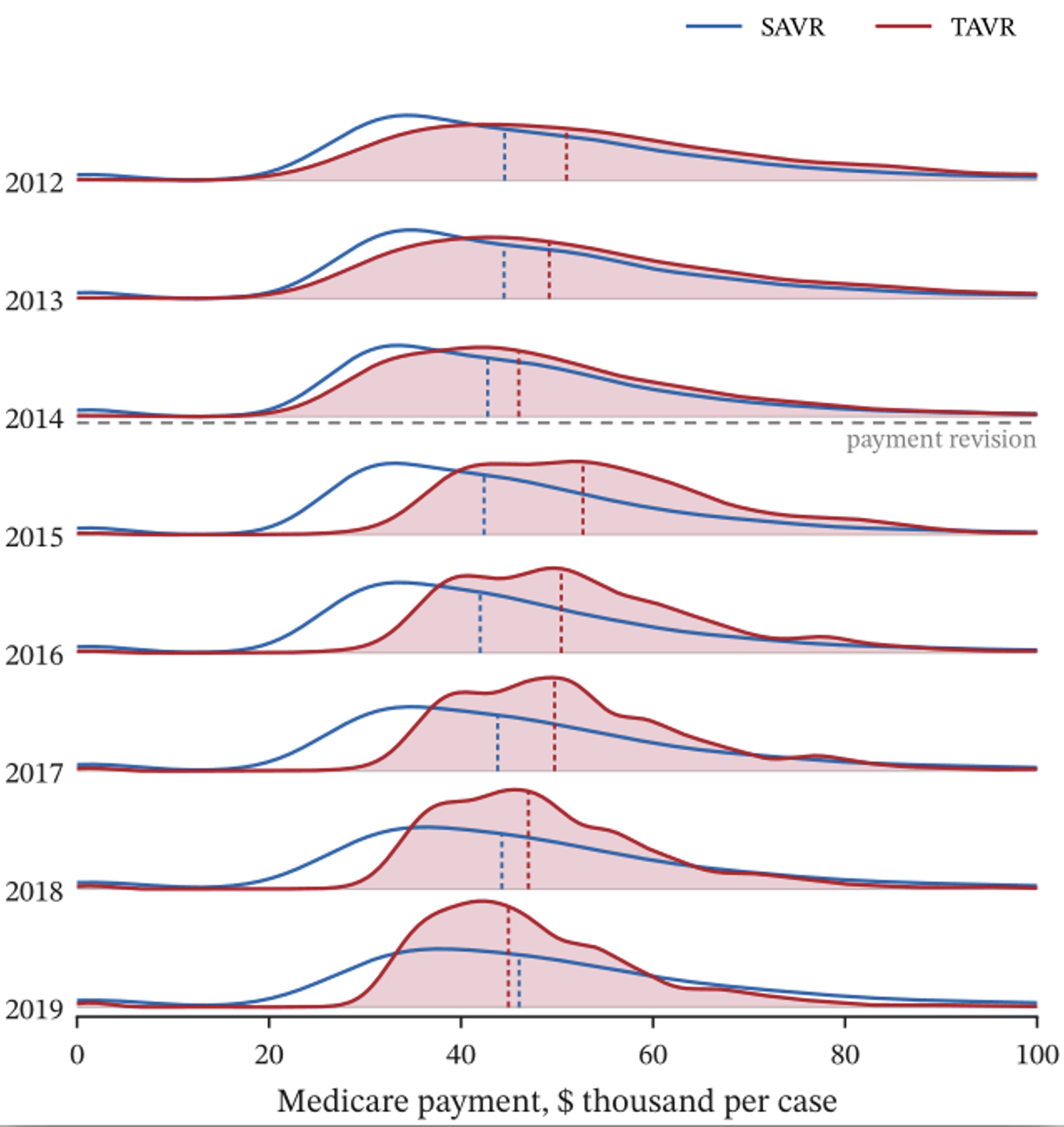}
    \caption{Medicare payment per case for TAVR and SAVR by fiscal year: kernel densities, one density scale for all years. TAVR shaded, SAVR as a line; dotted ticks mark the medians; the dashed rule marks the fiscal 2015 payment revision. Source: Medicare fee-for-service AVR episodes, FY2012--2019.}
    \label{fig:reimbursement}
\end{figure}

The claims report total charges and the Medicare payment for each stay but not the hospital's cost. We approximate cost following \citet{reddy2025contribution}, applying hospital-year cost-to-charge ratios from the cost reports; the device is the bulk of the variable cost of a transcatheter valve.\footnote{Cost-to-charge ratios are hospital-wide. They apply the hospital's average markup to the device, so where devices are marked up less than other inputs they understate the TAVR--SAVR cost difference; and they measure fully allocated average cost, which can exceed the cost of an additional admission; the adoption model of Section~\ref{sec:dynamic_model} removes fixed overhead from them.} Per-case reimbursement is the payment amount on each claim. Figures~\ref{fig:charges_density} and~\ref{fig:reimbursement} plot, year by year, the distribution of charges and of payments by procedure. Charges per TAVR case are higher than per SAVR case in every pre-revision year, a median of \$178 thousand against \$155 thousand in 2012, and estimated costs are higher until 2016 (Appendix Figure~\ref{fig:reimbursement_means}), yet Medicare paid the two procedures within a few thousand dollars of each other: the median TAVR payment is \$50 thousand in 2012 and \$45 thousand in 2014, against \$44 and \$42 thousand for SAVR. The 2015 revision raises the median TAVR payment to \$52 thousand, a gap of about \$10 thousand over SAVR; the gap then narrows as TAVR moves to lower-risk patients coded without major complications, and by 2019 the medians coincide at about \$44 thousand, while SAVR charges rise over the decade and overtake TAVR charges from 2016. In the pre-2015 years a TAVR case was therefore paid about what a SAVR case was paid while costing more, so its margin was below the surgical margin. The revision widened the median payment gap from about \$3 thousand in 2014 to about \$10 thousand in 2015, and the gap is largest in the years right after it. The densities show that the revision moved the whole TAVR payment distribution, not only its median: from 2015 the TAVR density lies to the right of the SAVR density and is more concentrated, and by 2019 the two overlap. These are distributions across episodes; Section~\ref{sec:dynamic_model} builds the hospital-, year- and risk-specific margins the adoption model uses.

\section{Descriptive Evidence}\label{sec:descriptive}

The claims record every hospital's adoption date and every patient's choice of hospital. This section documents two puzzles in the timing of entry: more than two fifths of the hospitals that chose to adopt entered while TAVR was paid the SAVR rate, and entry did not jump when the rate rose. A two-stage model of entry under a delayed rate gives four tests of what explains them. First, hospitals respond to margins, but through the procedure rather than through entry: at the revision, programs already operating shift patients from SAVR to TAVR and their TAVR volume rises by about 10 percent. Second, learning is not the main reason for early entry: a hospital with no rival in its HRR, which keeps all of the volume over which experience pays, enters no earlier than others. Third, patients sort toward adopters: SAVR patients bypass their nearest hospital for one that offers TAVR. Fourth, after the revision a hospital's adoption hazard falls by about 15 percent for every ten percentage points of its patient catchment already served by rival adopters.

\subsection{Two adoption puzzles}\label{sec:two-puzzles}

\begin{figure}[htb]
\centering
\includegraphics[width=0.55\linewidth]{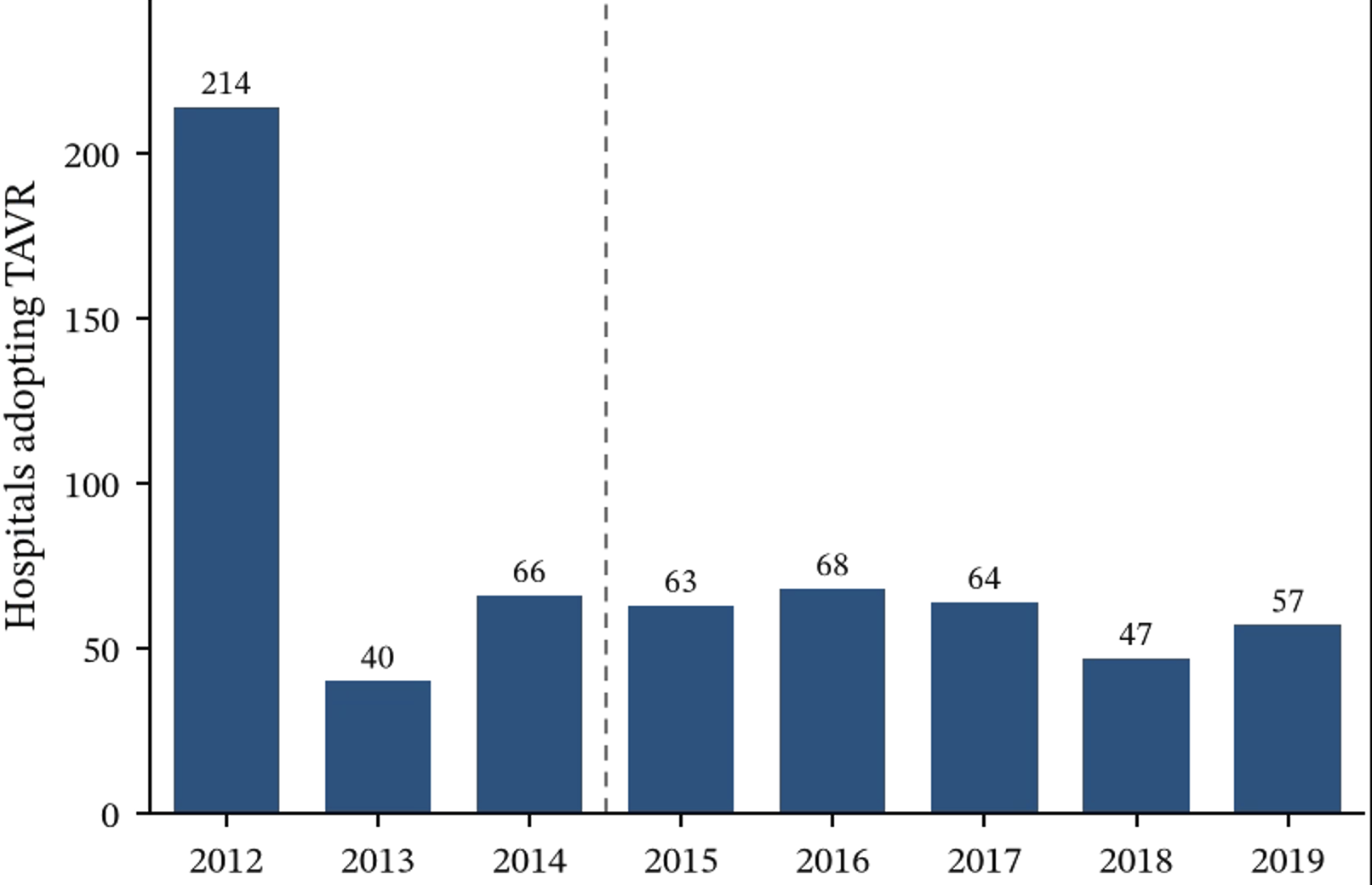}
\caption{Hospitals by fiscal year of TAVR adoption. The 2012 bar includes the 97 pivotal-trial sites, which ran programs before Medicare coverage; 117 hospitals entered in fiscal 2012.}
\label{fig:entrance_all}
\end{figure}

Of the $946$ hospitals in our sample, roughly $65\%$ adopted TAVR by the end of $2019$; the remainder had not adopted by the end of the sample window. The fiscal-year distribution of entrants (Figure~\ref{fig:entrance_all}) does not follow a smooth S-curve.\footnote{We use fiscal rather than calendar years because Medicare reimbursement claims are submitted on a fiscal-year basis.} Two features make it puzzling.

\emph{Puzzle 1: a pre-reimbursement wave.} Excluding the 97 pivotal-trial sites, which ran programs before coverage, 223 of the 523 hospitals that chose to adopt, 43 percent, entered in the three years before the rate revision, when a TAVR case earned less than a SAVR case. The wave is not a one-time response to coverage: 117 entered in fiscal 2012, the first year the procedure could be billed outside clinical trial sites, and entry continued in 2013 and 2014, at 40 and 66 hospitals.\footnote{The cost of the valve kept early TAVR margins low \citep{reddy2023economic}. In the margin predictions of Section~\ref{sec:dynamic_model}, which hold the hospital, the year, the risk score and the conditions fixed, a TAVR case at a hospital outside the pivotal trials earned \$0.7 thousand less than a SAVR case in 2012 and \$3.2 thousand less in 2013 and in 2014, and \$4 to 7 thousand more from 2015.} A model in which hospitals adopt when the procedure pays does not predict a discrete mass of entry at a relative loss.

\emph{Puzzle 2: a diffused post-reimbursement response.} Before 2015 a TAVR case was paid about what a SAVR case was paid, a median of \$3 thousand more, which did not cover its higher cost; the 2015 revision raised the median gap to about \$10 thousand per case (Section~\ref{sec:data}) and moved the TAVR margin above the SAVR margin. Entry did not jump again; it rose gradually over the following five years. A model in which the revision releases a queue of hospitals waiting for the rate predicts a discrete jump in 2015, which the data do not show. Start-up lags would spread such a response over adjacent years, but among the hospitals that choose when to enter, 66 adopt in 2014, 63 in 2015 and 68 in 2016. By the end of the sample roughly $35\%$ of hospitals had still not adopted. They are the margin on which a payer's policy can act, the subject of Section~\ref{sec:counterfactuals}.

\subsection{A two-stage model of entry under a delayed rate}\label{sec:stylized-framework}

To rationalize the two puzzles, why hospitals enter before the rate pays and why the rate, once it pays, draws in so few, we use a stylized model of the adoption decision under a delayed rate. Stage~1 is the window before the rate revision, $t<t_R$, in which TAVR is reimbursed at the surgical rate; stage~2 begins at $t_R$ and continues indefinitely at the revised rate, with stage-2 quantities in present value at $t_R$ and $\beta\in(0,1)$ discounting stage~2 to stage~1. Hospitals know $t_R$ and the revised rate.\footnote{This is the perfect-foresight assumption maintained in the adoption-timing model of Section~\ref{sec:dynamic_model}, which states it formally.}

Hospital $i$ in market $m$ pays a fixed cost $F_i$ to adopt. In stage $s$ it performs $v_{i,s}$ TAVR cases on its own AVR patients, each earning the TAVR margin over SAVR, $\mu_s$, with $\mu_1<0<\mu_2$: before the revision every TAVR case earns less than the surgical alternative, after it more. Two further primitives can raise the value of adopting. Learning: a hospital that operates in stage~1 enters stage~2 with lower per-case cost, raising its stage-2 margin by $\eta_i\ge0$. Steering: under regulated rates the only margin on which an adopter competes is attracting patients. Market $m$ contains a pool $S_m$ of AVR patients who will travel to a hospital that offers TAVR, whether or not they receive it,\footnote{Patients may prefer the minimally invasive procedure, or may read adoption as a signal of hospital quality. Section~\ref{sec:demand} estimates this preference.} and an adopter captures a share $\chi_i\in[0,1]$ of the pool available when it enters. Captured patients stay, so a stage-2 entrant draws on the residual pool $Q_m = S_m\bigl(1-\sum_{j\in E_m}\chi_j\bigr)\le S_m$, where $E_m$ is the set of stage-1 adopters in $m$, and each captured patient earns the SAVR margin $\mu^S_s>0$, since most receive the incumbent procedure. The incremental profit from adoption therefore has a steering component and a procedure-margin component,
\begin{align*}
\Delta\pi_{i,1}&=\chi_i S_m\,\mu^S_1+v_{i,1}\,\mu_1,\\
\Delta\pi^{E}_{i,2}&=\chi_i S_m\,\mu^S_2+v_{i,2}\,(\mu_2+\eta_i),\\
\Delta\pi^{L}_{i,2}&=\chi_i Q_m\,\mu^S_2+v_{i,2}\,\mu_2,
\end{align*}
where $E$ and $L$ index early (stage-1) and late (stage-2) adoption. The three options are worth
\[
V^E_i=-F_i+\Delta\pi_{i,1}+\beta\,\Delta\pi^{E}_{i,2},
\qquad
V^L_i=\beta\bigl(-F_i+\Delta\pi^{L}_{i,2}\bigr),
\qquad
V^N_i=0 .
\]

Hospital $i$ enters early rather than late when $V^E_i\ge V^L_i$,
\begin{equation}\label{eq:early}
\underbrace{\chi_i S_m\,\mu^S_1+\beta\,\chi_i\,(S_m-Q_m)\,\mu^S_2}_{\text{steering rent}}
\;+\;\underbrace{\beta\,v_{i,2}\,\eta_i}_{\text{learning gain}}
\;\ge\;
\underbrace{-\,v_{i,1}\,\mu_1}_{\text{stage-1 loss}}
\;+\;\underbrace{(1-\beta)\,F_i}_{\text{cost of paying early}} ,
\end{equation}
and enters late rather than never when $V^L_i\ge0$,
\begin{equation}\label{eq:late}
\chi_i Q_m\,\mu^S_2+v_{i,2}\,\mu_2\;\ge\;F_i .
\end{equation}
Both conditions are cutoffs in $F_i$. Write $R_i$ for the left side of~\eqref{eq:early}, the rent to early entry, $\ell_i=-v_{i,1}\mu_1$ for the stage-1 loss, and $L_i$ for the left side of~\eqref{eq:late}, the stage-2 return of a late entrant. Hospitals with $F_i$ below $F^E_i=(R_i-\ell_i)/(1-\beta)$ enter in stage~1, those between $F^E_i$ and $F^L_i=L_i$ enter at $t_R$, and the rest never enter; if $F^E_i$ exceeds $L_i$, no hospital enters at $t_R$ and the early entrants are those with $V^E_i\ge0$ (Appendix~\ref{app:planner}, Lemma~\ref{lem:sorting}). Fixed-cost heterogeneity sorts hospitals into cohorts, and the rents in~\eqref{eq:early} and the residual pool in~\eqref{eq:late} shift the cutoffs.

\begin{figure}[htb]
    \centering
    \includegraphics[width=0.8\linewidth]{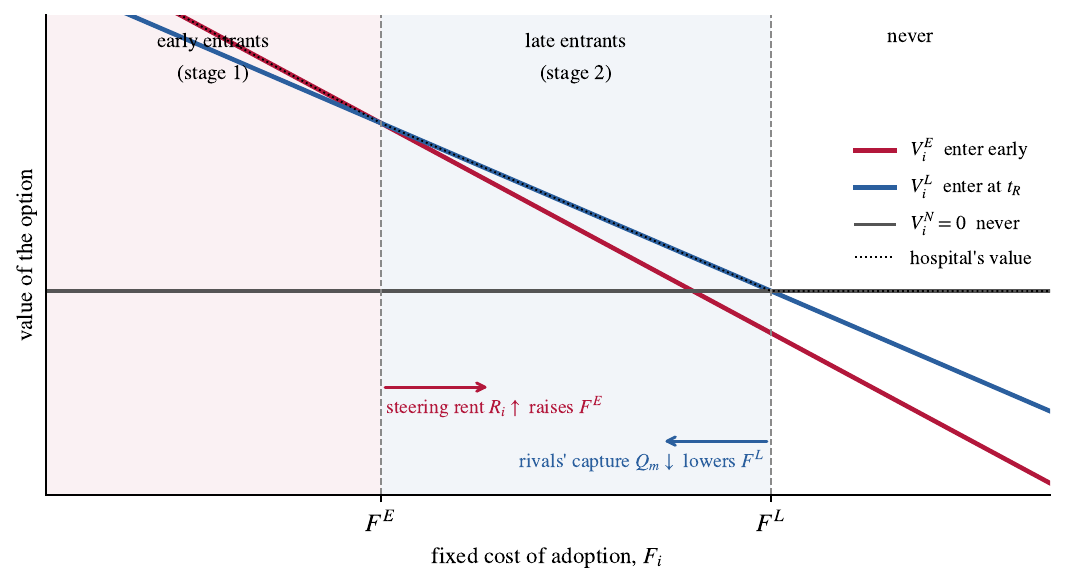}
    \caption{Entry cutoffs in the two-stage model. The lines are the values of entering early, $V^E_i$, entering at the revision, $V^L_i$, and never entering, against the fixed cost $F_i$; the dotted line is the hospital's value, the upper envelope. Hospitals below $F^E$ enter in stage~1, those between $F^E$ and $F^L$ at $t_R$, and the rest never. The steering rent $R_i$ shifts $V^E_i$ up and $F^E$ right; capture by stage-1 rivals lowers $Q_m$, shifts $V^L_i$ down and $F^L$ left. Illustrative parameters.}
    \label{fig:two_stage_cutoffs}
\end{figure}

The model says what early entry needs. With neither rent ($\eta_i=0$, $\chi_i=0$), a hospital enters before the revision only if $F_i\le-\ell_i/(1-\beta)<0$, that is, only if returns outside the Medicare margin exceed its setup cost. Within the Medicare margin, Puzzle~1 therefore requires learning or steering; returns outside that margin are a third source, which the adoption-timing model of Section~\ref{sec:dynamic_model} absorbs in the adoption cost $K$. The two rents differ in where they come from. The learning gain $\beta v_{i,2}\eta_i$ does not depend on rivals and grows with a hospital's volume, as does the stage-1 loss $v_{i,1}|\mu_1|$; because the per-case loss before the revision is small (Section~\ref{sec:data}), the learning gain is worth more to a hospital with no local competitor, which keeps all of its local volume. The steering rent requires patients who sort toward adopters and is larger the more of the pool a first mover can lock in before rivals enter. Puzzle~2 is the same primitive seen from the other side: every stage-1 capture lowers $Q_m$ in~\eqref{eq:late}, so where early entrants took much of the pool, fewer late hospitals clear their cutoff even though $\mu_2>0$. Steering rents for early entrants and residual-demand exhaustion for late entrants are the same mechanism.\footnote{Heterogeneous beliefs about the revised rate add little. A hospital that is optimistic about $\mu_2$ but faces neither rent has nothing to set against the stage-1 loss, so optimism alone cannot generate early entry; belief dispersion spreads entry dates among hospitals with positive rents, which in the reduced forms is indistinguishable from dispersion in $F_i$. The model is also silent on the pace of post-$t_R$ entry; that is what the adoption-timing model of Section~\ref{sec:dynamic_model} produces from the estimated distribution of $F_i$.}
Figure~\ref{fig:two_stage_cutoffs} draws the three options against $F_i$. The value of entering early falls faster in $F_i$ than the value of waiting, because an early entrant pays the fixed cost before discounting, so the two cross at $F^E$; the value of waiting crosses zero at $F^L$. The steering rent raises the early line and moves $F^E$ to the right, which is Puzzle~1; rivals' capture lowers the late line and moves $F^L$ to the left, which is Puzzle~2. The learning gain enters $V^E_i$ in the same way as the steering rent, so the figure does not separate the two. The market-structure test below is designed to tell them apart: learning predicts that a hospital with no rival in its HRR enters earlier, while the steering rent, which comes from patients rivals would otherwise take, gives it no reason to.

This gives four hypotheses, and each maps to an observable sign:
\begin{enumerate}\itemsep0.2em
\item[(H1)] \emph{Hospitals respond to margins.} The puzzles are puzzles only if the payment enters hospitals' decisions. If it did not, volume at existing programs would not respond to the rate revision.
\item[(H2)] \emph{Learning.} Hospital experience reduces per-patient resource use, the necessary condition for $\eta_i>0$; and if learning carries early entry, hospitals with no local rival enter earlier than others of the same size, since they keep all of the volume over which experience pays.
\item[(H3)] \emph{Steering, rent side.} Patients sort toward adopters, including patients who receive SAVR.
\item[(H4)] \emph{Steering, exhaustion side.} Post-$t_R$ adoption hazards decrease in cumulative within-market rival entry.
\end{enumerate}
H1--H3 concern early entry and are tested in Section~\ref{sec:puzzle1-tests}; H4 concerns late entry and is tested in Section~\ref{sec:puzzle2-tests}.

\subsection{Puzzle 1}\label{sec:puzzle1-tests}
\label{altruism}

We start with the first hypothesis. The test is whether hospitals that already offer TAVR change their volume when the rate rises, on the 214 programs operating in fiscal 2012, trial sites included, for which the 2015 cutoff precedes the expansion of eligibility in fiscal 2017 (Section~\ref{data}). We estimate
\begin{align}
\log(\text{TAVR volume})_{it}
&= \alpha_i
+ \beta\,(\text{year}_{t}-2012)
+ \tau\,\mathbf{1}_{t\ge 2015}
+ \delta\,\mathbf{1}_{t\ge 2015}(\text{year}_{t}-2012)
+ \varepsilon_{it}, \label{eq:model}
\end{align}
over fiscal years 2013--2018,\footnote{We exclude 2012 because adoption may occur mid-year, yielding partial-year volumes.} and test $\Delta_{2015}\equiv \tau + 3\delta = 0$. From Table~\ref{tab:tavr_feols}, $\widehat{\Delta}_{2015} = 0.577 + 3(-0.161)\approx 0.093$ (s.e.\ 0.041, $p=0.02$), an increase of roughly $10\%$ in TAVR volume at the cutoff ($e^{0.093}-1$); the program's TAVR share of its valve replacements rises by 2.9 percentage points (s.e.\ 1.0) against a 2014 mean of 38 percent.\footnote{New valve generations, SAPIEN~3 and Evolut~R, were approved in June 2015, the last quarter of fiscal 2015, so they were available for at most a quarter of the fiscal year.} Hospitals already offering TAVR respond to the rate, so the payment enters their decisions. The modality rule of Section~\ref{sec:dynamic_model}, estimated on every adopter, gives the same answer: a larger TAVR margin over SAVR raises the probability that a program assigns a patient to TAVR (Table~\ref{tab:supply_estimates}). Two further tests are reported in Appendix~\ref{app:altruism}. In fiscal 2015--2016, after the rate revision and before the FDA extended the label to intermediate-risk patients in August 2016, the SAVR patients of early adopters became less severe and younger relative to those of later adopters: the allocation of patients between the two procedures moved with the payment. Before the revision SAVR volume is about 10 percent higher once a hospital adopts, and its later decline at adopters dates from the wider indication rather than from the revision. These tests do not exclude other motives, since a hospital that valued patient benefit could also expand when a higher payment relaxes its budget; we maintain profit maximization in the models that follow.

\begin{table}[htb]
\centering
\caption{TAVR volume at 2012 adopters around the 2015 rate revision}
\label{tab:tavr_feols}
\begin{threeparttable}
\begin{tabular}{lc}
\toprule
 & $\log(\text{TAVR volume})$ \\
\midrule
Fiscal year $-$ 2012 & 0.324\sym{***} \\
 & (0.030) \\
Post-2015 & 0.577\sym{***} \\
 & (0.064) \\
Fiscal year $-$ 2012 $\times$ Post-2015 & $-0.161$\sym{***} \\
 & (0.031) \\
\midrule
Jump at 2015, $\tau+3\delta$ & 0.093\sym{**} \\
 & (0.041) \\
Hospital fixed effects & Yes (214) \\
Observations & 1,279 \\
$R^{2}$ & 0.883 \\
\bottomrule
\end{tabular}
\begin{tablenotes}[flushleft]\footnotesize
\item \textit{Notes:} Programs operating in fiscal 2012, pivotal-trial sites included, fiscal years 2013--2018. Standard errors in parentheses, clustered at the hospital level. $^{*} p<0.1$; $^{**} p<0.05$; $^{***} p<0.01$.
\end{tablenotes}
\end{threeparttable}
\end{table}

Learning is the first candidate. We estimate the within-hospital learning slope on length of stay and log hospital charges from
\begin{equation}
Y_{ihts} = \beta_0 + \beta_1 \cdot \text{Years since adoption}_{ht} + \mathbf{C}_{iht}'\boldsymbol{\theta} + \delta_h + \lambda_s + \varepsilon_{ihts},
\end{equation}
where $Y_{ihts}$ is length of stay in days or log total hospital charges for patient $i$ in hospital $h$, year $t$, and severity category $s$. Hospital fixed effects $\delta_h$ absorb time-invariant hospital heterogeneity and DRG-based severity fixed effects $\lambda_s$ absorb clinical case mix. The patient-level vector $\mathbf{C}_{iht}$ collects the Charlson Comorbidity Index and indicators for hypertension, congestive heart failure, chronic pulmonary disease, diabetes, renal disease, and prior cardiac surgery; this absorbs the secular shift toward lower-risk TAVR patients as eligibility expanded, the leading confound for $\hat\beta_1$. Standard errors are clustered at the hospital level.

\begin{table}[htb]
\centering
\caption{Learning: length of stay and charges by years since adoption}
\label{tab:tavr_utilization_charges}
\begin{threeparttable}
\begin{tabular}{lcc}
\toprule
 & Length of stay (days) & Log(total charges) \\
 & (1) & (2) \\
\midrule
Years since adoption & $-0.736$\sym{***} & $0.000$ \\
 & (0.027) & (0.004) \\
\midrule
Hospital and severity-category effects & Yes & Yes \\
Observations & 182,253 & 182,256 \\
$R^{2}$ & 0.328 & 0.675 \\
\bottomrule
\end{tabular}
\begin{tablenotes}[flushleft]\footnotesize
\item \textit{Notes:} TAVR patients at adopting hospitals (pivotal-trial sites included), fiscal years 2012--2019, from the year of adoption on. Patient-level controls: Charlson Comorbidity Index and indicators for hypertension, congestive heart failure, chronic pulmonary disease, diabetes, renal disease and prior cardiac surgery (valve surgery or coronary bypass in the 12 months before admission). Standard errors in parentheses, clustered at the hospital level. $^{*} p<0.1$; $^{**} p<0.05$; $^{***} p<0.01$.
\end{tablenotes}
\end{threeparttable}
\end{table}

Table~\ref{tab:tavr_utilization_charges} reports the estimates. Each additional year since adoption reduces length of stay by about three quarters of a day (Column~1), about $15\%$ of the mean stay of $4.8$ days, while total hospital charges per patient do not change (Column~2). With hospital effects, years since adoption moves one for one with calendar time within a hospital, so the slope also absorbs improvement in the procedure common to all programs and overstates own-program learning. Flat charges do not imply flat costs, since hospitals' cost-to-charge ratios fall from 0.27 to 0.22 over the period (Section~\ref{sec:dynamic_model}). Learning shortens stays, but it cannot lower the price of the prosthetic valve, which bounds the part of a case's cost that experience can reduce, $\eta_i$.

A second check uses market structure. If learning carried early entry, a hospital that keeps more of its local volume would enter earlier than others, since it keeps more of the volume over which experience pays. Using hospitals with AVR claims in fiscal 2012 and no earlier program, we estimate the linear probability model
\begin{equation*}
\mathbf{1}\{\tau_h < 2015\} = \alpha + \beta\,M_h + \gamma\,\text{Beds}_h + \delta_{\text{type}(h)} + \varepsilon_h,
\end{equation*}
for two measures of local market power $M_h$. The first, $\text{Monopoly}_h$, equals one if $h$ is the only hospital in its HRR, in HRRs with fewer than ten hospitals. Conditional on size and ownership type, monopoly status is not associated with earlier adoption (Table~\ref{tab:monopoly_lpm}, Column~1): the estimate is $-2.1$ percentage points (s.e.\ 4.7), and the 95 percent interval excludes a monopoly advantage larger than 7 points. HRR boundaries are crossed by patients (Section~\ref{sec:data}), so the second measure does not rely on them. It is the share of the patients of the hospital's own catchment that the hospital itself treated, $\sum_z w_{hz} f_{hz}$, where $w_{hz}$ is the share of $h$'s fiscal 2009--2011 AVR patients who live in ZIP code $z$ and $f_{hz}$ the share of $z$'s AVR patients that $h$ treated. On this measure local dominance predicts early entry (Column~2): ten percentage points more of its own catchment raise the probability of entering before the revision by 2.7 points (s.e.\ 0.9), against a mean of 34 percent. This is the sign learning predicts. It is also the sign of the defensive side of steering: a hospital that holds most of its catchment has the most patients to lose to a rival program, and entering first protects them. The market-structure test therefore does not separate the two channels; the demand model of Section~\ref{sec:demand} measures the patient channel directly.

\begin{table}[htb]
\centering
\small
\def\sym#1{\ifmmode^{#1}\else\(^{#1}\)\fi}
\caption{Local market power and adoption before the rate revision}
\label{tab:monopoly_lpm}
\begin{threeparttable}
\begin{tabular}{lcc}
\toprule
 & (1) & (2) \\
\midrule
Monopoly (only hospital in HRR) & $-0.021$ &  \\
 & (0.047) &  \\
Own share of catchment &  & $\phantom{-}0.272$\sym{***} \\
 &  & (0.091) \\
Beds (hundreds) & $\phantom{-}0.095$\sym{***} & $\phantom{-}0.087$\sym{***} \\
 & (0.007) & (0.006) \\
\midrule
95\% interval, market-power term & [$-0.113$, 0.070] & [0.094, 0.451] \\
Ownership-type FE & Yes & Yes \\
Hospitals & 766 & 935 \\
Mean of outcome & 0.351 & 0.341 \\
\bottomrule
\end{tabular}
\begin{tablenotes}[flushleft]\footnotesize
\item \textit{Notes:} Linear probability model of adoption in fiscal 2012--2014; hospitals with AVR claims in fiscal 2012 and no earlier programme. (1): HRRs with fewer than ten hospitals. (2): all HRRs; own share of catchment is the share of the fiscal 2009--2011 AVR patients of the hospital's catchment ZIP codes that it treated itself, weighted by its own patients. Standard errors clustered by HRR. $^{*}$ $p<0.10$; $^{**}$ $p<0.05$; $^{***}$ $p<0.01$.
\end{tablenotes}
\end{threeparttable}
\end{table}

\begin{figure}[htb]
\centering
\includegraphics[width=0.5\linewidth]{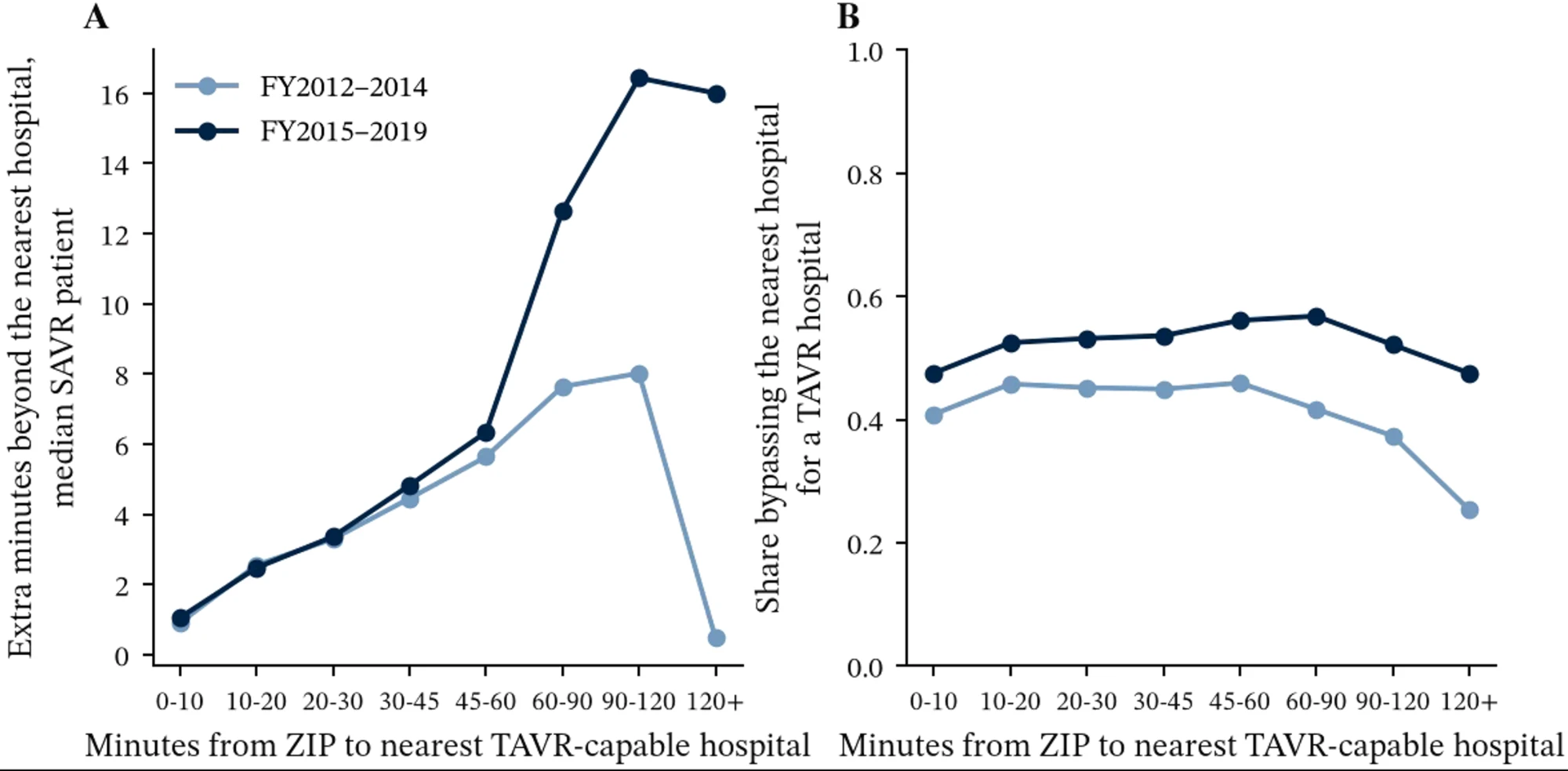}
\caption{SAVR patients' travel toward TAVR hospitals, by driving time from the patient's ZIP code to the nearest TAVR-capable hospital, fiscal 2012--2014 and 2015--2019. Panel A: median extra minutes beyond the nearest hospital. Panel B: share bypassing the nearest hospital for a TAVR hospital.}
\label{fig:savr_travel_more}
\end{figure}

Steering requires $S_m>0$: patients who travel to a hospital because it offers TAVR. SAVR patients provide the test. SAVR is available at every hospital, so a SAVR patient who bypasses her nearest hospital for one that offers TAVR is not traveling for the procedure she receives. She may be traveling for the hospital, read through its program, or for the chance of TAVR before the heart team assigned her procedure; Section~\ref{sec:demand} shows that patients with almost no chance of TAVR also value an adopter. Figure~\ref{fig:savr_travel_more} shows that in 2012--2014, 41 to 46 percent of SAVR patients who live within an hour of a program bypass their nearest hospital for a TAVR hospital, and in 2015--2019, 47 to 57 percent at every distance; where the nearest program is 60 to 120 minutes away, the median SAVR patient travels 13 to 16 minutes beyond her nearest hospital after 2015, against about 8 before. Among all treated patients whose nearest hospital had no program but who were treated at an adopter, 68 percent received SAVR in 2012--2014, about 2.1 SAVR cases for every TAVR case, and 48 percent over 2012--2019. Measured against the years before adoption, the event study (Appendix Figure~\ref{fig:es_admissions}) and the demand model, which conditions on hospital effects, attribute the flow to the program. Under regulated prices this is steering: adoption is a non-price attribute on which hospitals compete, and the demand model in Section~\ref{sec:demand} measures how much patients will travel for it.

\subsection{Puzzle 2}\label{sec:puzzle2-tests}

Condition~\eqref{eq:late} says that a hospital's value of entering at or after $t_R$ falls with the share of the contestable pool that rivals captured earlier (H4). Fixed-cost heterogeneity alone spreads post-$t_R$ entry over time but predicts no relation between a market's early entry and its later hazards; exhaustion does. We test it on the hospitals at risk after the revision. For each hospital $h$ that has not adopted by the end of fiscal year $t-1$, $t=2015,\dots,2019$, we model adoption in $t$ as a discrete-time hazard,
\begin{equation}\label{eq:coxModel}
\Pr\bigl(\tau_h=t \mid \tau_h\ge t\bigr)
\;=\;
G\bigl(\beta\,R_{h,t-1}+\mathbf{X}_h'\boldsymbol{\gamma}+\lambda\bigr),
\end{equation}
with $G$ complementary log-log or linear, and $\lambda$ year or HRR-by-year effects. $R_{h,t-1}$ measures rival adoption at the end of $t-1$ in two ways. The HRR competitor rate is the share of the other hospitals in $h$'s HRR that had adopted. It takes the same value for every hospital at risk in an HRR-year, so it is identified only from variation across HRRs or within an HRR over time, and it cancels from any model stratified by HRR and year. Catchment exposure is the share of $h$'s own pre-TAVR patient catchment already served by rival adopters, $\sum_z w_{hz}\sum_{k\neq h,\,\tau_k\le t-1} f_{kz}$, with $w_{hz}$ and $f_{kz}$ the fiscal 2009--2011 patient shares defined above. It differs across hospitals of the same HRR and year, so it is identified with HRR-by-year effects, which absorb any common local shock to the value of entry. $\mathbf{X}_h$ collects log beds, teaching status and ownership type; standard errors are clustered by HRR. The conventional confound, that rival entry proxies a common positive local shock, would bias $\beta$ upward.

Table~\ref{hazard_model} reports the estimates. Across HRRs, with year effects only (Column~1), the competitor rate is negative and imprecise, $-0.13$ (s.e.\ 0.23), consistent with common positive local shocks offsetting exhaustion. Within HRRs (Column~2) it is $-0.40$ (s.e.\ 0.04): ten percentage points more rival adoption lower the annual hazard by 4 points, against an at-risk adoption rate of 11.6 percent. The catchment measure compares hospitals of the same HRR in the same year: ten percentage points more exposure lower the hazard by 1.7 points (s.e.\ 0.8), about 15 percent (Column~3); in the complementary log-log model with year effects, which also uses variation across HRRs, the decline is 19 percent (Column~4). Earlier entry by the rivals that serve a hospital's own patients lowers the value of entering for those that remain.

Three qualifications bound this reading. A hospital still at risk despite heavy rival entry around it is more likely to have a high setup cost, so dynamic selection alone biases $\beta$ downward; the hospital controls absorb part of this selection, not all of it. Within an HRR, the hospitals still at risk are mechanically those whose rivals entered first, which pushes Column~2 negative even with independent hazards; the catchment measure, which varies across hospitals of the same HRR-year, is less exposed to this. And a negative $\beta$ does not single out the exhaustion of patients from other forms of rival crowding, such as consolidated referrals or competition for cardiac staff; Section~\ref{sec:demand} measures the patient channel directly.

\begin{table}[htb]
\centering
\small
\def\sym#1{\ifmmode^{#1}\else\(^{#1}\)\fi}
\caption{Adoption hazard after the rate revision and prior rival adoption}
\label{hazard_model}
\begin{threeparttable}
\begin{tabular}{lcccc}
\toprule
 & (1) & (2) & (3) & (4) \\
\midrule
HRR competitor rate$_{t-1}$ & $-0.126$ & $-0.402$\sym{***} &  &  \\
 & (0.230) & (0.044) &  &  \\
Catchment exposure$_{t-1}$ &  &  & $-0.170$\sym{**} & $-2.129$\sym{***} \\
 &  &  & (0.082) & (0.412) \\
log beds & $\phantom{-}0.866$\sym{***} & $\phantom{-}0.100$\sym{***} & $\phantom{-}0.095$\sym{***} & $\phantom{-}0.984$\sym{***} \\
 & (0.168) & (0.022) & (0.021) & (0.169) \\
Teaching (GME) & $\phantom{-}0.313$\sym{**} & $\phantom{-}0.036$\sym{*} & $\phantom{-}0.035$\sym{*} & $\phantom{-}0.353$\sym{***} \\
 & (0.129) & (0.019) & (0.019) & (0.127) \\
\midrule
Estimator & cloglog & LPM & LPM & cloglog \\
Fixed effects & Year & HRR, year & HRR $\times$ year & Year \\
Hospital-years at risk & 2,292 & 2,289 & 1,953 & 2,292 \\
Adoptions & 265 & 265 & 265 & 265 \\
Hospitals & 568 & 568 & 568 & 568 \\
HRRs & 204 & 204 & 204 & 204 \\
\bottomrule
\end{tabular}
\begin{tablenotes}[flushleft]\footnotesize
\item \textit{Notes:} Discrete-time hazard of TAVR adoption, fiscal 2015--2019; a hospital is at risk in year $t$ if it has not adopted by the end of $t-1$. HRR competitor rate: share of the other hospitals in the hospital's HRR that adopted by the end of $t-1$. Catchment exposure: share of the hospital's fiscal 2009--2011 AVR patient catchment, by ZIP code, served by hospitals that adopted by the end of $t-1$; it varies across hospitals within an HRR-year. Columns (1) and (4) are complementary log-log models with year effects; (2) and (3) linear probability models. All columns control for ownership type. Column (3) drops HRR-years with a single hospital at risk. Standard errors clustered by HRR. $^{*}$ $p<0.10$; $^{**}$ $p<0.05$; $^{***}$ $p<0.01$.
\end{tablenotes}
\end{threeparttable}
\end{table}

Larger and teaching hospitals adopt sooner in every column of Table~\ref{hazard_model}, consistent with heterogeneity in $F_i$. At this stage such sorting is indistinguishable from variation in beliefs, and the covariates should be read as an unconditional sorting fact rather than as estimates of cost differences. The adoption-timing model in Section~\ref{sec:dynamic_model} recovers the adoption cost from adoption dates directly; its estimate is the fixed cost net of returns to a program that accrue outside the Medicare margin, the counterpart of $F_i$ here.

Learning cannot fully account for early entry: a hospital with no rival in its HRR, which keeps all of its local volume, enters no earlier than others, and the learning gain does not depend on rivals, whereas entry by rivals into a hospital's own catchment lowers its hazard. Dominance of its own catchment predicts early entry, but that sign fits steering as well as learning. The reduced forms therefore point to competition for patients as a return to early adoption alongside learning, and they establish signs, not sizes; Section~\ref{sec:mechanism} finds that the patients a program draws carry nine tenths of the value of adopting. Section~\ref{sec:demand} estimates how much demand adoption shifts toward a hospital; Section~\ref{sec:dynamic_model} converts that shift into dollars and recovers the adoption cost that rationalizes observed entry dates.

\section{Patient demand for hospitals and technology}
\label{sec:demand}

The descriptive evidence shows that adoption attracts patients. To uncover the mechanism behind this, we estimate patient demand for hospitals on the population of Medicare beneficiaries diagnosed with aortic valve disease. The design has two features. First, we estimate demand for the hospital, not for the procedure, because that is the choice the patient makes: the heart team assigns the procedure after admission \citep{CMS2012TAVRNCD}. Second, the value of an adopter varies with the patient's risk score and clinical conditions, and by year as coverage expanded by risk tier. The estimates give the change in a hospital's admissions when it adopts, and because patients who almost surely receive SAVR also travel to an adopter, the change reaches well beyond TAVR recipients: in 2012 a program adds 21 patients a year to an adopter, 9 of whom receive SAVR (Section~\ref{sec:mechanism}). Section~\ref{sec:demand-results} reports the estimates.

\subsection{Demand set-up}
\label{sec:utility}

Throughout, $i$ denotes a patient-year, $j$ a hospital, $t$ a fiscal year, and $j=0$ the outside option of no aortic valve replacement in year $t$. Let $a_{jt}\in\{0,1\}$ indicate whether hospital $j$ performs TAVR in year $t$, and let $r_i$ denote patient $i$'s predicted operative mortality in percent, the risk measure defined in Section~\ref{sec:data}. Medicare regulates prices and standardizes cost sharing, so the only price a patient pays to reach a hospital is travel time, and no price term enters. The patient chooses a hospital or no treatment; the hospital then assigns the procedure, which Section~\ref{sec:dynamic_model} models.

The market in year $t$, $\mathcal{I}_t$, is the population defined in Section~\ref{sec:data}: fee-for-service beneficiaries aged 65 or older diagnosed with aortic valve disease that year, whether or not they are treated.\footnote{Treated patients are counted in the fiscal year of their first procedure.} Its size $|\mathcal{I}_t|$ is the pool a hospital can draw on when it adopts. Choice sets are based on travel time rather than on administrative market boundaries, since SAVR patients routinely bypass their nearest hospital for one that offers TAVR (Figure~\ref{fig:savr_travel_more}). Each patient's consideration set $\mathcal{J}_{it}$ consists of the forty hospitals nearest the patient's residential ZIP code by driving time and the outside option; treated patients whose chosen hospital lies outside that set are excluded. The rule excludes 4.8 percent of treated patients over 2012--2019, between 4.4 and 5.7 percent in each year, and affects TAVR and SAVR patients alike (4.7 and 4.9 percent). The median treated patient travels about 25 minutes.

We model the choice with a mixed logit, to allow patients to differ in the disutility of travel and in their propensity to be treated at all. The utility of patient $i$ from hospital $j\in\mathcal{J}_{it}$ in year $t$ is
\begin{equation}
\label{eq:Uijt}
\begin{split}
U_{ijt}
={}&
\delta_{j}
+ \gamma_{t}
+ \beta_{1,i}\,\text{tt}_{ij}
+ \beta_{2}\,\text{tt}_{ij}^{2}
+ \beta_{c}\,\mathbf{1}\{j=c(i)\}\\
&+ \kappa_{t}\, r_i
+ \textstyle\sum_{k}\kappa_{kt}\, s_{ik}
+ r_i\,\bigl(\omega_1\,\text{size}_j+\omega_2\,\text{teach}_j\bigr)\\
&+ a_{jt}\,\Bigl(\alpha_{0t}+\alpha_{1t}\, r_i+\textstyle\sum_{k}\alpha_{kt}\, s_{ik}+(\psi_0+\psi_1 r_i)\,E_{jt}\Bigr)
+ \sigma_0\,\zeta_i
+ \varepsilon_{ijt},
\qquad U_{i0t}=\varepsilon_{i0t},
\end{split}
\end{equation}
where $\text{tt}_{ij}$ is the driving time in minutes from $i$'s ZIP code to hospital $j$, $c(i)$ is the hospital nearest to $i$, $s_{ik}\in\{0,1\}$ indicates condition $k$ of the eight defined in Section~\ref{sec:data}, $E_{jt}$ is the number of years since $j$ opened its program,\footnote{Tenure is capped at seven years, the longest the data cover for programs that opened at coverage in fiscal 2012.} $\text{size}_j$ and $\text{teach}_j$ are the hospital's standardized log beds and teaching status, and $\varepsilon_{ijt}$ is i.i.d.\ type-I extreme value. The outside option is normalized to zero.

The terms play five roles. First, the fixed effect $\delta_j$ absorbs hospital attributes common to all patients and years, and the year effect $\gamma_t$ shifts the value of treatment at any hospital relative to no treatment, with $\gamma_{2012}=0$. Second, travel time and the nearest-hospital indicator are the price of reaching a hospital; the travel coefficient is random, $\beta_{1,i}=\beta_1+\sigma\eta_i$. Third, risk and the conditions shift the value of being treated anywhere, through $\kappa_t r_i$ and $\kappa_{kt}s_{ik}$, since sicker patients are, in the data, less likely to undergo a valve replacement at all; together with the random component $\sigma_0\zeta_i$ of the value of treatment, they govern substitution to the outside option, and $\sigma_0$ makes hospitals closer substitutes for one another than for no treatment. Fourth, the TAVR term is the additional value of a hospital that offers TAVR. It varies with risk, the conditions and the year, because neither risk nor the conditions alone predicts who can receive TAVR (Section~\ref{sec:data}) and coverage changed over time, and it is common to all patients who consider the hospital, including those who go on to receive SAVR; its tenure part $(\psi_0+\psi_1 r_i)E_{jt}$ lets a program's value grow as the patients it draws build its referrals, outcomes and reputation. Fifth, the interactions of risk with size and teaching let sicker patients sort toward large and teaching hospitals whether or not they offer TAVR, so that this sorting is not attributed to the program. The draws $\eta_i$ and $\zeta_i$ are independent standard normal.

With the extreme-value error, the choice probability is
\begin{equation}
\label{eq:Prob}
P_{ijt} \;=\;
\int
\frac{\exp\!\big(U_{ijt}(\eta,\zeta)\big)}
{1+\sum_{k\in \mathcal{J}_{it},\,k\neq 0}\exp\!\big(U_{ikt}(\eta,\zeta)\big)}
\ dF(\eta)\,dF(\zeta),
\end{equation}
where $k=0$ is the outside option, whose utility is normalized to zero, and $U_{ijt}(\eta,\zeta)$ is utility net of the extreme-value error; predicted flows are $Q_{jt}=\sum_{i\in \mathcal{I}_t} P_{ijt}$ over the market. The change in a hospital's admissions when it adopts, holding rivals' adoption fixed, is
\begin{equation}
\Delta Y_{j,t}^{\text{steer}}\;:=\;Q_{j,t}\bigl(a_{j,t}=1,\,a_{-j,t}\bigr)\;-\;Q_{j,t}\bigl(a_{j,t}=0,\,a_{-j,t}\bigr),
\qquad Q_{j,t}=\sum_{i\in\mathcal{I}_t} P_{ijt},
\label{eq:DeltaY-steer}
\end{equation}
the empirical counterpart of the captured patients $\chi_i S_m$ of Section~\ref{sec:stylized-framework}, computed hospital by hospital.

\subsection{Identification}
\label{sec:identification-demand}

The parameter of interest is the value of a TAVR hospital and how it varies with the patient's risk. It is identified from the choices of patients of different risk between hospitals with and without a program, within hospitals over time and across years as coverage widened by risk tier. We do not separate a preference for the hospital from a preference for the procedure: the procedure is assigned by the heart team after admission, jointly with the hospital's modality choice, so the chance of TAVR is not a patient characteristic on which demand can condition. Patients with aortic stenosis are referred by their cardiologist to a heart team, and Medicare's coverage rule requires that two cardiac surgeons evaluate each candidate for surgery \citep{CMS2012TAVRNCD}; a program therefore draws patients for evaluation, and those found suitable for surgery are operated on at the same hospital. The demand estimates measure where diagnosed patients are treated, by risk, and do not separate the patient's choice from her cardiologist's referral. The counterfactuals need only the value by risk, which determines the patients a program draws; the modality rule of Section~\ref{sec:dynamic_model} determines the procedure they receive. The counterfactuals hold this value fixed when a payment change alters the chance of TAVR, so they exclude any response of patients or referrers to that chance.

The technology terms are identified within hospitals and within markets. When hospital $j$ opens a program, its patient flow from every ZIP code changes while $\delta_j$ is fixed; 405 hospitals adopt between fiscal 2013 and 2019 in the estimation sample. Within a year, an adopter draws patients of different risk and conditions in different proportions than a non-adopter in the same market, which identifies $\alpha_{1t}$ and $\alpha_{kt}$ net of $\kappa_t$ and $\kappa_{kt}$, and programs of different ages identify the tenure terms. The travel parameters are identified by variation in patients' locations, the nearest-hospital term by the excess mass on the closest hospital, $\gamma_t$, $\kappa_t$ and $\kappa_{kt}$ by the share of the market treated at all and how it falls with risk and the conditions, and $\sigma_0$ by how that share responds when a hospital in a patient's set adopts.

Adoption is not random: large, high-volume hospitals adopt first, and a hospital may time its program to demand it expects. Hospital fixed effects absorb permanent differences in quality, scale and referral relationships, and year effects absorb aggregate changes in treatment, so the maintained assumption is that, conditional on them, the timing of adoption is unrelated to hospital-specific demand shocks. The threat is a shock other than the program that draws patients to a hospital at the date it adopts, such as a capacity expansion, a marketing campaign or new referral arrangements. The event study of hospital admissions around adoption in Appendix Figure~\ref{fig:es_admissions} argues against it \citep{SunAbraham2021}. Admissions grow slowly before adoption, by about 0.05 log points a year, as Medicare's volume requirements for opening a program imply, and jump by 0.38 log points in the adoption year, about seven years of the prior trend. The risk mix of admissions, flat before adoption, shifts upward at the same date, toward the patients TAVR serves; a shock to capacity or marketing that is not targeted at sicker patients would raise admissions across risk levels and load on the program's level rather than on $\alpha_{1t}$, which is identified from the composition of a hospital's patients rather than their number. A pre-existing trend would load on the tenure terms, which may therefore overstate how fast a program's value grows with its age. Within a program tenure moves with calendar time, which the year-specific adopter terms absorb, so the tenure terms are identified from programs of different ages in the same year and from the seven-year cap, and they also absorb differences between adoption cohorts.

\subsection{Estimation}
\label{sec:estimation-demand}

The parameters are estimated by simulated maximum likelihood on the market defined above, over fiscal years 2012 to 2019, 2.70 million patient-years and 933 hospitals. Choice probabilities are simulated with two-dimensional scrambled Halton draws, 15 per patient during optimization and 100 at the converged estimates. For a candidate parameter vector, the hospital effects are concentrated out: $\delta$ solves the condition that predicted and observed patient shares coincide for every hospital, with the outside option held at zero, and we maximize the log likelihood over the remaining parameters by a quasi-Newton method. With random coefficients, share matching is not the likelihood's first-order condition in $\delta$, so the estimator is not exact maximum likelihood. It is consistent, because at the true parameters expected and population shares coincide, and it is the standard approach when the number of fixed effects is large \citep{Berry1995, GoolsbeePetrin2004}.

\section{Demand results}
\label{sec:demand-results}

The estimates give three main results (Table~\ref{tab:demand_results}). First, travel is costly and patients prefer the nearest hospital, by amounts in line with other hospital-choice settings. Second, sicker patients are less likely to be treated at all, in line with the clinical evidence. Third, and central to the paper, the value of a TAVR hospital rises with the patient's risk and chance of receiving TAVR, and it reaches patients who almost always receive SAVR.

\begin{table}[htb]
\centering
\small
\caption{Patient demand for hospitals: estimated parameters}
\label{tab:demand_results}
\begin{threeparttable}
\begin{tabular*}{\linewidth}{@{\extracolsep{\fill}} l r r @{}}
\toprule
\multicolumn{3}{@{}l}{\textit{A. Travel and the preference for treatment}} \\
\midrule
Travel time (minutes), $\beta_1$ & \multicolumn{2}{r}{$-$0.0223} \\
Travel time squared ($\times 10^{6}$), $\beta_2$ & \multicolumn{2}{r}{3.94} \\
Nearest hospital, $\beta_c$ & \multicolumn{2}{r}{1.308} \\
Std.\ dev.\ of the travel coefficient, $\sigma$ & \multicolumn{2}{r}{0.0051} \\
Std.\ dev.\ of the preference for treatment, $\sigma_0$ & \multicolumn{2}{r}{0.0000} \\
\midrule
\multicolumn{3}{@{}l}{\textit{B. Value of a TAVR hospital}} \\
Fiscal year & Level, $\alpha_{0t}$ & $\times$ risk, $\alpha_{1t}$ \\
\midrule
2012 & $-$0.018 & 0.127 \\
2013 & $-$0.290 & 0.253 \\
2014 & $-$0.395 & 0.285 \\
2015 & $-$0.390 & 0.318 \\
2016 & $-$0.455 & 0.394 \\
2017 & $-$0.398 & 0.461 \\
2018 & $-$0.648 & 0.604 \\
2019 & $-$0.436 & 0.614 \\
\addlinespace
Tenure, $\psi_0$ & \multicolumn{2}{r}{0.0753} \\
Tenure $\times$ risk, $\psi_1$ & \multicolumn{2}{r}{0.0045} \\
\midrule
Log likelihood & \multicolumn{2}{r}{$-$1,674,627.8} \\
Patient-years & \multicolumn{2}{r}{2,696,108} \\
Hospitals & \multicolumn{2}{r}{933} \\
\bottomrule
\end{tabular*}
\begin{tablenotes}[flushleft]\footnotesize
\item \textit{Notes:} Mixed logit with hospital fixed effects, concentrated out, and an outside option normalized to zero. Panel B: the terms of a hospital's TAVR program, $a_{jt}(\alpha_{0t}+\alpha_{1t}r_i+(\psi_0+\psi_1r_i)E_{jt})$, with $r_i$ the STS predicted mortality in percentage points and $E_{jt}$ years since the program opened, capped at seven. The program terms for the eight conditions are in Table~\ref{tab:demand_conditions}; the year effects, the risk and condition terms of the value of treatment, and the risk interactions with hospital size and teaching status are in Appendix Table~\ref{tab:demand_baseline}. Standard errors to be added.
\end{tablenotes}
\end{threeparttable}
\end{table}

\begin{table}[htb]
\centering
\footnotesize
\caption{Value of a TAVR hospital by patient condition}
\label{tab:demand_conditions}
\begin{threeparttable}
\setlength{\tabcolsep}{3pt}
\begin{tabular*}{\linewidth}{@{\extracolsep{\fill}} l rrrrrrrr @{}}
\toprule
 & 2012 & 2013 & 2014 & 2015 & 2016 & 2017 & 2018 & 2019 \\
\midrule
Balloon valvuloplasty & 1.910 & 2.330 & 2.373 & 2.574 & 2.542 & 3.040 & 2.105 & 1.955 \\
Coronary bypass & 0.418 & 0.703 & 0.760 & 0.894 & 1.094 & 1.324 & 1.343 & 1.341 \\
Home oxygen or respiratory failure & 0.322 & 0.420 & 0.385 & 0.360 & 0.551 & 0.481 & 0.426 & 0.583 \\
Peripheral artery disease & 0.185 & 0.261 & 0.265 & 0.311 & 0.311 & 0.343 & 0.383 & 0.661 \\
Aortic atherosclerosis & 0.435 & 0.575 & 0.627 & 0.641 & 0.736 & 0.683 & 0.790 & 0.475 \\
Skilled-nursing or home-health use & 0.056 & 0.015 & 0.110 & 0.176 & 0.197 & 0.157 & 0.117 & 0.050 \\
Dialysis or CKD stage 4--5 & $-$0.076 & $-$0.078 & $-$0.124 & $-$0.219 & $-$0.056 & $-$0.307 & $-$0.216 & $-$0.405 \\
Prior valve surgery & 0.157 & 0.341 & 0.352 & 0.523 & 0.618 & 0.662 & 0.792 & 0.736 \\
\bottomrule
\end{tabular*}
\begin{tablenotes}[flushleft]\footnotesize
\item \textit{Notes:} Coefficients $\alpha_{kt}$ on the indicator of condition $k$ interacted with the hospital's TAVR program, by fiscal year, from the demand model of Table~\ref{tab:demand_results}. A coefficient of 0.0223 is worth one minute of travel. Conditions are recorded from the patient's claims in the twelve months before the admission (Section~\ref{sec:data}). Standard errors to be added.
\end{tablenotes}
\end{threeparttable}
\end{table}

Every additional ten minutes of driving lowers the odds of choosing a hospital by about 20 percent.\footnote{The travel coefficient is $-0.0223$ per minute, so ten minutes change the log-odds by $-0.223$, $e^{-0.223}\approx0.80$. With price fixed by Medicare, travel time is the price patients pay, and all other hospital attributes are measured against it in minutes.} The quadratic term is positive, so the disutility of travel flattens at long distances: it offsets about 2 percent of the linear term at 100 minutes and 4 percent at 200. The option of the closest hospital is worth about 59 minutes of travel on its own, and the standard deviation of the travel coefficient is 0.0051, about a quarter of the mean. Together, the premium and the gradient leave a hospital 20 minutes farther than the nearest with about 17 percent of its odds, against about 12 percent with the distance coefficient for childbirth in \citet{HoPakes2014}, at 30 miles per hour, and 9 percent with the standard-logit coefficient in \citet{RavalRosenbaum2021}.

Sicker patients are less likely to be treated. The risk gradient $\kappa_t$ is negative throughout (Appendix Table~\ref{tab:demand_baseline}), $-0.32$ per point of the risk score in 2012 and between $-0.39$ and $-0.69$ after: holding hospitals fixed, a patient one point riskier is less likely to undergo a valve replacement at all. The clinical evidence has the same sign: among patients with severe aortic stenosis, those who are not operated have higher surgical risk, a median STS score of 2.7 against 1.8 percent for those operated \citep{Bach2009AS}, and older patients and those with poor ventricular function are more often denied surgery \citep{Iung2005Decision}. Sicker patients who are treated, however, prefer a hospital that offers TAVR, as the third result shows. The dispersion of the preference for treatment is estimated at zero.\footnote{The inside-option baseline varies by year with the risk score and each of the eight conditions, and the data show no normal dispersion in the propensity to be treated beyond them; other forms of unobserved heterogeneity are not ruled out, and the random travel coefficient still mixes the inside--outside margin. We keep the term so that the specification nests the random-coefficient model; at $\sigma_0=0$ the inside--outside margin is a logit in the observed characteristics given the travel draw.}

\begin{figure}[tbp]
 \centering
 \includegraphics[width=\linewidth]{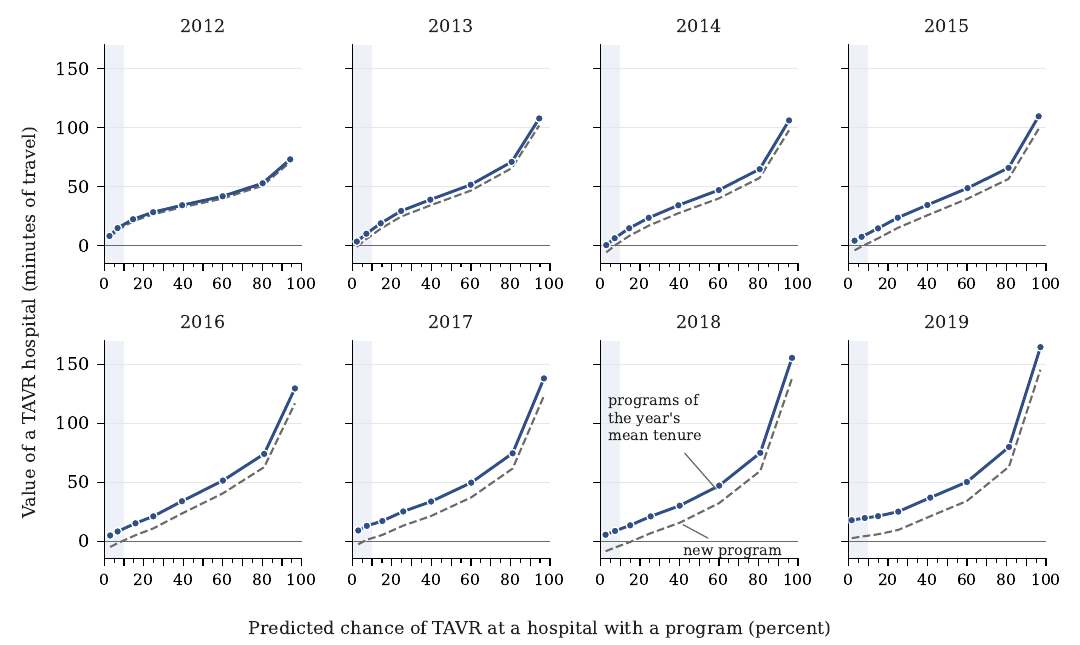}
 \caption{The value of a TAVR hospital and the chance of TAVR, by fiscal year. The chance of TAVR is predicted for each diagnosed patient by a logit, estimated separately for each year among patients treated at a hospital with a program, of TAVR on age (linear spline with knots at 75, 80 and 85), the logarithm of the STS risk score (linear spline with knots at 1, 2, 4 and 8 percent) and the eight conditions. The vertical axis is the value of a TAVR hospital, in minutes of travel, averaged over patients within bins of the predicted chance (0--4, 4--10, 10--20, 20--30, 30--50, 50--70, 70--90, 90--100 percent). Solid lines: at the mean tenure of the programs operating that year; dashed lines: at a new program. A bin is shown if it has at least 11 patients and at least 11 of them are treated at a hospital with a program. The shaded band marks a predicted chance below 10 percent.}
 \label{fig:demand_shift}
\end{figure}

Figure~\ref{fig:demand_shift} describes the value of an adopter against the patient's chance of TAVR, year by year. The chance is predicted, separately for each year, by a logit of TAVR on age, the risk score and the eight conditions among patients treated at a hospital with a program, and every diagnosed patient is given her predicted chance; the prediction matches the share actually receiving TAVR within a few percentage points in nearly every bin. The figure plots the average value of an adopter, in minutes of travel, within bins of that chance, at the mean tenure of the programs operating that year and, for reference, at a new program. The chance reflects the heart team's assignment, so the figure describes which patients value an adopter; it does not separate a preference for the hospital from one for the procedure. First, the value rises with the chance of TAVR, by 0.5 to 1 minute of travel per percentage point below a chance of 50 percent and more steeply above it. At a given chance below 50 percent the value is nearly the same in every year, between 30 and 39 minutes at a chance near 40 percent, so the rise in the value of an adopter over the sample comes mostly from the rise in the chance of TAVR as the indication widened; at a chance above 90 percent the value rose from 73 minutes in 2012 to 165 minutes in 2019. Second, patients who almost surely receive SAVR also travel to an adopter. At a predicted chance below 4 percent, patients value an adopter at between 0.4 and 18 minutes of travel, and at a chance near 7 percent at between 6 and 20 minutes, in every year, although roughly nine in ten of them or more receive SAVR. This value rises with the program's age: at a new program it is close to zero or negative for the lowest-risk patients in 2013--2018, and each year of operation adds about 3.5 minutes. That patients with almost no chance of TAVR value an adopter suggests that they value more than the procedure, the hospital or its program, although the estimates do not separate the two. The conditions matter because the score does not determine the chance alone: Table~\ref{tab:demand_conditions} reports the condition terms, and every condition other than dialysis raises the value of an adopter in every year, by as much as an hour of travel for prior bypass surgery in 2019. Prior bypass surgery, home oxygen, peripheral artery disease and aortic atherosclerosis raise both the probability of TAVR and the value of an adopter, and dialysis lowers both (Appendix Figure~\ref{fig:conditions}). A prior valve operation changes the probability of TAVR by between $-2.4$ and $2.8$ percentage points in every year, yet raises the value of an adopter by 15 to 39 minutes of travel.

Over time, as coverage widened by risk tier, the value of an adopter to the average patient receiving a valve replacement, TAVR or SAVR, rose from 13 minutes of travel in 2012 to 25 in 2015 and 67 in 2019 for a patient with none of the eight conditions, and from 27 to 96 minutes with one, and its slope in risk steepened, from $\alpha_{1t}=0.13$ per point in 2012 to $0.61$ in 2019. Within a year, older programs are worth more: each year of operation adds 3.4 minutes of travel for a patient at zero risk and 0.2 more per point of risk, about 4 minutes a year at 2 percent risk and 26 minutes after seven years. An early entrant therefore keeps the advantage its draw builds, the demand counterpart of the captured pool of Section~\ref{sec:stylized-framework}, and the reason a rival's early entry lowers a late entrant's return. The late adopters therefore entered a market in which the technology mattered to a larger share of patients than it had for the 2012 cohort, the demand-side counterpart of the diffusion documented in Section~\ref{sec:two-puzzles}.

\subsection{Use in counterfactuals}
\label{sec:demand-counterfactuals}

The demand system enters the supply model and the counterfactuals through each hospital's volume and the risk mix of its patients, with and without a program, since adoption shifts both. Because the market includes the untreated, about 370,000 patient-years in 2019 of which 56,000 are treated, part of a program's gain in admissions comes from patients who would otherwise have gone untreated. In estimation, rivals are held at their observed adoption status. In the counterfactuals, under any candidate adoption path $\{T_j^*(\mathcal{P})\}$ we recompute $\{a_{jt}(\mathcal{P})\}$, the choice probabilities $P_{ijt}(\mathcal{P})$ from~\eqref{eq:Prob}, and the flows $Q_{jt}(\mathcal{P})$. Each hospital's rivals adopt on the dates they choose under the policy, not on their observed dates. The steering primitive~\eqref{eq:DeltaY-steer} is recomputed at these flows and feeds each hospital's adoption value. The travel-voucher policy enters through the same system. The voucher pays a patient's extra travel to a TAVR hospital when her nearest hospital has no program: it adds to that hospital's utility the subsidized minutes valued at her own travel coefficient, $|\beta_{1,i}|$, times her chance of receiving TAVR, so its effect on choice is set in minutes, and the consumer-surplus change is computed from the same utilities in travel-time units.

\section{Adoption timing}\label{sec:dynamic_model}

The demand estimates give each hospital's volume with and without a program in every year, given rivals' adoption. Combined with reimbursement and costs from the claims, they give the present value of adopting at each candidate date, and observed adoption dates identify the cost of adoption. The model recovers a distribution of adoption costs in dollars, a median of \$6.5M for adopters against \$11.0M for hospitals that never adopt, reproduces the 2012 wave and the absence of a second one in 2015, and attributes nine tenths of the value of adopting to the patients a program draws.

\subsection{Demand, modality and operating profit}

Each hospital chooses an adoption year once, irreversibly, under perfect foresight of the reimbursement path and given its rivals' adoption paths, among the years in which it meets the volume requirement. At $t_0$ it draws a shock for each candidate year and chooses the year that maximizes the present value of adopting net of the shock; because all shocks are known at $t_0$, waiting has no option value, and a hospital delays only because a later date is worth more. In each year after adoption, patients choose hospitals, the hospital assigns each admitted patient to TAVR or SAVR, and profits are realized.

Let $\theta$ denote a patient's type: the predicted operative mortality $r$ and the conditions $\mathbf{s}$ of Section~\ref{sec:data}. The market in year $t$, the diagnosed beneficiaries of Section~\ref{sec:demand}, has size $M_t$ and type distribution $F_t(\theta)$. Demand gives the probability that a patient of type $\theta$ is treated at $h$,
\[
s_{h,t}(\theta; a_h, A_{-h,t}) \equiv \Pr(\text{treated at }h \mid \theta,t; a_h, A_{-h,t}),
\]
where $a_h\in\{0,1\}$ indicates whether $h$ has adopted and $A_{-h,t}$ stacks rivals' adoption states, held at their observed values. Expected admissions are
\begin{equation*}
Y_{h,t}(a_h)=M_{t}\int s_{h,t}(\theta; a_h,A_{-h,t})\,dF_{t}(\theta).
\end{equation*}

For adopters, let $p_{h,t}(\theta)=\Pr(\mathrm{TAVR}\mid h,\theta,t)$ be the probability that an admitted patient receives TAVR; for non-adopters $p_{h,t}(\theta)=0$. We estimate $p$ as a logit on the claims of adopting hospitals,
\begin{equation}\label{eq:modality}
\begin{split}
p_{h,t}(\theta)=\Lambda\Bigl(&\phi_h+\beta_{\theta,t}\,r+\mathbf{s}'\boldsymbol{\rho}_{q(t)}+g\,(t-2012)\\
&+d\,\mathbf{1}\{t\ge 2015\}+d_{14}\mathbf{1}\{t\ge 2014\}+d_{16}\mathbf{1}\{t\ge 2016\}+\gamma\,\Delta_{h,t}(\theta)\Bigr),
\end{split}
\end{equation}
where $\Lambda$ is the logistic function, $\phi_h$ a hospital effect, $\beta_{\theta,t}$ a risk slope that differs by year, $\boldsymbol{\rho}_{q(t)}$ the effects of the eight conditions, with dialysis and prior valve surgery combined, which differ by period $q(t)$, and $\Delta_{h,t}(\theta)$ the per-case margin gap between the two procedures defined below. The common time term is a linear trend $g$, a step $d$ from fiscal 2015 and steps $d_{14}$ and $d_{16}$ from 2014 and 2016, when new devices and indications were approved, rather than a set of year effects, for the reason given in Section~\ref{sec:supply-estimates}; only $d$ moves with the payment in the counterfactuals. Patients above the 99th percentile of SAVR patients' risk in that year receive TAVR wherever a program exists; at a hospital without one they earn it no margin, which affects about 1 percent of SAVR patients. Per-case margins are
\[
m^m_{h,t}(\theta)=\text{reimb}^m_{h,t}(\theta)-\text{cost}^m_{h,t}(\theta),\qquad
\Delta_{h,t}(\theta)=m^T_{h,t}(\theta)-m^S_{h,t}(\theta),\qquad m\in\{T,S\},
\]
and period-$t$ operating profit is
\begin{equation*}
\pi^{op}_{h,t}
=M_{t}\int s_{h,t}(\theta; a_h,A_{-h,t})\Big[m^S_{h,t}(\theta)+p_{h,t}(\theta)\,\Delta_{h,t}(\theta)\Big]\,dF_{t}(\theta).
\end{equation*}
Charges and reimbursements per case are predicted from regressions on the claims, by modality, with year effects, risk-bucket effects, a quadratic in risk and indicators for the conditions, with dialysis and prior valve surgery combined. Hospital effects are estimated on SAVR claims and imposed on TAVR claims, so a hospital's TAVR level is its SAVR level plus the common TAVR terms. The 97 pivotal-trial sites account for 83 percent of TAVR claims in 2012 and share the common TAVR year effects; estimating separate year effects for them changes the value of adopting by at most \$0.1M per cohort. TAVR charges also depend on the hospital's years of experience with the procedure, $E_{h,t}=\max\{0,\,t-\tau_h\}$, entered as indicators. Because the hospital effect is held at its SAVR value, the experience profile is identified from differences in experience across hospitals within a year. The assumption is that a hospital's TAVR charges relative to its SAVR charges do not vary with its adoption date except through experience. The hospital observes a margin shock when it assigns the modality, with the variance of the regression residuals, and it is integrated out. For hospitals without TAVR claims, the modality hospital effect is not identified, and we set it, and their TAVR charge and reimbursement levels, to fitted values from a regression of the identified effects on standardized log beds and region.\footnote{Giving these hospitals their own SAVR charge and payment effects as TAVR levels instead, and re-estimating the adoption stage, raises the median adoption cost from \$6.5M to \$6.7M for adopters and from \$11.0M to \$11.7M for hospitals that never adopt, and the scale $\hat\sigma_\varepsilon$ from \$2.63M to \$2.81M; the modality estimates are unchanged and $\hat\sigma_\nu$ stays near zero.} Costs follow the cost-to-charge approach of \citet{reddy2025contribution},
\begin{equation}\label{eq:costfn}
\text{cost}^m_{h,t}(\theta)\;=\;c_{h,t}\cdot ch^m_{h,t}(\theta),\qquad m\in\{T,S\},
\end{equation}
where $ch^m_{h,t}(\theta)$ is charges per case and $c_{h,t}$ is the hospital's ratio of costs to charges in fiscal year $t$ from its Medicare cost report, net of fixed overhead. The ratio of total costs to total charges averages 0.25 and falls from 0.27 in 2012 to 0.22 in 2019. An additional admission does not change the cost of buildings, equipment, administration or plant operation, so we multiply the ratio by the share of the patient-care departments' cost that is not allocated from those departments (Worksheet B of the cost report). The share averages 0.76 and barely moves over the sample, so the ratio we use averages 0.19; missing hospital-years take the year's median.\footnote{The ratio is hospital-wide, so the implied TAVR cost applies the hospital's average markup to a charge that includes the device.} Experience enters the charge predictions and the risk composition of admitted patients.

Because adoption changes which patients choose $h$, both admissions and the composition of admitted patients depend on $a_h$. Let $H^{a_h}_{h,t}$ denote the type distribution of the patients hospital $h$ admits under state $a_h$, the market distribution reweighted by $M_t\,s_{h,t}(\theta;a_h,A_{-h,t})$, and let the per-admission profit be
\[
\pi^{\mathrm{adm}}_{h,t}(\theta;a_h)
\;=\;
m^S_{h,t}(\theta)\;+\;p_{h,t}(\theta;a_h)\,\Delta_{h,t}(\theta),
\qquad p_{h,t}(\theta;0)\equiv 0 .
\]
Operating profit under $a_h$ is admissions times the mean per-admission profit over the admitted mix,
\[
\pi^{op}_{h,t}(a_h)
\;=\; Y_{h,t}(a_h)\cdot
\underbrace{\mathbb E_{\theta\sim H^{a_h}_{h,t}}
\big[\pi^{\mathrm{adm}}_{h,t}(\theta;a_h)\big]}_{\displaystyle \mu^{a_h}_{h,t}},
\]
so the period-$t$ incremental profit from adoption decomposes into a volume effect and a composition effect,
\begin{align}
\Delta\pi_{h,t}(A_{-h,t})
&\equiv \pi^{op}_{h,t}(1)-\pi^{op}_{h,t}(0) \nonumber\\
&=\underbrace{\big[Y_{h,t}(1)-Y_{h,t}(0)\big]\;\mu^{0}_{h,t}}_{\text{admissions at pre-adoption mix}}
\;+\;\underbrace{Y_{h,t}(1)\,\big[\mu^{1}_{h,t}-\mu^{0}_{h,t}\big]}_{\text{composition shift at adopted mix}}.
\label{eq:delta_pi_decomp}
\end{align}
The first term is the steering return of Section~\ref{sec:stylized-framework} in dollars. The second records that after adoption $h$ admits a different risk mix and earns different per-admission dollars through the modality rule $p_{h,t}(\theta;1)$ and the margins $\Delta_{h,t}(\theta)$.

\subsection{The timing problem}

Each hospital chooses $\tau_h\in\mathcal T_h\subseteq\{t_0,t_0{+}1,\ldots,T,\infty\}$, where $\infty$ denotes never, $t_0=2012$ and $T=2019$. The set $\mathcal T_h$ contains never and every year in which the hospital meets the volume requirement of Medicare's coverage rule, measured as at least ten fee-for-service AVRs in the year before (Section~\ref{data}); an observed adoption in a year in which the hospital falls short is kept in its set. The present value at $\tau$ of adopting at $\tau$, with annual discount factor $\delta=0.92$, is
\[
\begin{split}
\Delta\Pi_h(\tau)={}&\sum_{t=\tau}^{T}\delta^{\,t-\tau}\Delta\pi_{h,t}(A_{-h,t})\\
&+\sum_{s=1}^{\infty}\delta^{\,T+s-\tau}\max\bigl\{0,\,\Delta\pi_{h,T}\bigl(A_{-h,T};\min\{T+s-\tau,\bar E\}\bigr)\bigr\},\\
\Delta\Pi_h(\infty)={}&0,
\end{split}
\]
where $\Delta\pi_{h,T}(A_{-h,T};e)$ is the incremental flow in the last sample year of a program of tenure $e$ and $\bar E=7$ is the tenure cap of the demand model, so that after 2019 a program keeps the conditions of 2019 while its tenure keeps growing to the cap. Let $K_{h,\tau}$ be the net one-time cost of adopting in year $\tau$, specified below, and let $\varepsilon_{h\tau}$ be i.i.d.\ type-I extreme value with scale $\sigma_\varepsilon$. The adoption year solves
\[
\tau_h\in\arg\max_{\tau\in\mathcal T_h}
\Bigl\{\delta^{\,\tau-t_0}\bigl(\Delta\Pi_h(\tau)-K_{h,\tau}\bigr)+\varepsilon_{h\tau}\Bigr\},
\]
so that with $v_{h\tau}=\delta^{\,\tau-t_0}\bigl(\Delta\Pi_h(\tau)-K_{h,\tau}\bigr)/\sigma_\varepsilon$ and $v_{h,\infty}=0$ the timing probabilities are multinomial logit over the candidate years in $\mathcal T_h$ and the never-adopt option,
\begin{equation}\label{eq:timinglogit}
P_h(\tau)=\frac{\exp v_{h\tau}}{1+\sum_{\tilde\tau\in\mathcal T_h\setminus\{\infty\}}\exp v_{h\tilde\tau}}.
\end{equation}
Profits and the adoption cost are both discounted to $t_0$, so the bracket is the net present value at $t_0$ of a cost paid at $\tau$. We estimate the scale $\sigma_\varepsilon$; it converts adoption cost into dollars and is identified by how sharply adoption responds to cross-sectional variation in $\Delta\Pi-K$.

Estimation proceeds in two stages. The modality rule is estimated by maximum likelihood on the claims of adopting hospitals, and the profits $\Delta\pi_{h,t}$ are built from it, the margin predictions, and the demand shares. The timing parameters are then estimated by maximum likelihood on observed adoption years, with the profits held fixed.

\subsection{Adoption cost}\label{sec:supply-results}

We parameterize
\begin{equation}\label{eq:Kdet}
K_{h\tau}=\kappa_0+f(Z_h)+X_h\beta_X+\psi_{r(h)}+\lambda_\tau+\nu_h,
\qquad \nu_h\sim N(0,\sigma_\nu^2),
\end{equation}
where $Z_h$ is standardized log beds and $f$ is piecewise linear in it, with slope changes at the two tercile cuts, $X_h$ is an indicator for teaching status, $\psi_{r(h)}$ are region effects (states, with small states grouped into contiguous regions so that every region has both adopters and hospitals that never adopt), and $\lambda_\tau$ are adoption-year effects. The year terms are a level after 2012, a linear drift and shifts from 2014 and 2016, the years of new device approvals, $\lambda_\tau=c+g\,(\tau-t_0)+s_{14}\mathbf{1}\{\tau\ge2014\}+s_{16}\mathbf{1}\{\tau\ge2016\}$ for $\tau>t_0$ and zero in 2012; $f$ is the same in every year. The permanent effect $\nu_h$ is integrated out by Gauss--Hermite quadrature over the choice probabilities~\eqref{eq:timinglogit}, so the timing model is a mixed logit. The 97 pivotal-trial sites, which ran TAVR programs in the clinical trials before Medicare coverage, are treated as already in: they enter the demand side and the counterfactuals as adopters from the first year but make no timing choice, so the timing likelihood runs over the remaining 848 hospitals, of which 521 adopt by 2019.

$K_{h\tau}$ enters the index linearly, with free sign. Because $\Delta\Pi_{h\tau}$ counts only the incremental Medicare fee-for-service margin, $K_{h\tau}=F_{h\tau}-B_h$ is the setup cost net of returns the model does not include, from Medicare Advantage and commercial patients, referrals into the rest of the cardiac service line, or other objectives; a negative $K$ means that these returns exceed the setup cost. These returns are held fixed in the counterfactuals, including any part of them that depends on rivals' entry. Only the difference is identified, as in the entry literature \citep{BresnahanReiss1991, Berry1992, CilibertoTamer2009}. The return to entering before rivals that runs through Medicare patients is not in $K$: the tenure terms of demand give an early entrant a larger draw than a later rival, and $\Delta\Pi$ carries it. $K_{h\tau}$ is the empirical counterpart of the fixed cost $F_i$ of the two-stage model. Each hospital makes one adoption decision, so we impose no hospital fixed effect; $\nu_h$ is its random-effects counterpart.

\begin{table}[!tbp]
\centering
\footnotesize
\caption{Supply model estimates}
\label{tab:supply_estimates}
\begin{threeparttable}
\begin{tabular}{lr@{\hspace{2.2em}}lr}
\toprule
\multicolumn{2}{c}{Stage 1: modality} & \multicolumn{2}{c}{Stage 2: adoption timing} \\
\cmidrule(lr){1-2} \cmidrule(lr){3-4}
Parameter & Estimate & Parameter & Estimate \\
\midrule
$\gamma$ (margin gap) & 0.400 & $\sigma_\varepsilon$ & 26.282 \\
 & (0.190) & $\sigma_\nu$ & 0.047 \\
$\beta_\theta$ (risk) & 1.053 & $\kappa_0$ & 175.300 \\
 & (0.027) & size (log beds, standardized) & 1.927 \\
risk $\times$ 2013 & 0.297 & size, added above the first tercile cut & -70.211 \\
 & (0.036) & size, added above the second tercile cut & -4.177 \\
risk $\times$ 2014 & 0.413 & teaching, $X_1$ & -18.921 \\
 & (0.036) & $\lambda$ 2013 & 27.044 \\
risk $\times$ 2015 & 0.626 & $\lambda$ 2014 & 9.919 \\
 & (0.035) & $\lambda$ 2015 & 12.494 \\
risk $\times$ 2016 & 0.746 & $\lambda$ 2016 & 7.911 \\
 & (0.035) & $\lambda$ 2017 & 10.486 \\
risk $\times$ 2017 & 0.892 & $\lambda$ 2018 & 13.061 \\
 & (0.035) & $\lambda$ 2019 & 15.636 \\
risk $\times$ 2018 & 1.108 & hospitals & 945 \\
 & (0.035) & adopters 2012--2019 & 521 \\
risk $\times$ 2019 & 1.063 & log likelihood & -1,261.5 \\
 & (0.036) &  &  \\
trend, per year & 0.451 &  &  \\
 & (0.008) &  &  \\
step from 2014 & 0.143 &  &  \\
 & (0.031) &  &  \\
step from 2015 & 0.107 &  &  \\
 & (0.033) &  &  \\
step from 2016 & 0.152 &  &  \\
 & (0.027) &  &  \\
claims & 315,918 &  &  \\
hospitals & 613 &  &  \\
log likelihood & -143,417.9 &  &  \\
\bottomrule
\end{tabular}
\begin{tablenotes}[flushleft]\footnotesize
\item \textit{Notes:} Standard errors (Stage 1) in parentheses, conditional on the first-step charge, payment and risk estimates and the fixed costs; none for Stage 2. The margin gap $\Delta$ that multiplies $\gamma$ is in \$100{,}000. The common time term in Stage 1 is a linear trend in years since 2012 plus steps from fiscal 2014, 2015 and 2016. $\sigma_\varepsilon$, $\sigma_\nu$, $\kappa_0$ and the size, teaching and $\lambda$ terms are in units of \$100{,}000, so 8.10 is \$0.81 million. Size enters as a piecewise-linear function of standardized log beds whose slope changes at the two tercile cuts, the same in every year. The $\lambda$ terms follow a level after 2012, a linear trend and shifts from 2014 and 2016. The 97 pivotal-trial sites that performed TAVR before fiscal 2012 are treated as already in and make no timing choice; the adopter count excludes them.
\end{tablenotes}
\end{threeparttable}
\end{table}

\subsection{Estimates}\label{sec:supply-estimates}

Table~\ref{tab:supply_estimates} reports both stages. The adoption costs are of the order of industry estimates, and the model reproduces the timing of entry.

\begin{figure}[htb]
 \centering
 \includegraphics[width=0.72\linewidth]{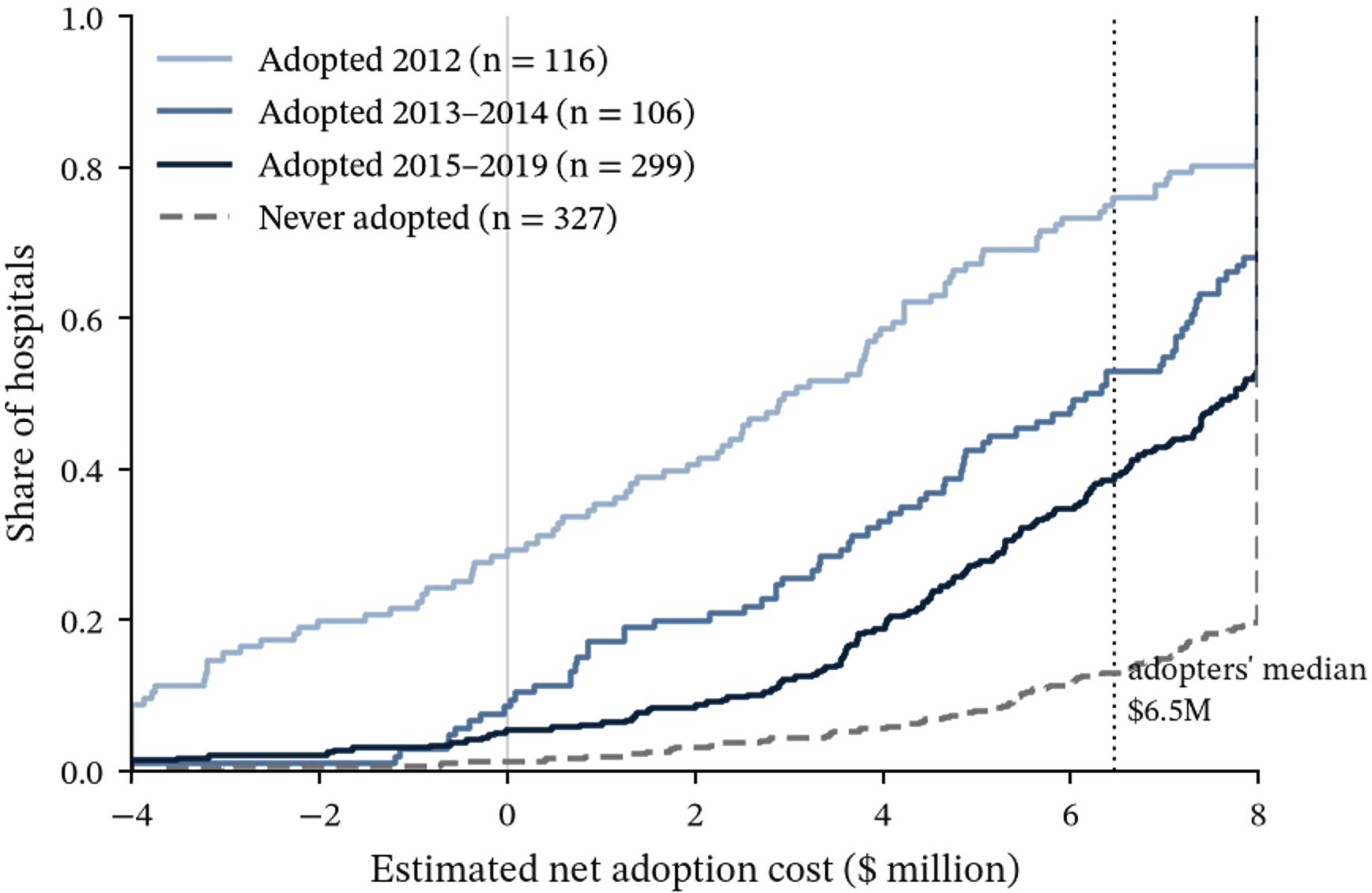}
 \caption{Estimated adoption cost by adoption cohort. Each curve is the cumulative distribution of $K_{h\tau}$ across hospitals, in millions of dollars, read at the adopter's own adoption year or, for hospitals that never adopt, at the average across candidate years. The dotted line marks the adopters' median, \$6.5M. The 2012 cohort excludes the 97 trial sites treated as in before the window.}
 \label{fig:fixed_costs}
\end{figure}

The median adopter's adoption cost is \$6.5M, with an interquartile range of \$3.4M to \$9.7M, of the order of industry estimates of the gross cost of a program, about \$1M, and of the hybrid operating room it requires, \$3--5M \citep{tangorra2015roi, ecri2013hybrid, fornell2011hybrid}. The scale $\hat\sigma_\varepsilon$ is \$2.63M and the standard deviation of the permanent hospital effect $\hat\sigma_\nu$ is \$0.005M: once size, teaching and region are accounted for, there is almost no permanent unobserved heterogeneity in adoption cost, and the remaining dispersion is the date-specific shock $\varepsilon_{h\tau}$, which varies across a hospital's candidate years. The distribution of $K$ separates by adoption status (Figure~\ref{fig:fixed_costs}). The median hospital that never adopts faces \$11.0M against \$6.5M for one that does. Adoption cost is negative for 11 percent of adopters, 59 of 521, and for 4 of the 327 never-adopters.\footnote{$K_{h\tau}$ is read at each adopter's own adoption year and, for hospitals that never adopt, at the average across candidate years, the measure the counterfactuals use. A negative estimate records $B_h>F_{h\tau}$ in~\eqref{eq:Kdet}: the returns to a program outside the incremental Medicare margin exceed its setup cost. Most such hospitals, 34 of the 59, adopted in 2012, where the year profile places the cost lowest: hospitals equipped before coverage had incurred much of the setup cost before the window. These hospitals are predicted to adopt, so they are never offered the lump sum, which targets predicted non-adopters, and they enter the counterfactuals mainly through the rate increment paid on their TAVR cases.}

Size carries most of the observable variation in adoption cost above the smallest third of hospitals: there a hospital one standard deviation larger in log beds faces a cost lower by about \$7M (\$6.8M between the tercile cuts, \$7.2M above the second), while among the smallest third size barely moves it. Teaching hospitals face a cost lower by a further \$1.9M. The adoption-year terms are a level of \$2.4M after 2012 and a drift of \$0.26M a year, with shifts of $-$\$2.0M from 2014 and $-$\$0.7M from 2016, so adopting costs \$2.7M more in 2013 than in 2012, \$1.0M more in 2014 and \$1.6M more in 2019. We treat the year profile as a nuisance term: it absorbs whatever made 2012 attractive beyond the margins in the model, including the release at coverage of hospitals that were equipped and willing but could not bill.

\begin{figure}[htb]
 \centering
 \includegraphics[width=\linewidth]{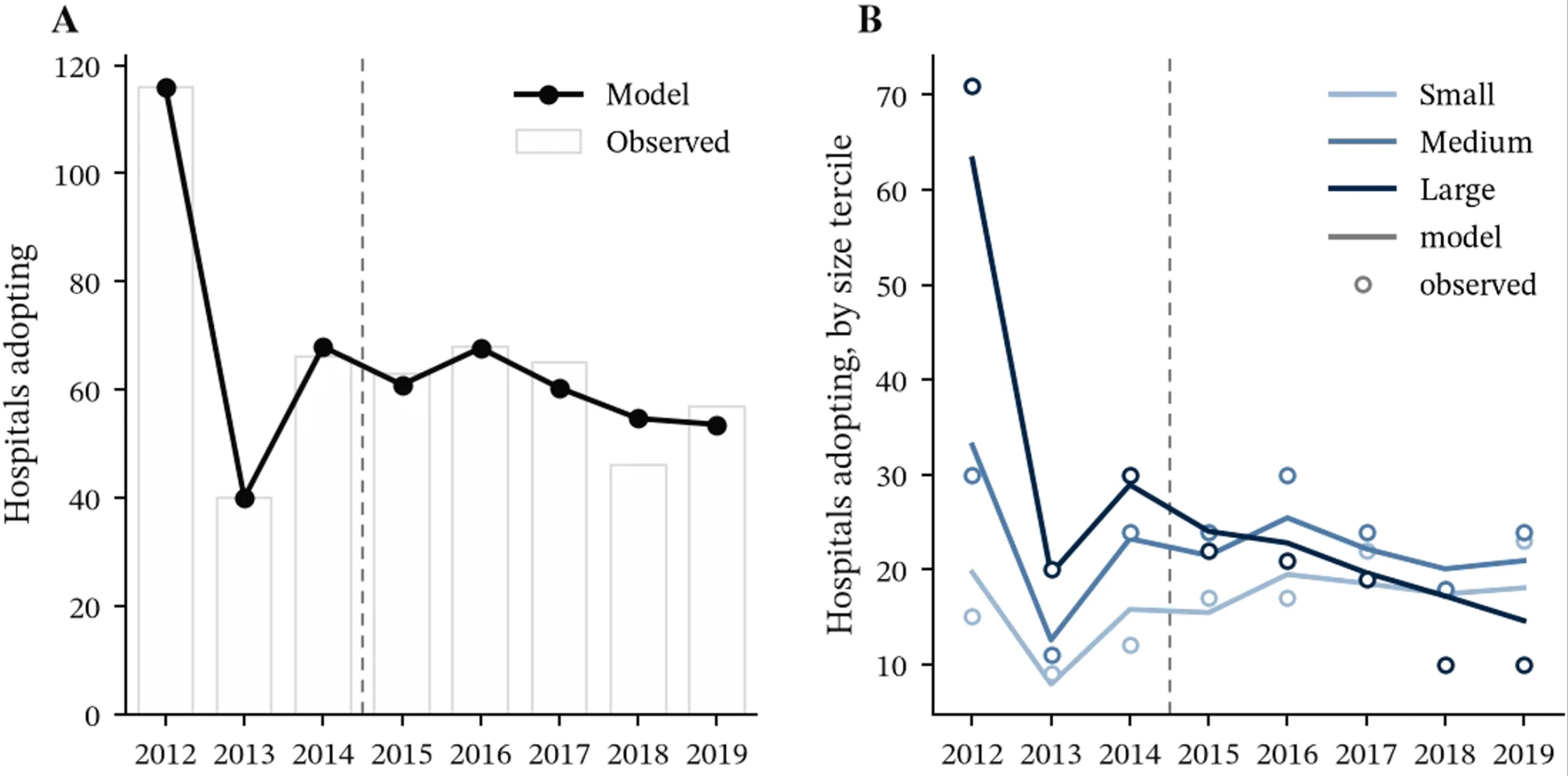}
 \caption{Model fit. Panel A shows hospitals adopting TAVR in each fiscal year, observed (bars) and implied by the estimated timing probabilities (line), for the 848 hospitals that choose when to enter, of which 521 adopt by 2019; the 97 trial sites in before the window are excluded. Panel B shows the same by tercile of hospital size, model (lines) and observed (circles). The dashed line marks the fiscal 2015 payment revision.}
 \label{fig:model_fit}
\end{figure}

Figure~\ref{fig:model_fit} compares observed and model-implied entry by fiscal year. The model matches the 2012 wave and the total, 521 adopters observed against 521.0 predicted, and reproduces the absence of a second wave in 2015. After 2012 it matches 2013, stays within three hospitals of the observed count in 2014 to 2016 and in 2019, under-predicts 2017 by five and over-predicts 2018 by nine; the size panel shows that the over-prediction in 2018 and 2019 is among large hospitals, offset in 2019 by small and medium ones.

\subsection{The mechanism of early entry}\label{sec:mechanism}

Early entry is financed by demand: the patients a program steers to its hospital and keeps as the program ages. Table~\ref{tab:steering} shows it in two steps. Panel A reports the value of adopting by cohort. In 2012 a program raises an adopter's valve patients from 42 to 63 a year, and 15 of the 21 it adds are expected to receive SAVR, in line with the 68 percent observed in the claims (Section~\ref{sec:puzzle1-tests}). Of its 12 TAVR cases, 5 are added patients and 6 are its own patients moved from surgery, so its surgery volume still rises by 9. Added patients carry 87 to 94 percent of the value of adopting in every cohort: those expected to receive TAVR 67 to 74 percent, those expected to receive SAVR 15 to 26 percent. Surgery patients dominate the early inflow, but most of the value accrues later, as the indication widens and the revised rate applies. The value falls from \$11.2M for the 2012 cohort to \$4.7M for the 2019 cohort, and a shrinking part of it accrues by 2019, as later cohorts operate fewer years inside the window. By 2019 a hospital without a program would keep 18 of the 67 patients it treats with one, because its rivals have adopted.

Panel B uses the recovered adoption costs to simulate entry when part of that value is removed, holding rivals at their observed dates. Without steering, $\Delta Y^{\text{steer}}=0$ in~\eqref{eq:DeltaY-steer}, a program keeps the patients the hospital would have had without one and gains only through the modality rule on them. The value of adopting then falls by nine tenths, and \emph{predicted entry falls by half before the revision} and by a sixth after it; the 2012 wave falls to 43 percent of its size. Removing only the added patients expected to receive SAVR lowers the value by a fifth and the 2012 wave by a fifth, and leaves entry after the revision essentially unchanged; removing only those expected to receive TAVR lowers the value by seven tenths and the 2012 wave by two fifths. Patients drawn to surgery are a material part of the return to early entry, but most of it comes from patients the program brings in for the new procedure. Entry falls less than value because adoption costs are dispersed and lowest in the 2012 cohort. Holding a program's patient draw at its first-year level, so that an early entrant gains nothing from tenure, removes 44 percent of the value and a third of the 2012 wave. The year profile of adoption cost matters less than steering: without the jump in cost between 2012 and 2013, the level $c$ in~\eqref{eq:Kdet}, the 2012 wave falls to 63 percent of its size, and because the jump also raises the cost in every later year, total entry rises by 11 percent.

\begin{table}[htb]
\centering
\caption{The value of adopting and entry without steering}
\label{tab:steering}
\small
\begin{threeparttable}
\begin{tabular}{lcccc}
\toprule
\multicolumn{5}{l}{\textit{A. Value of adopting at the adoption year, by cohort}}\\
 & Adopters & Value (\$M) & From TAVR (\%) & By 2019 (\%) \\
\midrule
2012 & 116 & 11.2 & 91 & 37 \\
2013 & 40 & 10.5 & 94 & 32 \\
2014 & 66 & 7.5 & 98 & 29 \\
2015 & 63 & 8.0 & 96 & 24 \\
2016 & 68 & 8.3 & 95 & 20 \\
2017 & 65 & 7.8 & 95 & 14 \\
2018 & 46 & 6.1 & 92 & 9 \\
2019 & 57 & 4.7 & 85 & 4 \\
\midrule
\multicolumn{5}{l}{\textit{B. Predicted entry, percent of the baseline}}\\
 & Value & \multicolumn{3}{c}{Hospitals adopting} \\
\cmidrule(lr){3-5}
 & of adopting & 2012 & 2012--2014 & 2015--2019 \\
\midrule
Baseline (level) & \$8.3M & 116 & 224 & 297 \\
No steering & 10 & 43 & 50 & 84 \\
No SAVR inflow & 80 & 80 & 86 & 102 \\
No TAVR inflow & 29 & 59 & 63 & 88 \\
No tenure gain & 56 & 66 & 73 & 99 \\
No 2013 cost step & 100 & 63 & 100 & 118 \\
No steering, no 2013 step & 10 & 28 & 55 & 105 \\
\bottomrule
\end{tabular}
\begin{tablenotes}[flushleft]\footnotesize
\item \textit{Notes:} Panel A: mean over the hospitals of each adoption cohort of the value of adopting at the adoption year, discounted to that year, with the share that comes from TAVR cases (the rest is the change in surgery profit) and the share that accrues by 2019 (the rest is the value after 2019). Panel B: value of adopting at the adoption year, mean over observed adopters, and expected number of hospitals adopting, from the timing probabilities at the estimated costs (predicted, so the baseline row differs slightly from the observed cohorts of Panel A), with rivals at their observed dates; the baseline row is in levels, the others in percent of it. No steering: $\Delta Y^{\text{steer}}=0$. No SAVR inflow, no TAVR inflow: the added patients expected to receive SAVR, or TAVR, removed from $\Delta Y^{\text{steer}}$, split by the procedure the modality rule is expected to assign. No tenure gain: a program's patient draw held at its first-year level. No 2013 cost step: $c=0$ in~\eqref{eq:Kdet}.
\end{tablenotes}
\end{threeparttable}
\end{table}

The modality rule shows the other side of this mechanism: the payment acts on the procedure. It is estimated on 315,918 claims at the 613 adopting hospitals whose post-adoption claims include both procedures; at the others every admitted patient receives the same procedure, so the hospital effect is not identified and is set to its projection on size and region, as for hospitals that never adopt. Assignment to TAVR rises steeply with risk, and more steeply each year as the indication widened. The common time term in~\eqref{eq:modality} is a linear trend with steps rather than a set of year effects. The revision changed the margin gap at every hospital at once, so year effects would absorb it and leave $\gamma$ identified only from the small differences in the gap across hospitals and patients within a year; the trend and step leave the year-to-year movement of the gap around that path to identify it as well. The margin coefficient is $\hat\gamma=0.40$ (s.e.\ 0.19): a larger TAVR margin over SAVR raises the probability that a program assigns an admitted patient to TAVR. The trend is 0.45 a year and the step from 2015 is 0.11 (s.e.\ 0.03), next to steps of 0.14 from 2014 and 0.15 from 2016. The counterfactuals move the 2015 step with the payment change, so a rate change shifts the TAVR share at programs already operating through both $\gamma$ and $d$; the step also absorbs other changes in fiscal 2015, such as the valve generations approved at its end, so this attribution is generous to the payment.

Section~\ref{sec:counterfactuals} takes three objects from these estimates: the timing probabilities $P_{h\tau}$, the adoption cost $K_{h\tau}$, and each hospital's predicted probability of never adopting, which the targeted transfer conditions on.

\section{Counterfactual policy design}\label{sec:counterfactuals}

The adoption-timing model of Section~\ref{sec:dynamic_model} delivers, for every hospital, the value of adopting at each candidate date and an adoption cost in dollars; the demand model of Section~\ref{sec:demand} delivers how patients move when access changes and what that movement is worth to them in travel time. Together they allow us to re-solve every hospital's entry decision and every patient's hospital choice under any payment rule. We use them to trace the policy frontier of a payer with three tools: the year in which the rate rises, a lump sum paid at adoption to hospitals predicted not to enter, and a travel voucher paid to patients who receive TAVR. Because a hospital's return to adoption depends on its rivals' entry, each policy is evaluated at the Nash equilibrium in adoption dates it induces, and the frontier gives, at each outlay, the most hospitals offering TAVR, the most patients treated and the most patient welfare that any combination of the three tools reaches in equilibrium. Section~\ref{sub:mechanism} defines the tools and says what each does, Section~\ref{sub:ic} the equilibrium, and Section~\ref{sub:cf-results} the frontier. At the outlay of the observed revision the frontier delays the rate rise and spends the saving on the voucher and, for coverage, on the targeted lump sum; it reaches 52 more hospitals than the observed policy, or 32 percent more TAVR patients for 9 percent less.

\subsection{Policy instruments}\label{sub:mechanism}

We consider three policy instruments; Appendix~\ref{app:planner} derives what each does in the two-stage model of Section~\ref{sec:stylized-framework}. The first instrument is the year $t_R$ in which the 2015 payment rise takes effect, at its observed size, any fiscal year from 2012 to 2018. One would expect an earlier rise to bring hospitals in sooner, since it raises the margin on every TAVR case. When the value of adopting comes mostly from the patients a program draws rather than from the margin on the procedure, the result is different: hospitals whose return from patients covers their adoption cost enter without the rise, so an earlier rise pays them on every case and recruits only hospitals whose shortfall at the revised rate is smaller than one stage of profit (Corollary~\ref{cor:accelerate}). It acts mainly on the procedure at programs already operating.

The second instrument is a lump sum $S$, from \$0.5 million to \$15 million, paid at adoption from $t_R$. One would expect a subsidy for entry to bring hospitals in. Offered to every hospital, however, it also pays those that would enter anyway and delays early entrants near their cutoff, which wait to collect it (Proposition~\ref{prop:gate}). We therefore pay it only to hospitals that the model predicts, with probability above one half, will not adopt by 2019 and that would treat at least ten TAVR patients a year, about 135 hospitals without the voucher and 90 with it. Targeted this way it screens out the hospitals the rise pays. The delay identifies the inframarginal recipients: hospitals that enter at the unrevised rate have a return to early entry, and because the lump sum is paid at adoption from $t_R$, they never collect it (Proposition~\ref{prop:targeting}). Targeting is imperfect: the permanent part of adoption cost is a function of observables ($\hat\sigma_\nu=\$0.005$M) but the date-specific shock is not ($\hat\sigma_\varepsilon=\$2.63$M), so some recipients would have entered unaided.

The third instrument is a travel voucher that pays a TAVR patient \$1 for each extra minute she travels when her nearest hospital has no program. One would expect a subsidy to patients to move patients but not hospitals. When the return to adoption runs through patients the result is different, and it moves both: it enlarges the pool every adopter draws on, so it draws patients to existing programs and raises the return to entry at both dates, without giving an early entrant a reason to wait. Patients choose a hospital before the heart team assigns the procedure, so the voucher enters choice at its expected value, the subsidized minutes times the share of patients like her who receive TAVR at adopters that year; the value of a minute in dollars scales only the voucher cash, not its effect on choice.

For the frontier we consider policies that combine the three instruments, triples $(t_R,S,k)$ with $k\in\{0,1\}$ for the voucher, over a grid of 434 policies; the observed policy is $(2015,0,0)$. Outlay is the present value at 2012 of the payment increment on TAVR cases from $t_R$, the lump sums and the voucher cash, through 2019; the change in base payments from a change in who receives TAVR is reported separately as induced reimbursement. The base payment for a patient a policy brings into treatment is not part of the outlay: treatment at the regulated rate is what the policy buys, not a cost of the instrument, and the change-year counterfactuals count only the increment for the same reason. In adoption values the rate and the voucher, once in force, stay in force after 2019; outlay counts both through 2019. The 97 trial sites operate under every policy and are never offered the lump sum, and cost parameters, the modality model and market sizes are held fixed. The counterfactuals take the mapping from observables to adoption cost as known, so they price the cost of delaying payment, not the value of the information the delay produced.

\subsection{Equilibrium}\label{sub:ic}

Adoption dates are solved as an open-loop Nash equilibrium in dates: each hospital's distribution over dates is its best response to the patient shares it expects when every other hospital draws its date from its own equilibrium distribution. Open-loop play rules out preemption, since a hospital cannot condition its date on rivals' realized entry, unlike the subgame-perfect equilibrium of \citet{schmidt2006timing}. Under the observed schedule the equilibrium leaves 618.8 programs by 2019 against 618 observed. Patient shares, the patients a program takes from existing programs, consumer surplus and the voucher cash are computed at each policy's equilibrium. We solve the equilibrium for every policy without a lump sum and for 109 policies with one, those best for each objective at each budget with rivals held at their observed dates and their neighbors on the lump-sum grid; 99 converge, and ten stop at the cap of 30 rounds with a gap in adoption probabilities of 0.002 to 0.04, among them the best coverage policy at twice the observed outlay, whose count is therefore approximate.

\subsection{Results}\label{sub:cf-results}

\begin{figure}[tbp]
\centering
\includegraphics[width=\linewidth]{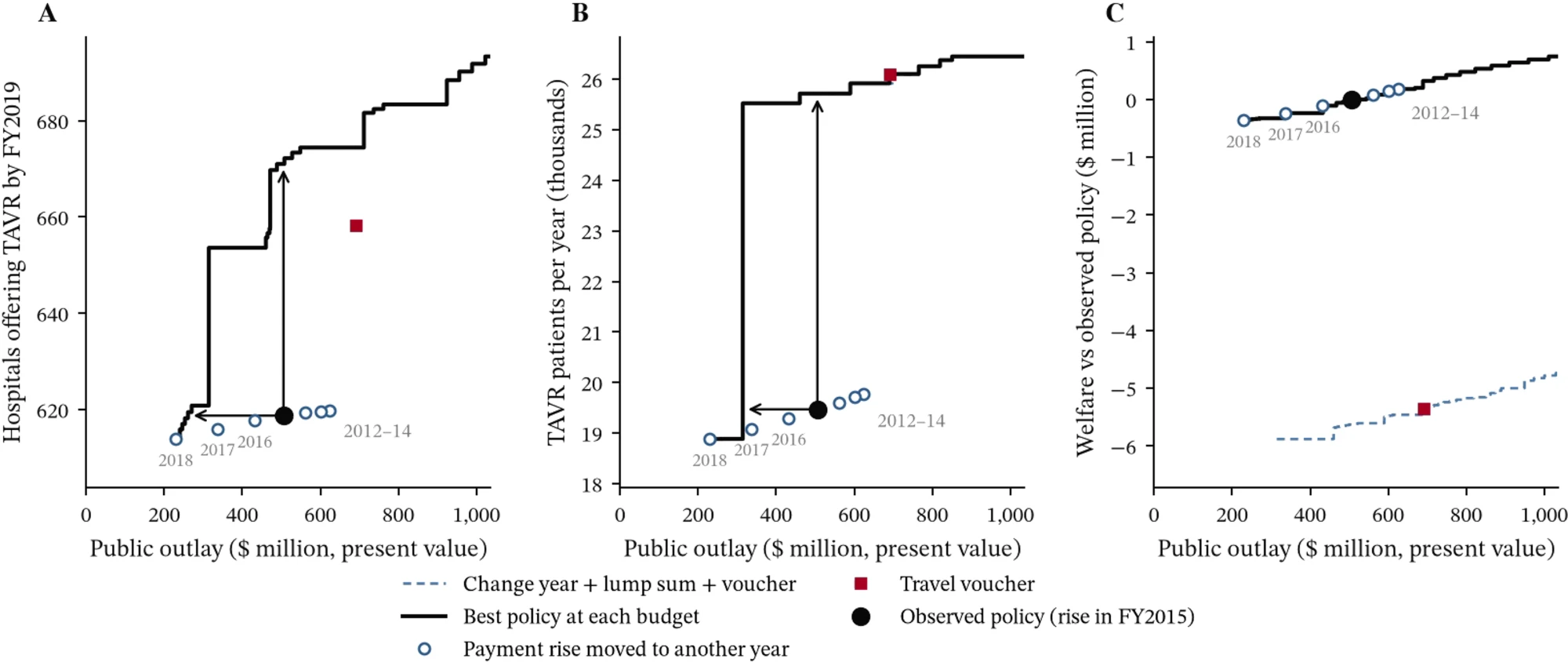}
\caption{Policy frontier. Each policy combines a year in which the 2015 payment rise takes effect, a lump sum paid at adoption to hospitals the model predicts will stay out, and a travel voucher that pays for the extra travel time to a TAVR hospital. Panel A reports hospitals offering TAVR by 2019, panel B TAVR patients per year, and panel C welfare $W$, patients' consumer surplus from hospital access relative to the observed policy, net of voucher cash, each against the public outlay in present value. Each policy is a Nash equilibrium in adoption dates. The solid line is the best policy at each budget, the dashed line the best policy that combines the voucher with another instrument, the circles move the payment rise alone, the square adds the voucher alone, and the dot is the observed policy.}
\label{fig:frontier}
\end{figure}

We first move each instrument alone. First, the rate rise is a weak lever (Table~\ref{tab:cf_change_year}). Advancing it from 2015 to 2013 costs \$95 million and adds 0.9 hospitals, about \$106 million per hospital, and advancing it to 2012 costs \$119 million and adds one. Delaying it to 2017 saves \$169 million and loses 2.7 hospitals, and delaying it to 2018 saves \$276 million and loses 4.9. Each year of delay therefore releases money that other instruments convert into more coverage than the rise bought. This is the sorting result in numbers: hospitals with a return to early entry enter at the unrevised rate, and the rise pays them without changing their decision.

Second, a lump sum targeted on predicted non-adopters buys entry at a fraction of that price. With the rise left in 2015, a \$4 million lump sum adds 24 hospitals for \$163 million, \$6.7 million each, and an \$8.5 million lump sum adds 54 for \$546 million, \$10.1 million each. The price per hospital rises with the lump sum because every targeted hospital that adopts collects it, including those that would have adopted for less. The lump sum adds few patients, 0.1 to 0.3 thousand a year, because the hospitals it brings in treat few. At the observed budget the coverage policy pays \$9 million at adoption to the 82 hospitals the model predicts will stay out under the rise in 2018 and the voucher, somewhat below the median adoption cost of hospitals that never adopt, \$11.0 million (Section~\ref{sec:supply-results}).

Third, the voucher is the instrument for patients, and it brings in hospitals as well. On its own, with the rise left in 2015, it adds 6.6 thousand TAVR patients a year and 40 hospitals for \$184 million, \$4.7 million per added hospital, less than any lump sum. Its cash is \$12 million; the rest of its cost is the payment increment on the TAVR cases it creates. It is in every best policy at the observed outlay and above, for both objectives.

Combining the three tools traces the frontier in Figure~\ref{fig:frontier}, and Table~\ref{tab:cf_best_policy} reports the best policy at four budgets. The observed policy, the payment rise of 2015, costs \$506 million in present value and leaves 619 hospitals offering TAVR by 2019 and 19.5 thousand TAVR patients a year. At the same outlay the frontier reaches 671 hospitals, 52 more or 8.5 percent of observed coverage, with the rise moved to 2018, a lump sum of \$9 million for targeted hospitals and the voucher, for \$489 million. For patients it reaches 25.7 thousand a year, 6.2 thousand or 32 percent more, with the rise moved to 2017 and the voucher, for \$460 million, \$46 million less than the observed revision. At half the outlay, the rise moved to 2018 with a \$2 million lump sum keeps coverage within one hospital of the observed policy, 618.2 against 618.8, and loses 0.6 thousand patients a year.

Whether the rise is delayed depends on the objective. The lump sum is paid only for adoption from $t_R$, so $t_R$ also dates the entry it buys. Coverage counts a hospital that enters in 2019 as fully as one that enters in 2012, so the coverage policy delays the rise to 2018 at the observed outlay and at one and a half times it. Patients treated per year count every year a program operates and every patient it draws, so the patient policy delays less, to 2017 at the observed outlay, keeps the rise in 2015 at one and a half times it and advances it to 2012 at twice it. It pays no lump sum at the observed outlay or above, because entry bought late in the window adds few patient-years.

Most of the extra TAVR cases are patients who would otherwise not have been treated. We split the change in TAVR cases over 2012--2019 against the observed policy into patients newly treated, patients who would have received SAVR at a hospital without a program, and patients a program would have treated with SAVR, the rest being moves between programs with different TAVR rates. For the coverage policy at the observed outlay the three shares are 87.3, 7.5 and 5.3 percent, for the patient policy 85.3, 8.1 and 7.2 percent, and for the voucher alone 82.2, 9.5 and 8.8 percent. These policies bring patients into surgery as well: the coverage policy treats 73 thousand more patients over the period with either procedure, against 46 thousand more TAVR cases. An earlier rate rise works differently. Of the 2 thousand extra TAVR cases from advancing it to 2013, 42 percent are patients switched from SAVR at programs already operating, 36 percent newly treated and 23 percent drawn from hospitals without a program: the rate changes the procedure more than the hospital. Because the market includes every diagnosed patient (Section~\ref{sec:data}), the newly treated share may overstate how many untreated patients a program brings into treatment.

The two objectives also respond differently to the budget. Coverage reaches 671 hospitals at the observed outlay, 682 at one and a half times it with a \$14.5 million lump sum, and 692 at twice it with the rise in 2016 and a \$10 million lump sum: the last \$500 million buys 21 hospitals. Patients reach 25.7 thousand a year at the observed outlay, 26.1 thousand at one and a half times it and 26.5 thousand at twice it, where the best policy spends \$850 million of the \$1,011 million available, because the voucher at full strength is the strongest patient instrument we consider. The remaining margin for patients is earlier entry, which the rise in 2012 buys at about \$119 million a hospital.

Patient welfare measured in travel time moves little. Across budgets the best policy changes consumer surplus by at most \$0.75 million at \$1 a minute, and it never includes the voucher, whose cash is a transfer and does not count in welfare (Figure~\ref{fig:frontier}, panel C). The measure is narrow by construction: it values only access, the travel a patient saves or incurs to reach the hospital she chooses, net of voucher cash, and it omits the clinical return to TAVR over SAVR and to treatment over none. Patients treated is therefore the objective that carries the health gain. Induced reimbursement, the TAVR--SAVR payment gap at the schedule without the increment summed over TAVR cases, is negative for every policy that raises TAVR, \$296 million for the coverage policy and \$308 million for the patient policy at the observed budget, because without the increment a TAVR case is paid less than a SAVR case. Because the modality rule responds to the payment, an earlier rise also raises the TAVR share at programs already operating, and the patient counts include that response.

The introduction asked whether a planner with the observed budget can reach more hospitals, or treat more patients, than the observed policy. Once rivals respond, the answer is yes for both: delay the rise and spend the saving on the voucher and, for coverage, on a lump sum targeted on predicted non-adopters.

\begin{table}[!tbp]
\centering
\footnotesize
\caption{Counterfactual: the 2015 payment change moved to another year}
\label{tab:cf_change_year}
\begin{threeparttable}
\begin{tabular}{lrrrrrr}
\toprule
\begin{tabular}[b]{@{}c@{}}Change\\ year\end{tabular} & Outlay & \begin{tabular}[b]{@{}c@{}}Induced\\ reimb.\end{tabular} & \begin{tabular}[b]{@{}c@{}}Hospitals\\ by 2019\end{tabular} & \begin{tabular}[b]{@{}c@{}}TAVR\\ 2012--19 (k)\end{tabular} & \begin{tabular}[b]{@{}c@{}}TAVR\\ 2019 (k)\end{tabular} & \begin{tabular}[b]{@{}c@{}}CS\\ vs SQ\end{tabular} \\
\midrule
\multicolumn{7}{l}{\textit{Infinite horizon (perpetuity after 2019)}} \\
2012 & 624.7 & -10.8 & 619.8 & 158.17 & 37.10 & +0.2 \\
2013 & 600.9 & -8.4 & 619.7 & 157.66 & 37.07 & +0.1 \\
2014 & 560.5 & -4.9 & 619.3 & 156.85 & 37.02 & +0.1 \\
2015 (status quo) & 505.6 & +0.0 & 618.8 & 155.77 & 36.95 & +0.0 \\
2016 & 432.2 & +6.1 & 617.7 & 154.36 & 36.85 & -0.1 \\
2017 & 336.8 & +14.0 & 616.1 & 152.70 & 36.73 & -0.2 \\
2018 & 229.8 & +22.2 & 613.9 & 151.06 & 36.60 & -0.4 \\
\bottomrule
\end{tabular}
\begin{tablenotes}[flushleft]\footnotesize
\item \textit{Notes:} \$ million, PV at 2012. Each row is a Nash equilibrium in adoption dates. Outlay: payment increment on TAVR cases, lump sums and voucher cash (\$1 per minute). Induced reimb.: extra base payment from the change in who gets TAVR, vs status quo; not in the outlay.
\end{tablenotes}
\end{threeparttable}
\end{table}


\begin{table}[!tbp]
\centering
\footnotesize
\caption{Counterfactual: best policy at each budget}
\label{tab:cf_best_policy}
\begin{threeparttable}
\setlength{\tabcolsep}{4.5pt}
\begin{tabular}{lcccrrr}
\toprule
Objective & Budget & Best policy & Outlay & Value & vs SQ & Induced \\
 & ($\times$SQ) & ($t_R$; $S$, \$M; $k$) & (\$M) & & & reimb. \\
\midrule
\multicolumn{7}{l}{\textit{Infinite horizon (perpetuity after 2019)}} \\
Hospitals offering TAVR by 2019 & 0.5 & 2018; 2.00; --- & 252.4 & 618.2 & $-0.5$ & $+21.9$ \\
 & 1 & 2018; 9.00; 1.00 & 488.8 & 671.0 & $+52.2$ & $-296.3$ \\
 & 1.5 & 2018; 14.50; 1.00 & 736.2 & 682.4 & $+63.7$ & $-293.4$ \\
 & 2 & 2016; 10.00; 1.00 & 988.4 & 691.8 & $+73.1$ & $-303.5$ \\
\addlinespace
TAVR patients per year, 2012--19 (k) & 0.5 & 2018; 2.00; --- & 252.4 & 18.88 & $-0.59$ & $+21.9$ \\
 & 1 & 2017; 0.00; 1.00 & 459.6 & 25.72 & $+6.24$ & $-307.7$ \\
 & 1.5 & 2015; 0.00; 1.00 & 690.0 & 26.10 & $+6.63$ & $-320.2$ \\
 & 2 & 2012; 0.00; 1.00 & 850.3 & 26.45 & $+6.98$ & $-332.1$ \\
\addlinespace
TAVR patients in 2019 (k) & 0.5 & 2018; 2.00; --- & 252.4 & 36.66 & $-0.28$ & $+21.9$ \\
 & 1 & 2017; 0.00; 1.00 & 459.6 & 47.12 & $+10.18$ & $-307.7$ \\
 & 1.5 & 2015; 0.00; 1.00 & 690.0 & 47.28 & $+10.33$ & $-320.2$ \\
 & 2 & 2012; 0.00; 1.00 & 850.3 & 47.44 & $+10.49$ & $-332.1$ \\
\addlinespace
Consumer surplus vs SQ (\$M) & 0.5 & 2018; 2.00; --- & 252.4 & -0.3 & $-0.34$ & $+21.9$ \\
 & 1 & 2015; 0.00; --- & 505.6 & 0.0 & $+0.00$ & $+0.0$ \\
 & 1.5 & 2012; 2.50; --- & 748.4 & 0.4 & $+0.43$ & $-19.6$ \\
 & 2 & 2012; 5.50; --- & 1,010.7 & 0.7 & $+0.75$ & $-31.5$ \\
\bottomrule
\end{tabular}
\begin{tablenotes}[flushleft]\footnotesize
\item \textit{Notes:} \$ million, PV at 2012; SQ is the observed policy, outlay \$505.6M. Each policy is a Nash equilibrium in adoption dates. Outlay: payment increment on TAVR cases, lump sums and voucher cash (\$1 per minute). Induced reimb.: extra base payment from the change in who gets TAVR, vs status quo; not in the outlay. $S$: lump sum per adopting hospital (\$M); $k$: share of the extra travel paid; --- means the instrument is not used.
\end{tablenotes}
\end{threeparttable}
\end{table}

\section{Conclusion}\label{sec:conclusion}

Under a regulated price that paid TAVR no more than the procedure it replaced, adoption was financed through patients. The patients a program draws account for nine tenths of the value of adopting, a fifth of it from patients who receive surgery, and without them half of the entry before the rate revision does not occur. A TAVR program raised a hospital's value to patients across the risk distribution, and in the first years most of the patients it drew received the incumbent procedure: in 2012, 15 of the 21 patients a program added were expected to receive SAVR, in line with the 68 percent of patients who passed over a nearest hospital without a program to be treated at an adopter. The return to adoption therefore accrued on the hospital's whole program, and a program's value grew with its age, which is why more than two fifths of the hospitals that chose to adopt entered before the rate paid and why their entry lowered the return to those that waited.

The timing of the rate update is a weak lever for diffusion, and paying at the margin is what works. In this setting, a payer spending what the rate revision cost could have reached 52 more hospitals, or treated 32 percent more TAVR patients for 9 percent less outlay, by delaying the rate rise and spending the saving on a travel voucher for TAVR patients and, for coverage, on a lump sum at adoption targeted on predicted non-adopters. The rate rise is almost entirely inframarginal for entry: it pays every case at every adopter, and most adopters enter without it. The delayed rate also identifies the inframarginal recipients: hospitals that enter at the unrevised rate have a private return to early adoption, and a lump sum paid from the revision never reaches them. How far to delay depends on the objective: at the observed outlay coverage favors a rise in 2018 and patients a rise in 2017; with more money the patient policy moves the rise earlier, to 2012 at twice the outlay, while the coverage policy keeps it at 2016 or later. Three limits deserve statement alongside the recommendation. The voucher's effect on hospital choice is set in minutes of travel; the cash a minute costs, which the demand model does not identify, scales only the voucher cash in its outlay. The outlay counts what the instruments pay above the unrevised schedule; the base payment for the patients a policy brings into treatment is the price of the treatment it buys and is not counted. And how far this carries beyond TAVR depends on whether adoption elsewhere also earns a return outside the regulated price, which is the primitive the argument rests on.

\appendix
\section{Additional descriptive evidence}\label{app:altruism}

This appendix collects the descriptive exhibits cited in the text and reports the two further tests of the hospital response to margins referred to in Section~\ref{altruism}.

\begin{figure}[htb]
    \centering
    \includegraphics[width=0.95\linewidth]{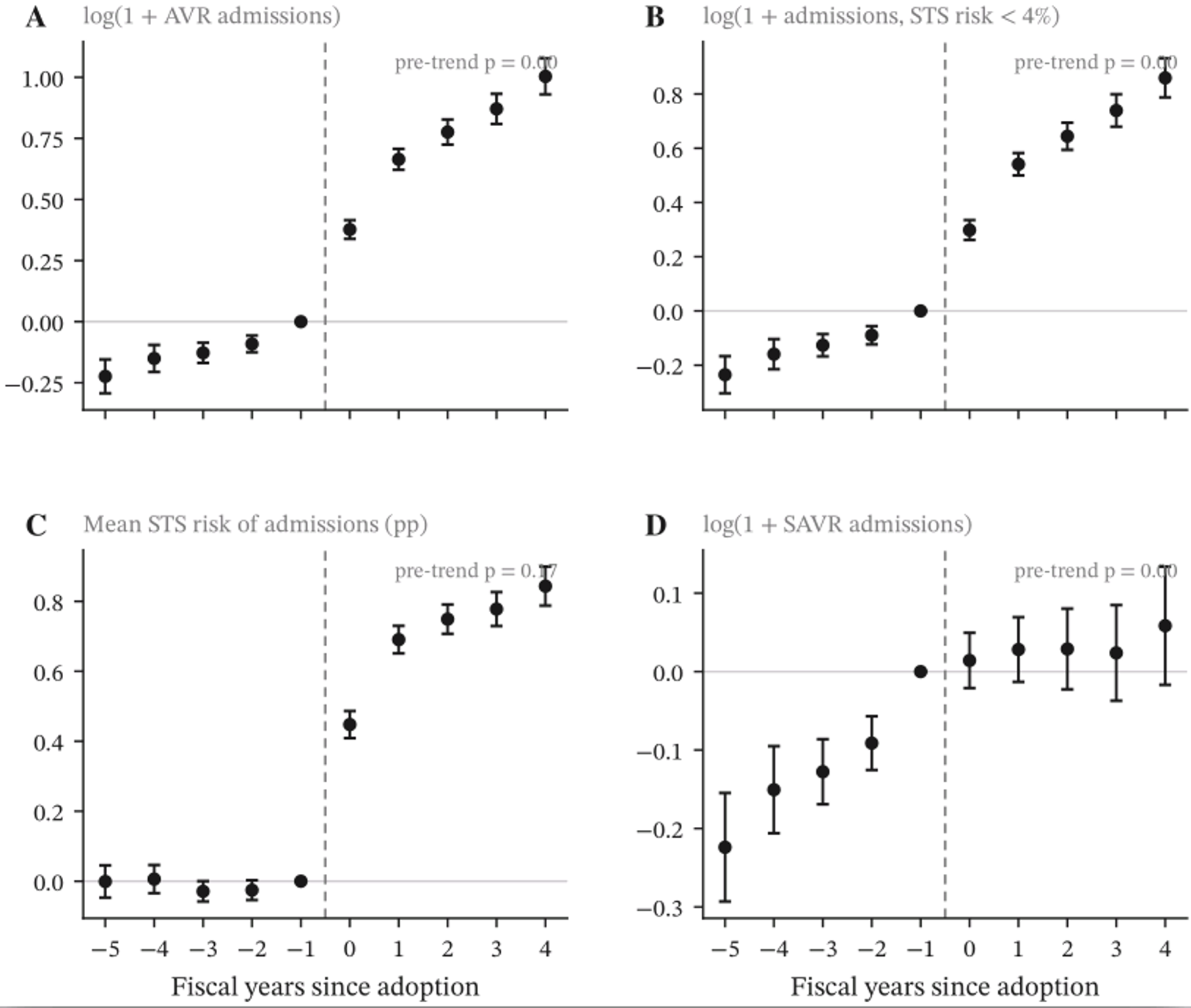}
    \caption{AVR admissions around a hospital's TAVR adoption. Interaction-weighted event-study coefficients \citep{SunAbraham2021} with hospital and year effects, never-adopters as the control cohort and adoption cohorts 2012--2019 (pivotal-trial sites excluded); relative year $-1$ omitted; 95 percent confidence intervals clustered by hospital. Panels: (A) $\log(1+\text{AVR admissions})$; (B) $\log(1+\text{admissions with STS risk below 4 percent})$; (C) mean STS risk of admissions, in percentage points; (D) $\log(1+\text{SAVR admissions})$. The pre-trend $p$-value is the joint test that the coefficients for relative years $-5$ to $-2$ are zero.}
    \label{fig:es_admissions}
\end{figure}

\begin{figure}[tbp]
 \centering
 \includegraphics[width=\linewidth]{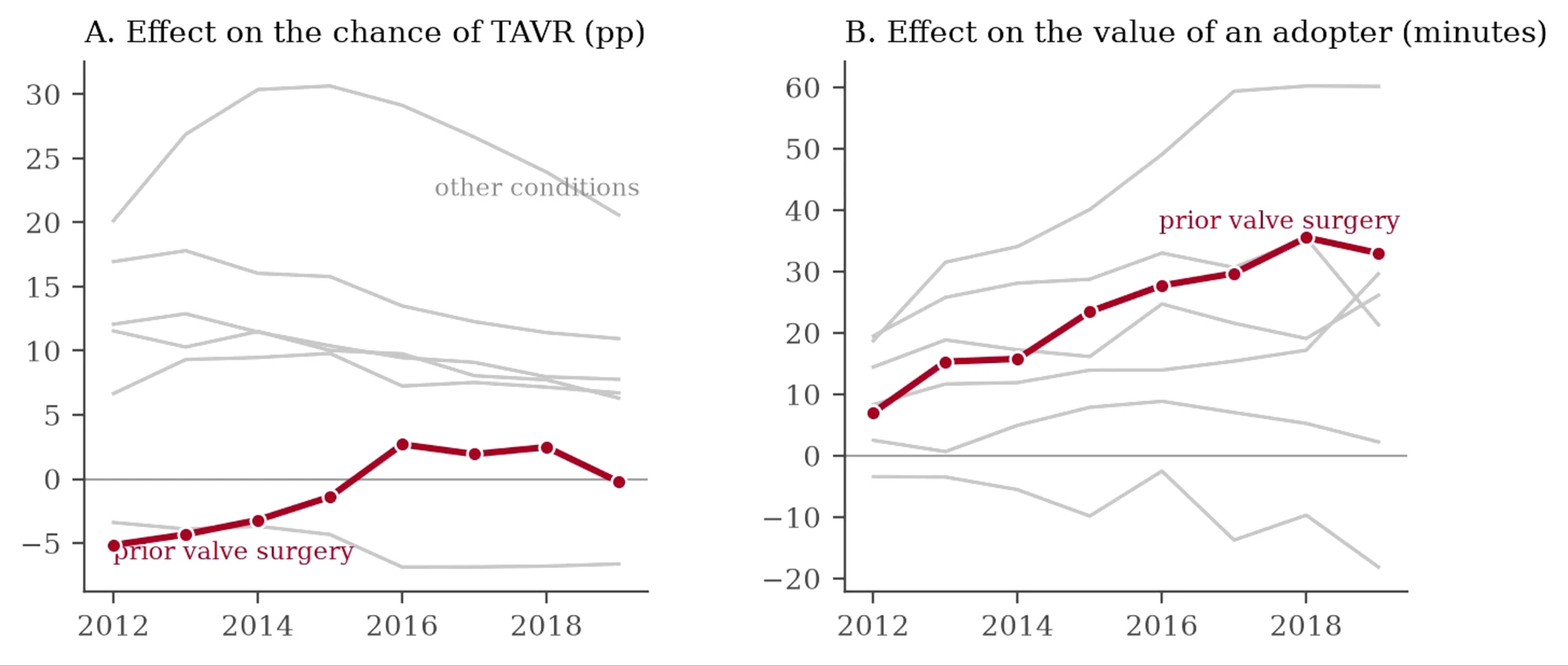}
 \caption{Conditions and the value of an adopter. Panel A: effect of each condition on the probability of TAVR, in percentage points, from a linear probability model on patients treated at an adopter with STS-tier effects, by fiscal year. Panel B: the condition's term in the value of an adopter, in minutes of travel. Red: prior valve surgery; grey: prior coronary bypass, home oxygen, peripheral artery disease, aortic atherosclerosis, skilled-nursing or home-health use, dialysis.}
 \label{fig:conditions}
\end{figure}

\begin{table}[htb]
\centering
\footnotesize
\caption{Patient demand: value of treatment at any hospital}
\label{tab:demand_baseline}
\begin{threeparttable}
\setlength{\tabcolsep}{3pt}
\begin{tabular*}{\linewidth}{@{\extracolsep{\fill}} l rrrrrrrr @{}}
\toprule
 & 2012 & 2013 & 2014 & 2015 & 2016 & 2017 & 2018 & 2019 \\
\midrule
\multicolumn{9}{@{}l}{\textit{A. Year effects and risk}} \\
Year effect, $\gamma_t$ & 0 & 0.144 & 0.176 & 0.098 & 0.130 & $-$0.033 & 0.074 & $-$0.245 \\
Risk, $\kappa_t$ & $-$0.316 & $-$0.393 & $-$0.416 & $-$0.425 & $-$0.482 & $-$0.545 & $-$0.676 & $-$0.687 \\
\addlinespace
\multicolumn{9}{@{}l}{\textit{B. Conditions, $\kappa_{kt}$}} \\
Balloon valvuloplasty & 0.286 & 0.199 & 0.193 & 0.048 & 0.063 & $-$0.656 & 0.137 & 0.139 \\
Coronary bypass & $-$0.660 & $-$0.818 & $-$0.842 & $-$1.021 & $-$1.179 & $-$1.441 & $-$1.418 & $-$1.438 \\
Home oxygen or respiratory failure & $-$0.518 & $-$0.650 & $-$0.655 & $-$0.619 & $-$0.795 & $-$0.689 & $-$0.672 & $-$0.795 \\
Peripheral artery disease & $-$0.176 & $-$0.138 & $-$0.168 & $-$0.182 & $-$0.167 & $-$0.153 & $-$0.174 & $-$0.498 \\
Aortic atherosclerosis & 0.780 & 0.765 & 0.677 & 0.657 & 0.445 & 0.456 & 0.302 & 0.608 \\
Skilled-nursing or home-health use & $-$1.674 & $-$1.649 & $-$1.657 & $-$1.723 & $-$1.729 & $-$1.660 & $-$1.579 & $-$1.445 \\
Dialysis or CKD stage 4--5 & $-$0.043 & $-$0.080 & $-$0.055 & 0.011 & $-$0.123 & 0.085 & 0.057 & 0.282 \\
Prior valve surgery & $-$0.643 & $-$0.745 & $-$0.735 & $-$0.709 & $-$0.591 & $-$0.561 & $-$0.578 & $-$0.470 \\
\bottomrule
\end{tabular*}
\begin{tablenotes}[flushleft]\footnotesize
\item \textit{Notes:} Terms that shift the value of treatment at any hospital relative to no treatment, from the demand model of Table~\ref{tab:demand_results}; the 2012 year effect is normalized to zero. Not shown: the risk interactions with hospital size and teaching status, $\omega_1=-0.0030$ and $\omega_2=-0.0053$. Standard errors to be added.
\end{tablenotes}
\end{threeparttable}
\end{table}


\begin{table}[htb]
\centering
\small
\caption{AVR episodes by procedure: resource use and payment}
\label{tab:episode_outcomes}
\begin{threeparttable}
\begin{tabular}{lcc}
\toprule
 & TAVR & SAVR \\
\midrule
Episodes (N) & 182,327 & 238,387 \\
Age & 81.5 & 73.7 \\
Female (share) & 0.46 & 0.37 \\
STS predicted mortality (percent) & 3.2 & 1.7 \\
DRG weight & 6.90 & 6.79 \\
Length of stay (days) & 4.9 & 10.0 \\
In-hospital mortality (share) & 0.02 & 0.04 \\
Travel time to treating hospital, mean (minutes) & 95.7 & 96.4 \\
Bypassed the nearest hospital (share) & 0.65 & 0.57 \\
Travelled more than 60 minutes (share) & 0.29 & 0.27 \\
Total charges (\$) & 217,741 & 232,081 \\
Medicare payment (\$) & 50,187 & 49,271 \\
\bottomrule
\end{tabular}
\begin{tablenotes}[flushleft]\footnotesize
\item \textit{Notes:} Medicare fee-for-service AVR episodes, FY2012--2019. Means unless noted. STS predicted mortality from the age--sex--comorbidity grid. Travel time from the patient's residential ZIP to the treating hospital.
\end{tablenotes}
\end{threeparttable}
\end{table}

\begin{figure}[htb]
    \centering
    \includegraphics[width=0.7\linewidth]{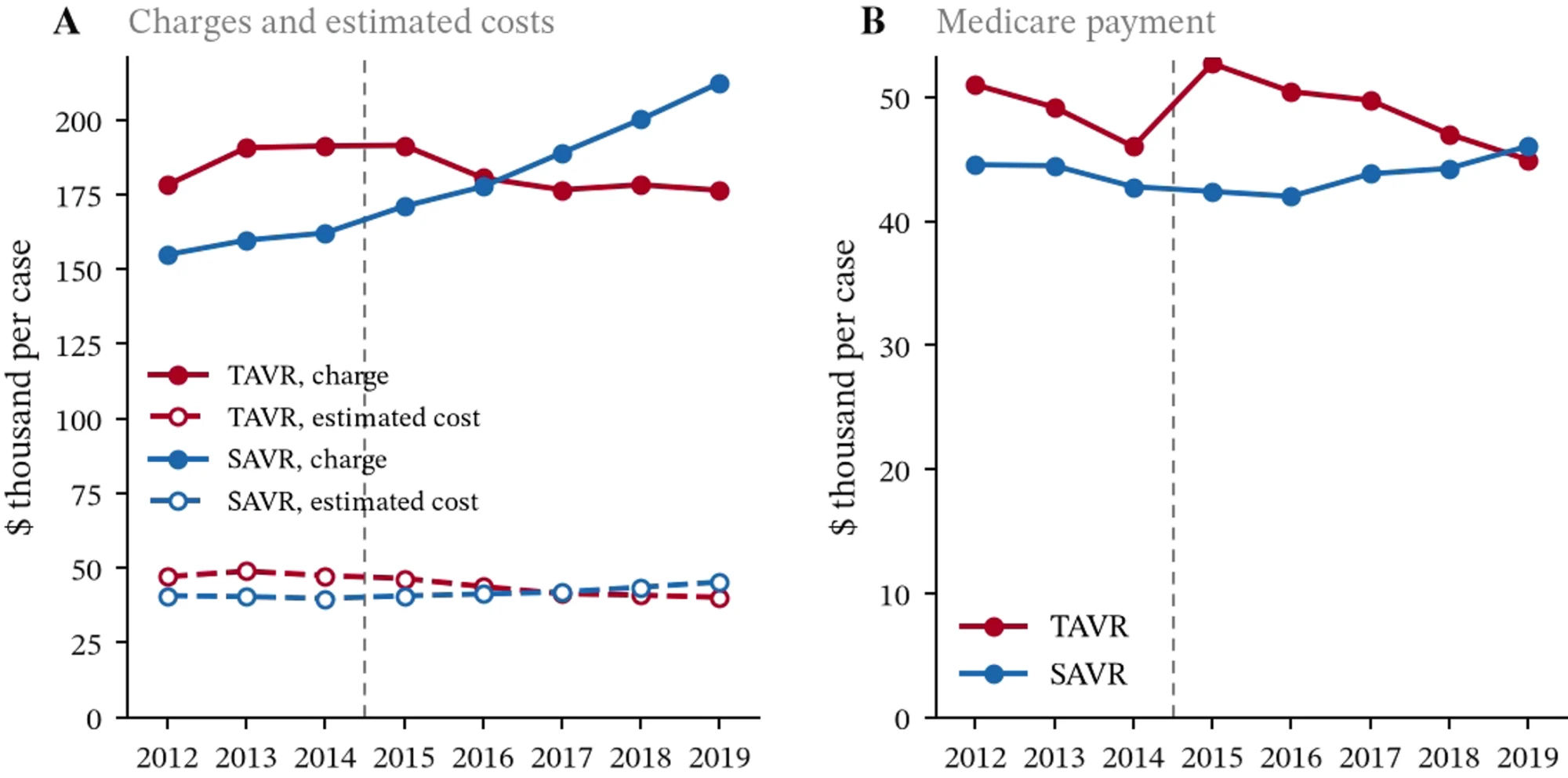}
    \caption{Charges, estimated costs and Medicare reimbursement per case for TAVR and SAVR by fiscal year. Medians across episodes. Panel A shows charges and costs estimated with hospital-year cost-to-charge ratios; Panel B the Medicare payment.}
    \label{fig:reimbursement_means}
\end{figure}

The two tests of the hospital response to margins follow.

\subsection{Selection of high-benefit patients}

If early adopters placed greater weight on patient benefit than later entrants, they would concentrate TAVR on patients with higher expected benefit (older or more severe cases). A direct comparison of TAVR patients across hospitals is not feasible because non-adopters have no TAVR patients. We use SAVR, the main substitute, as the benchmark outcome: if early adopters channel high-benefit patients into TAVR more aggressively, SAVR severity and age should fall more sharply at early adopters after reimbursement improves. We control for hospital-year case mix using a leave-one-out mean: for each SAVR patient $i$, the control variable is the mean age and severity across all \emph{other} AVR patients at the same hospital-year. Outcomes are in logs. We estimate
\begin{align}
\log(Y_{iht}) = \beta_0 + \beta_1 \cdot \text{Early Adopter}_h + \beta_2 \cdot \text{Post Reimb}_t +  \notag\\
\beta_3 \cdot (\text{Early Adopter}_h \cdot \text{Post Reimb}_t) + \mathbf{Z}_{ht}'\theta + \delta_s + \epsilon_{iht}
\end{align}
where $Y_{iht}$ is DRG weight or age for SAVR patient $i$ at hospital $h$ in year $t$; $\text{Early Adopter}_h$ indicates pre-2015 TAVR adoption; $\text{Post Reimb}_t$ indicates fiscal years after the reimbursement change; $\mathbf{Z}_{ht}$ collects hospital-level case-mix and structural controls; and $\delta_s$ are city fixed effects. Selection of high-benefit patients by early adopters predicts $\beta_3<0$. The sample is SAVR patients at adopting hospitals in fiscal years 2012--2019; the comparison group adopts in fiscal 2015 or later, and $\mathbf{Z}_{ht}$ collects the leave-one-out means, log beds and teaching status. Because the FDA extended the TAVR label to intermediate-risk patients in August 2016, we also split the post period into fiscal 2015--2016, when only the payment had changed, and fiscal 2017 on. With city rather than hospital effects, the estimate also reflects which hospitals contribute SAVR patients, and pre-existing differences in TAVR ramp-up are not excluded; because the leave-one-out case-mix control includes TAVR cases, whose payment groups changed in 2015, the age result is the cleaner of the two.

\begin{table}[htb]
\centering
\caption{SAVR patient characteristics and TAVR adoption timing: difference in differences}
\label{tab:savr_did}
\begin{threeparttable}
\begin{tabular}{lcccc}
\toprule
 & \multicolumn{2}{c}{log(DRG weight)} & \multicolumn{2}{c}{log(age)} \\
\cmidrule(lr){2-3}\cmidrule(lr){4-5}
 & (1) & (2) & (3) & (4) \\
\midrule
Early adopter & $0.007$ & $0.007$\sym{*} & $0.002$ & $0.001$ \\
 & (0.004) & (0.004) & (0.002) & (0.002) \\
Post reimbursement & $-0.006$\sym{**} & & $-0.024$\sym{***} & \\
 & (0.002) & & (0.001) & \\
Early adopter $\times$ post reimbursement & $-0.006$ & & $-0.008$\sym{***} & \\
 & (0.004) & & (0.002) & \\
Post, fiscal 2015--2016 & & $-0.015$\sym{***} & & $-0.012$\sym{***} \\
 & & (0.002) & & (0.001) \\
Post, fiscal 2017 on & & $0.003$ & & $-0.036$\sym{***} \\
 & & (0.003) & & (0.002) \\
Early adopter $\times$ post, fiscal 2015--2016 & & $-0.019$\sym{***} & & $-0.009$\sym{***} \\
 & & (0.004) & & (0.002) \\
Early adopter $\times$ post, fiscal 2017 on & & $0.004$ & & $-0.005$\sym{***} \\
 & & (0.005) & & (0.002) \\
$p$-value, equal interactions & & $<0.001$ & & $0.050$ \\
\midrule
City fixed effects & Yes & Yes & Yes & Yes \\
Observations & 205,508 & 205,508 & 206,434 & 206,434 \\
$R^{2}$ & 0.043 & 0.045 & 0.033 & 0.036 \\
\bottomrule
\end{tabular}
\begin{tablenotes}[flushleft]\footnotesize
\item \textit{Notes:} SAVR patients at adopting hospitals, fiscal years 2012--2019. Early adopter: adoption before fiscal 2015; the comparison group adopts in fiscal 2015 or later. Controls: leave-one-out hospital-year mean age and DRG weight of the other AVR patients, log beds and teaching status. The $p$-value tests that the two interactions are equal. Standard errors in parentheses, clustered at the city level. $^{*} p<0.1$; $^{**} p<0.05$; $^{***} p<0.01$.
\end{tablenotes}
\end{threeparttable}
\end{table}

In Table~\ref{tab:savr_did}, SAVR patients become younger after 2015 at all adopters, and the pooled interactions are small: $-0.006$ for DRG weight, not significant, and $-0.008$ for age (Columns~1 and~3). The split shows when the shift occurs. In fiscal 2015--2016 the DRG weight of SAVR patients at early adopters falls by $1.9$ percent and their age by $0.9$ percent relative to later adopters (Columns~2 and~4); from fiscal 2017 the gap in DRG weight closes and the gap in age halves. The shift coincides with the rate revision rather than with the wider indication, and it has the sign of the prediction: early adopters moved their most severe surgical candidates to TAVR when the payment rose. The test shows that the allocation of patients between the procedures moved with the payment. It does not by itself separate a profit motive from a weight on patient benefit: until August 2016 the label confined TAVR to high-risk patients, so any expansion of TAVR draws on the most severe SAVR candidates whatever the hospital maximizes. Two features bias the estimate toward zero or blur its timing: hospitals in the comparison group that open programs in fiscal 2015--2016 also lose severe SAVR patients, and the new valve generations approved in June 2015 fall in the same window. SAVR severity remains an indirect proxy for the unobserved TAVR allocation decision.

\subsection{Within-hospital modality reallocation}

\begin{figure}[htb]
    \centering
    \includegraphics[width=0.5\textwidth]{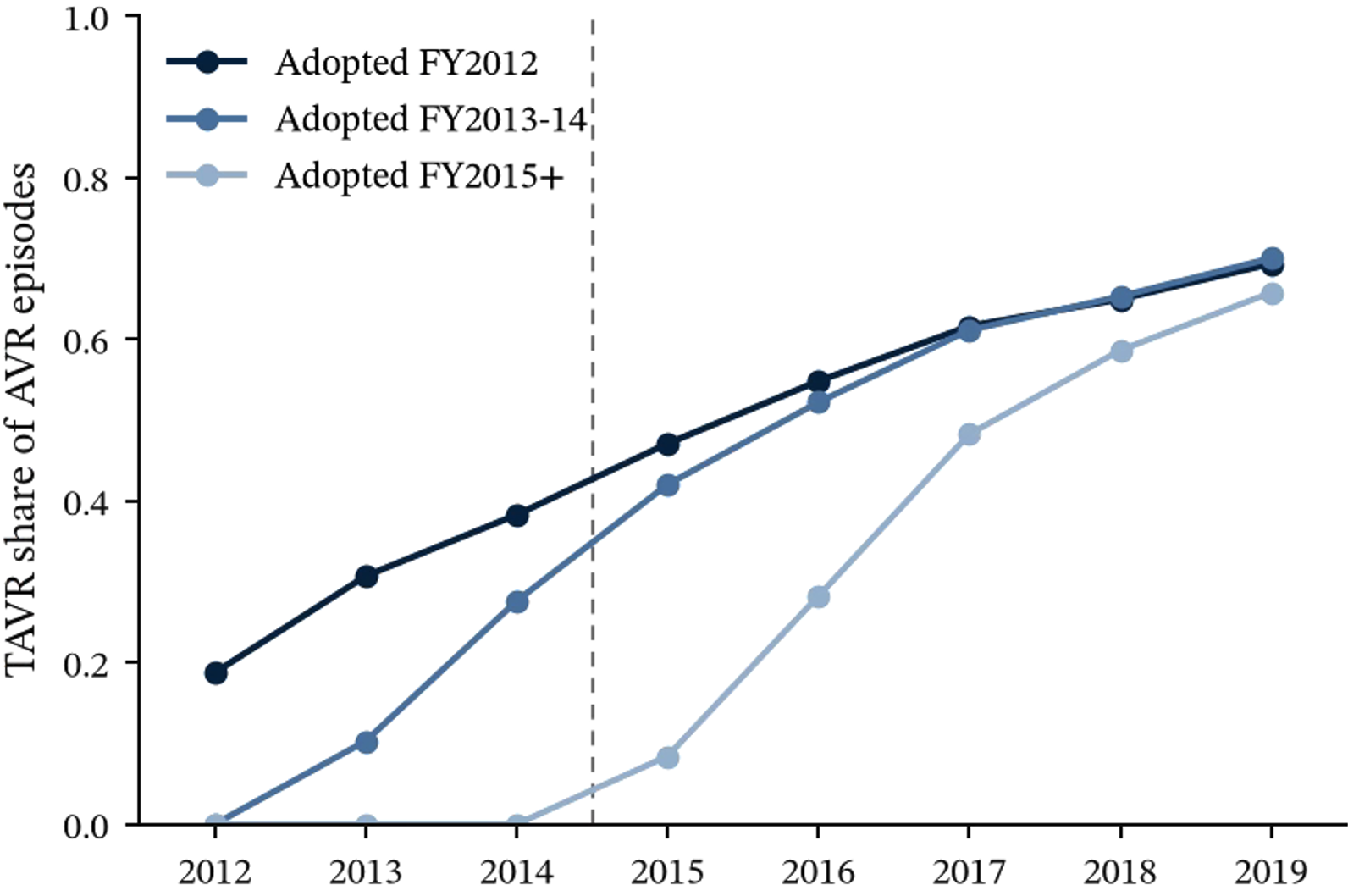}
    \caption{Within-hospital TAVR shares by fiscal year (early adopters)}
    \label{fig:dencitymove}
\end{figure}

A purely clinical decision rule should not reallocate the modality mix at a given hospital when relative reimbursement changes, holding the patient pool fixed. Figure~\ref{fig:dencitymove} shows the within-hospital TAVR share shifting right after 2014. We estimate
\begin{equation}
\label{eq:savr}
\log(\text{SAVR}_{h,t}) \;=\; \alpha
\;+\; \beta_1\,\text{Adopt}_{h,t}
\;+\; \beta_2\,\text{Adopt}_{h,t}\times \mathbf{1}\{t \ge 2015\}
\;+\; \gamma_h \;+\; \delta_{t\times \mathrm{HRR}} \;+\; \varepsilon_{h,t},
\end{equation}
where $\text{Adopt}_{h,t}=\mathbf{1}\{t\ge T_h\}$ indicates that hospital $h$ is offering TAVR by year $t$, $\gamma_h$ are hospital fixed effects, and $\delta_{t\times \mathrm{HRR}}$ are fiscal-year by HRR fixed effects. The panel covers hospital-years with at least one SAVR case in fiscal 2012--2019. As in Table~\ref{tab:savr_did}, we also split the post period at fiscal 2017, and we add the hospitals that never adopt, with $\text{Adopt}_{h,t}=0$, so that the HRR-year effects are estimated against hospitals without a program.

\begin{table}[htb]
\centering
\caption{Effect of TAVR adoption on SAVR volumes}
\label{tab:savr_reg}
\begin{threeparttable}
\begin{tabular}{lcccc}
\toprule
 & \multicolumn{4}{c}{$\log(\text{SAVR volume})$} \\
\cmidrule(lr){2-5}
 & \multicolumn{2}{c}{Adopters} & \multicolumn{2}{c}{All hospitals} \\
 & (1) & (2) & (3) & (4) \\
\midrule
Adopt & $0.099$\sym{***} & $0.100$\sym{***} & $0.084$\sym{***} & $0.081$\sym{***} \\
 & (0.032) & (0.032) & (0.031) & (0.031) \\
Adopt $\times$ $\mathbf{1}$(fiscal year $\geq$ 2015) & $-0.087$\sym{***} & & $0.030$ & \\
 & (0.032) & & (0.029) & \\
Adopt $\times$ fiscal 2015--2016 & & $-0.053$\sym{*} & & $0.004$ \\
 & & (0.032) & & (0.027) \\
Adopt $\times$ fiscal 2017 on & & $-0.166$\sym{***} & & $0.051$ \\
 & & (0.057) & & (0.038) \\
$p$-value, equal interactions & & $0.041$ & & $0.148$ \\
\midrule
Hospital fixed effects & Yes & Yes & Yes & Yes \\
Fiscal year $\times$ HRR fixed effects & Yes & Yes & Yes & Yes \\
Observations & 3{,}970 & 3{,}970 & 6{,}771 & 6{,}771 \\
$R^{2}$ & 0.902 & 0.902 & 0.900 & 0.900 \\
\bottomrule
\end{tabular}
\begin{tablenotes}[flushleft]\footnotesize
\item \textit{Notes:} Hospital-years with at least one SAVR case, fiscal years 2012--2019. Columns~1--2: hospitals that adopt by 2019, pivotal-trial sites included; Columns~3--4 add the hospitals that never adopt. Hospital-years alone in their fixed-effect cell are dropped. The $p$-value tests that the two interactions are equal. Standard errors in parentheses, clustered at the hospital level. $^{*} p<0.1$; $^{**} p<0.05$; $^{***} p<0.01$.
\end{tablenotes}
\end{threeparttable}
\end{table}

Table~\ref{tab:savr_reg} shows that before the revision SAVR volume is about 10 percent higher once a hospital adopts TAVR, among adopters and against all hospitals (Columns~1 and~3): a program draws patients to the hospital beyond those it treats with TAVR, consistent with the value patients place on an adopter in the demand estimates. Among adopters the differential shrinks after 2015 ($\hat\beta_2=-0.087$, leaving 0.012), and the decline concentrates in fiscal 2017 and later, when the label extended to intermediate-risk patients (Column~2), and it disappears once hospitals without a program enter the comparison (Columns~3--4). The within-hospital decline in SAVR volume therefore follows eligibility more than the payment; the response of the modality choice to the margin is estimated in the supply model (Table~\ref{tab:supply_estimates}). The event study in Figure~\ref{fig:es_admissions}, which is robust to heterogeneous effects, finds a small and imprecise SAVR response at the adoption date, so the static estimate is best read as the pre-revision differential.

\section{The planner's problem in the two-stage model}\label{app:planner}

This appendix states and proves the results summarized in Section~\ref{sub:mechanism}. Hospitals, markets and payoffs are those of Section~\ref{sec:stylized-framework}. For hospital $i$ in market $m$ write
\begin{align*}
R_i&:=\chi_i S_m\,\mu^S_1+\beta\,\chi_i\,(S_m-Q_m)\,\mu^S_2+\beta\,v_{i,2}\,\eta_i,\\
\ell_i&:=-\,v_{i,1}\,\mu_1>0,\\
L_i&:=\chi_i Q_m\,\mu^S_2+v_{i,2}\,\mu_2,
\end{align*}
for the rent to early entry (steering plus learning), the stage-1 loss, and the stage-2 return of a late entrant. The values of entering early, late, and never are
\[
V^E_i=V^L_i+R_i-\ell_i-(1-\beta)F_i,\qquad V^L_i=\beta\,(L_i-F_i),\qquad V^N_i=0 .
\]

\begin{assumption}\label{ass:types}
Within a market, hospitals differ in the fixed cost $F_i$, which is private information with continuous density $g$ on $[0,\bar F]$, and in observables $x_i$. The planner observes adoption dates and $x_i$, not $F_i$.
\end{assumption}

\begin{assumption}\label{ass:interior}
$(R_i-\ell_i)/(1-\beta)\le L_i$, and $g$ is positive at $F^E_i$ and $L_i$.
\end{assumption}

Under Assumption~\ref{ass:interior}, $F^E_i=(R_i-\ell_i)/(1-\beta)$ and $F^L_i=L_i$ in Lemma~\ref{lem:sorting}, so the late cohort $\mathcal M=(F^E_i,L_i]$ is an interval that can be nonempty. This is the case in which the delayed rate sorts hospitals into three groups, and Propositions~\ref{prop:gate} and~\ref{prop:targeting} are stated under it.

The planner's instruments are: a stage-1 rate increment $\rho\in\{0,\mu_2-\mu_1\}$, paid on every stage-1 TAVR case at every operating hospital, since a regulated price cannot differ across hospitals; a lump sum $s\ge0$ paid at $t_R$ to a hospital that enters in stage~2, possibly conditioned on $x_i$; and a voucher $\theta\ge0$ per minute of travel, paid to a patient whose nearest hospital has no program for her travel to an adopter. She evaluates a policy on coverage $N$, the number of hospitals that ever adopt; access $A$, the discounted number of TAVR cases; welfare $W$, patients' consumer surplus from hospital access in travel-time units, net of voucher cash; and outlay $C$, the present value of rate increments, lump sums and voucher cash. For a budget $B$ the frontier is the maximum of each outcome over policies with $C\le B$. The observed policy is the delayed rate with no transfers, $\rho=s=\theta=0$.

\subsection{Sorting under a delayed rate}

\begin{lemma}[Sorting]\label{lem:sorting}
Under the observed policy the adoption date is nondecreasing in $F_i$: there are cutoffs $F^E_i\le F^L_i$ with
\[
F^E_i=\min\Bigl\{\tfrac{R_i-\ell_i}{1-\beta},\;\Delta\pi_{i,1}+\beta\Delta\pi^{E}_{i,2}\Bigr\},
\qquad
F^L_i=\max\{L_i,\,F^E_i\},
\]
such that hospital $i$ enters in stage~1 if $F_i\le F^E_i$ ($\mathcal E$), at $t_R$ if $F^E_i<F_i\le F^L_i$ ($\mathcal M$), and never if $F_i>F^L_i$ ($\mathcal S$).
\end{lemma}

\begin{proof}
$V^E_i-V^L_i$, $V^L_i$ and $V^E_i$ are linear in $F_i$ with slopes $-(1-\beta)$, $-\beta$ and $-1$. Early entry requires $V^E_i\ge V^L_i$ and $V^E_i\ge0$, which hold for $F_i$ below the smaller of the two roots, $F^E_i$. Late entry requires $V^L_i\ge0$, i.e.\ $F_i\le L_i$, and $V^L_i\ge V^E_i$. For $F_i>F^E_i$ either $V^E_i<V^L_i$ or $V^E_i<0$, and in the latter case $V^L_i\ge0$ implies $V^L_i>V^E_i$; so the late set is $(F^E_i,L_i]$, empty if $L_i\le F^E_i$, which is $(F^E_i,F^L_i]$. The rest never enter.
\end{proof}

\begin{corollary}[Earlier update]\label{cor:accelerate}
Holding rivals' entry, and hence $Q_m$, fixed, paying the revised rate in stage~1 ($\rho=\mu_2-\mu_1$) replaces the stage-1 loss $\ell_i$ by a gain $v_{i,1}\mu_2$ and leaves $L_i$ unchanged. It moves hospitals in $\mathcal M$ with $F_i\le (R_i+v_{i,1}\mu_2)/(1-\beta)$ from $t_R$ to stage~1; it recruits from $\mathcal S$ only hospitals with $L_i<F_i\le \Delta\pi_{i,1}+v_{i,1}(\mu_2-\mu_1)+\beta\Delta\pi^{E}_{i,2}$, those whose shortfall at the revised rate is smaller than one stage of profit at that rate plus the early-entry rent; and it pays $v_{i,1}(\mu_2-\mu_1)$ to every hospital in $\mathcal E$, which changes no decision. The payment to $\mathcal E$ is increasing in stage-1 volume.
\end{corollary}

Corollary~\ref{cor:accelerate} and Proposition~\ref{prop:gate} hold rivals' capture fixed; the equilibria of Section~\ref{sec:counterfactuals} let it respond.

\subsection{Screening on the adoption date}

\begin{proposition}[Gate transfer]\label{prop:gate}
Under Assumption~\ref{ass:interior}, holding rivals' entry, and hence $Q_m$, fixed, offer $s>0$ to every hospital that enters at $t_R$. Then:
\begin{enumerate}\itemsep0.1em
\item[(i)] hospitals in $\mathcal R=\{L_i<F_i\le L_i+s\}\subset\mathcal S$ are recruited and enter at $t_R$;
\item[(ii)] hospitals in $\mathcal D=\{F^E_i-\tfrac{\beta}{1-\beta}s<F_i\le F^E_i\}\subset\mathcal E$ delay entry from stage~1 to $t_R$;
\item[(iii)] $\mathcal D$ is non-empty for every $s>0$, so no positive transfer is exactly incentive compatible;
\item[(iv)] to first order in $s$, the mass delayed is $\frac{\beta}{1-\beta}\,\frac{g(F^E_i)}{g(L_i)}$ times the mass recruited, and the TAVR cases lost in stage~1 from $\mathcal D$ exceed those gained in stage~2 from $\mathcal R$ when $v_{i,1}\,g(F^E_i)\,\beta\ge v_{i,2}\,g(L_i)\,(1-\beta)$.
\end{enumerate}
\end{proposition}

\begin{proof}
With the transfer, $V^L_i=\beta(L_i-F_i+s)$. A hospital in $\mathcal S$ enters iff $F_i\le L_i+s$, which is (i). A hospital in $\mathcal E$ keeps entering early iff $V^E_i\ge V^L_i$, i.e.\ $R_i-\ell_i-(1-\beta)F_i\ge\beta s$, i.e.\ $F_i\le F^E_i-\beta s/(1-\beta)$, which is (ii) and, since $g>0$, (iii). For (iv), the masses are $g(L_i)\,s$ and $g(F^E_i)\,\beta s/(1-\beta)$ to first order; each recruit adds $\beta v_{i,2}$ discounted cases and each delayed hospital loses $v_{i,1}$ stage-1 cases.
\end{proof}

Suppose $F_i=\varphi(x_i)+\nu_i$, with $\nu_i$ independent of $x_i$, mean zero and scale $\sigma_\nu$, and $\varphi(x_i)$ continuously distributed. A \emph{targeted} transfer is offered only to hospitals with $\varphi(x_i)>L_i$, those predicted never to enter.

\begin{proposition}[Targeting]\label{prop:targeting}
Under Assumption~\ref{ass:interior} and the targeted transfer, the delayed set is $\mathcal D\cap\{\varphi(x_i)>L_i\}$. Its mass converges to zero as $\sigma_\nu\to0$, for any $s$. In the limit, the personalized transfer $s(x_i)=\varphi(x_i)-L_i$ recruits all of $\mathcal S$ with no rent and no delay.
\end{proposition}

\begin{proof}
A delayed hospital has $F_i\le F^E_i\le L_i$ and $\varphi(x_i)>L_i$, so $\nu_i<F^E_i-\varphi(x_i)<0$ with $|\nu_i|>\varphi(x_i)-L_i>0$. As $\sigma_\nu\to0$ this event has vanishing probability because $\varphi(x_i)$ has no atom at $L_i$. At $\sigma_\nu=0$, $s(x_i)$ leaves every hospital in $\mathcal S$ exactly indifferent and is offered to no hospital outside it.
\end{proof}

\bibliographystyle{aer}
\bibliography{export-4.bib}
\end{document}